\documentclass[aps,prd,eqsecnum,11pt,showpacs,preprintnumbers,nofootinbib]{revtex4-1}
\pdfoutput=1
\def\nbox#1#2{\vcenter{\hrule \hbox{\vrule height#2in
\kern#1in \vrule} \hrule}}
\def\sq{\,\raise.5pt\hbox{$\nbox{.09}{.09}$}\,}
\def\sqb{\,\raise.5pt\hbox{$\overline{\nbox{.09}{.09}}$}\,}

\newcommand{\bea}{\begin{eqnarray}}
\newcommand{\eea}{\end{eqnarray}}
\newcommand{\be}{\begin{equation}}
\newcommand{\ee}{\end{equation}}
\newcommand{\bes}{\begin{subequations}}
\newcommand{\ees}{\end{subequations}}
\def\lag{\langle}
\def\rag{\rangle}

\def\nn{\nonumber \\}
\newcommand{\ups}{\upsilon}
\newcommand{\lam}{\lambda}
\newcommand{\sech}{\textnormal{sech}}

\usepackage{amssymb,amsmath}
\usepackage{graphicx}
\usepackage{graphics}
\usepackage{accents}
\newcommand*{\dt}[1]{%
  \accentset{\mbox{\large\bfseries .}}{#1}}
\newcommand*{\ddt}[1]{%
  \accentset{\mbox{\large\bfseries .\hspace{-0.25ex}.}}{#1}}
\newcommand*{\dddt}[1]{%
  \accentset{\mbox{\large\bfseries .\hspace{-0.25ex}.\hspace{-0.25ex}.}}{#1}}

\begin{document}

\preprint{LA-UR-13-26990}

\title{\hspace{1cm}  \\ \Large
Quantum Vacuum Instability of `Eternal' de Sitter Space\vspace{1cm}}

\author{Paul R. Anderson}
\affiliation{Department of Physics,\\
Wake Forest University, \\
Winston-Salem, NC 27109 USA}
\email{anderson@wfu.edu}
\author{Emil Mottola}
\affiliation{Theoretical Division, MS B285\\
Los Alamos National Laboratory\\ Los Alamos, NM 87545 USA}
\email{emil@lanl.gov\\}

\begin{abstract}

\vspace{5mm}\noindent
The Euclidean or Bunch-Davies $O(4,1)$ invariant `vacuum' state of quantum fields in global de Sitter
space is shown to be unstable to small perturbations, even for a massive free field with no self-interactions.
There are perturbations of this state with arbitrarily small energy density at early times that is exponentially
blueshifted in the contracting phase of `eternal' de Sitter space, and becomes large enough to
disturb the classical geometry through the semiclassical Einstein eqs. at later times. In the closely
analogous case of a constant, uniform electric field, a time symmetric state equivalent to the de Sitter
invariant one is constructed, which is also not a stable vacuum state under perturbations. The role of
a quantum anomaly in the growth of perturbations and symmetry breaking is emphasized in both cases.
In de Sitter space, the same results are obtained either directly from the renormalized stress tensor
of a massive scalar field, or for massless conformal fields of any spin, more directly from the effective
action and stress tensor associated with the conformal trace anomaly. The anomaly stress tensor shows
that states invariant under the $O(4)$ subgroup of the de Sitter group are also unstable to perturbations
of lower spatial symmetry, implying that both the $O(4,1)$ isometry group and its $O(4)$ subgroup are broken
by quantum fluctuations. Consequences of this result for cosmology and the problem of vacuum energy are
discussed.

\end{abstract}
\maketitle

\topmargin -1.5cm
\textheight 9in
\textwidth 6.8in

\section{`Vacuum' States in de Sitter Space}

The existence of a ground state as the state of lowest energy is fundamental to all quantum mechanical
systems. For quantum field theory (QFT) in flat Minkowski spacetime the vacuum state is defined
as the eigenstate of the Hamiltonian operator of the system with the lowest eigenvalue. The existence
of a Hamiltonian generator of time translational symmetry, with a non-negative eigenvalue spectrum,
bounded from below is crucial to the existence and determination of the vacuum ground state.

This definition of the vacuum in flat spacetime makes use of an essential property of the Poincar\'e group,
namely that positive and negative (particle and antiparticle) halves of the Hamiltonian spectrum do not mix,
remaining distinct under any of the continuous generators of the group. Hence the vacuum state
in flat space QFT is the same for all inertial frames related to each other by translations, rotations and
Lorentz boosts, and the vacuum enjoys complete invariance under Poincar\'e symmetry.

As is well known, none of these properties hold in a general curved spacetime, in time dependent background
fields, nor even in flat spacetime under general coordinate transformations which are not Poincar\'e
symmetries. In these circumstances the definitions of `vacuum' and `particles' become much more subtle.
Related to this, whereas the infinite zero point energy associated with the QFT vacuum may be
disregarded as unobservable in flat space QFT, the energy of the quantum vacuum in curved spacetime
cannot be neglected when the coupling to gravity is taken into account.

\begin{figure}
\begin{center}
\vspace{-7mm}
\includegraphics[height=6cm,width=6cm]{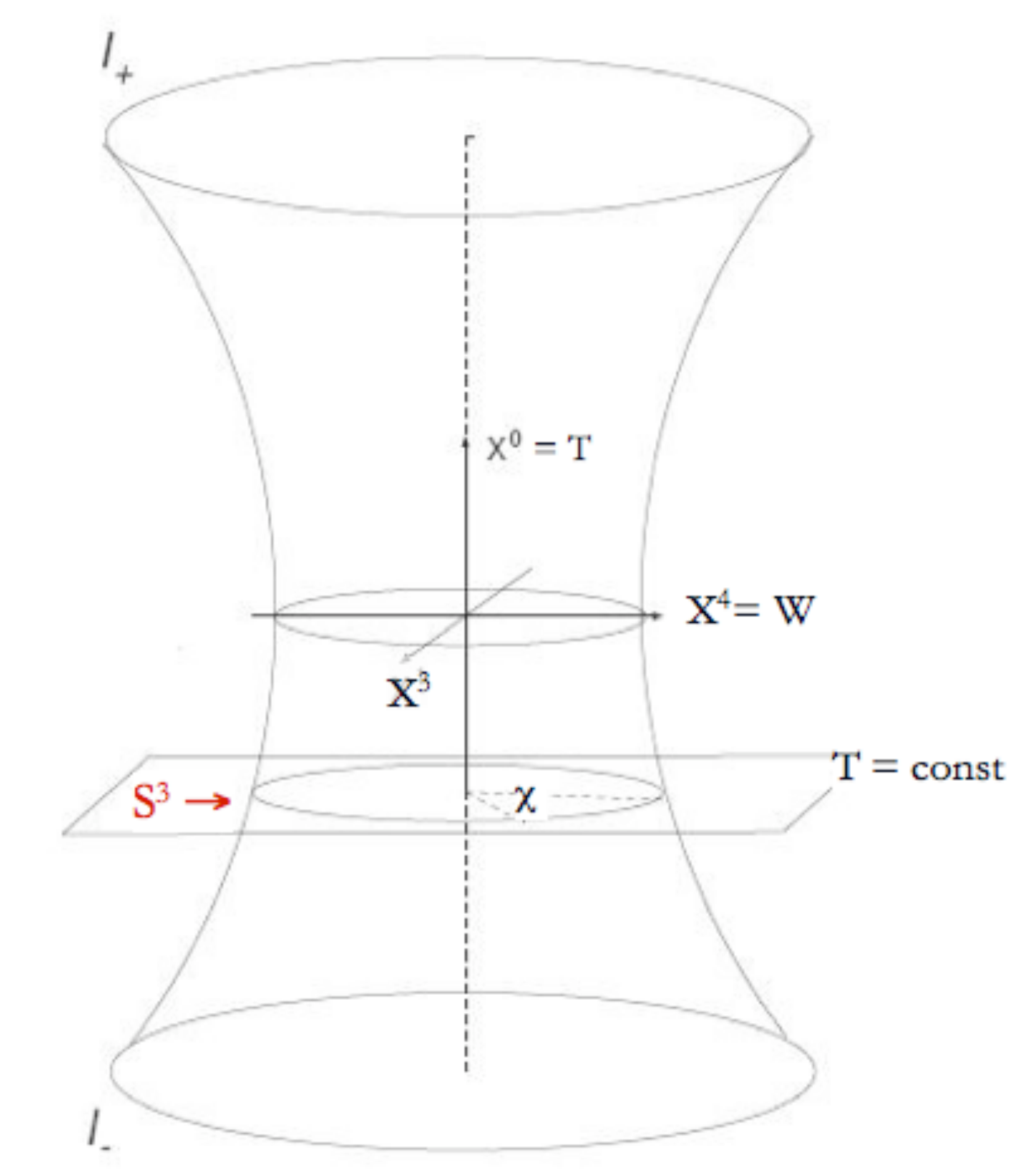}
\caption{The de Sitter manifold represented as a single sheeted hyperboloid of revolution
about the $T=X^0$ axis, embedded in five dimensional flat spacetime $(X^0, X^a), \,a= 1,\dots 4$, in which the
$X^1$, $X^2$ coordinates are suppressed. The hypersurfaces at constant $T=X^0= H^{-1} \sinh u$ are
three-spheres, $\mathbb{S}^3$. The $\mathbb{S}^3$ at $T= \pm \infty$ are denoted by $I_{\pm}$.}
\label{Fig:deShyper}
\end{center}
\vspace{-7mm}
\end{figure}

These issues come to the fore in the important special case of de Sitter space, the classical spacetime
with a positive cosmological constant $\Lambda > 0$, which itself may be regarded as a vacuum
energy density uniformly curving space. The geodesically complete full de Sitter manifold may
be represented as a single sheeted hyperboloid of revolution embedded in five dimensional flat
Minkowski spacetime, {\it c.f.} Fig. \ref{Fig:deShyper} \cite{HawEll}. It has the isometry group $O(4,1)$
with $10$ continuous symmetry generators, the same number as the Poincar\'e group of Minkowski space,
and the maximal number possible for any solution of the vacuum Einstein field equations in $4$ spacetime
dimensions. This maximal symmetry is evident from the constant and uniform Riemann and Ricci curvature
tensors and scalar of de Sitter space, which are respectively
\vspace{-2mm}
\bes
\bea
&&\hspace{-6mm}R^{ab}_{\ \ cd} = H^2 \left(\delta^a_{\ c}\,\delta^b_{\ d} - \delta^a_{\ d}\,\delta^b_{\ c}\right),\\
&&R^a_{\ b} = 3 H^2\, \delta^a_{\ b} = \Lambda\, \delta^a_{\ b}\,,\\
&& \hspace{3mm}R = 12 H^2 = 4\Lambda\,.
\eea
\label{curvdS}\ees

\vspace{-9mm}\noindent
A natural attempt to generalize the QFT vacuum of flat space to de Sitter space makes use of this geometrical
symmetry of de Sitter space to define the de Sitter invariant `vacuum' $\vert \ups\rag$ as the state possessing
the same maximal $O(4,1)$ symmetry in the Hilbert space of states. Introduced by Chernikov and Tagirov (CT) \cite{CherTag},
this state is commonly known also as the Bunch-Davies (BD) state \cite{BunDav,BirDav}, or the Euclidean `vacuum,'
because its Green's functions are those obtained by analytic continuation from the Euclidean ${\mathbb S}^4$,
at least for massive fields where no obvious infrared issues arise \cite{DowCrit}.

It is important to recognize that unlike in flat space, the construction of the CTBD state is not
based on diagonalization of any Hamiltonian nor any minimization of energy. In fact no suitable Hamiltonian
operator with a spectrum bounded from below exists at all in de Sitter space, even for free QFT. In the
globally complete coordinates of the de Sitter hyperboloid
\be
ds^2 = H^{-2} \left( -du^2 + \cosh^2\!u\,  d\Sigma^2 \right)
\label{hypermet}
\vspace{-4mm}
\ee
with \vspace{-1mm}
\be
d\Sigma^2 \equiv d\hat N\cdot d\hat N = d\chi^2 + \sin^2\!\chi  \, d\hat n \cdot  d\hat n
= d\chi^2 + \sin^2\!\chi\, (d\theta^2 + \sin^2\!\theta\, d\phi^2)
\label{S3}
\ee
the standard round metric on ${\mathbb S}^3$, the de Sitter metric is dependent on the time $u$. Thus
translation in $u$ is not a symmetry of de Sitter space and the generator of $u$ time translations is not
conserved. As a consequence, the `vacuum' defined by Hamiltonian diagonalization at one instant of $u$ time
will contain `particles' at any other $u$ time. This is equally true in the flat spatial slicing of de Sitter space
\be
ds^2 = -d\tau^2 + e^{2H\tau} \, d\vec x\cdot d\vec x
\label{flatFRW}
\ee
used most frequently in cosmology, which is similarly dependent on the time $\tau$.

\begin{figure}
\begin{center}
\vspace{-7mm}
\includegraphics[height=8cm,width=8cm]{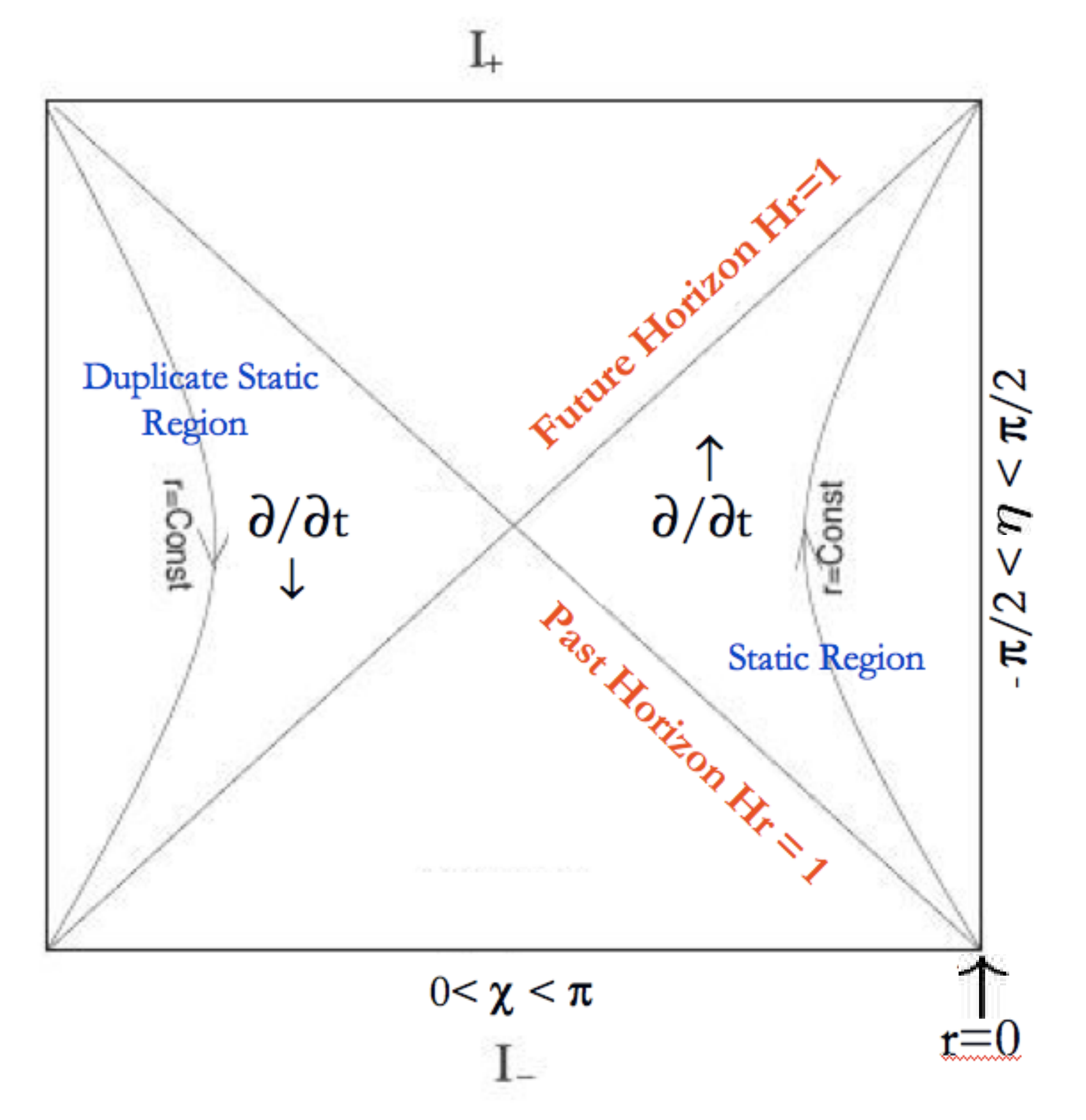}
\caption{The Carter-Penrose conformal diagram for de Sitter space, in which light rays emanating
from any point are at $45^{\circ}$, and the angular coordinates $\theta,\phi$ are suppressed.
The quarter of the diagram labeled as the static region are covered by the static coordinates of (\ref{static}).
The orbits of the static time Killing field $\partial/\partial t$, $r=0$ and curves of constant $r> 0$ are shown.
In contrast, the surfaces of constant $u$ time coordinate of (\ref{hypermet}) are horizontal straight
lines across the diagram with $\chi \in [0,\pi]$, labelled on the right by the conformal time coordinate
$\eta \equiv \sin^{-1} ({\rm tanh}\,u) \in (-\frac{\pi}{2}, \frac{\pi}{2})$ as $u$ ranges from $-\infty$ to $+\infty$
at past ($I_-$) and future ($I_+$) infinity respectively.}
\label{Fig:dSCarPen}
\end{center}
\vspace{-8mm}
\end{figure}

The non-existence of a conserved Hamiltonian generator bounded from below in de Sitter space
is a consequence of the de Sitter symmetry group $O(4,1)$ itself. Unlike the Poincar\'e group, no
invariant separation into positive and negative, particle and antiparticle subspaces exists in de Sitter
space, and any de Sitter symmetry generator chosen for the role of the Hamiltonian has a spectrum of both
positive and negative eigenvalues which are mixed by the action of other generators of the group \cite{Nacht}.
One of the $4$ non-compact Lorentz boost generators of the $O(4,1)$ symmetry group may
be selected (arbitrarily) as the Hamiltonian of the system, generating time translations
$t \rightarrow t + \Delta t$ in the static coordinates of de Sitter space, where the line element takes the form
\be
ds^2 = -(1-H^2r^2)\, dt^2 + \frac{dr^2}{1-H^2r^2} + r^2 (d\theta^2 + \sin^2\!\theta\, d\phi^2)\,.
\label{static}
\ee
In these coordinates the geometry is independent of the time $t$. However the event horizon
at $r = H^{-1}$ relative to the origin $r=0$ is now manifest, and the static coordinates cover only one quarter
of the full de Sitter manifold. The Killing symmetry $\partial/\partial t$ is not globally timelike, and changes its
orientation from one quadrant to another, as may be seen from the Carter-Penrose conformal diagram
of de Sitter space: Fig. \ref{Fig:dSCarPen}. A direct consequence of this is that the corresponding
Hamiltonian symmetry generator across any complete Cauchy surface is not positive definite, but
rather unbounded from below, as Lorentz boosts are. Hence its eigenstates or expectation values
cannot be used to select a global minimum energy vacuum state. The choice of $\partial/\partial t$
is also arbitrary and the separation into positive and negative energies with respect to $\partial/\partial t$
is non-invariant under de Sitter group transformations. The particle concept is likewise affected, as the
CTBD de Sitter invariant `vacuum' state $\vert \ups\rag$ is actually a state with a thermal distribution
of `particles' with respect to the Killing Hamiltonian generator $\partial_t$ of (\ref{static}) with the
Hawking de Sitter temperature \cite{GibHawLap}
\vspace{-1mm}
\be
T_{_H} = \frac{\hbar H}{\,2 \pi k_{_B}}
\label{HawkdS}
\ee
and in that sense is not a vacuum state at all.

The horizon and causal structure of de Sitter space raises questions of how a vacuum state can
be prepared operationally even in principle. Inspection of the conformal diagram in Fig.\,\ref{Fig:dSCarPen} 
shows that points on a Cauchy surface at $u_0 <0$ with widely different $\hat N$ (for example at $\chi =0$ 
and its antipodal point $\chi=\pi$) could never have been linked by any causal signal in the past. As the initial
time $u_0$ is taken earlier and earlier, this causal disconnection affects more and more of the initial
$u=u_0$ Cauchy surface. Since past infinity $I_-$ is {\it spacelike}, as $u_0 \rightarrow -\infty$ in this
limit no two different points on ${\mathbb S}^3$ could have been in any causal contact whatsoever.
Thus any global initial data on ${\mathbb S}^3$, including that necessary to construct the CTBD state
$\vert \ups\rag$ cannot have been provided at any initial time $u_0 < 0$ by any causal process
within de Sitter space itself. Instead initial data has simply to be posited over the full spacelike
${\mathbb S}^3$, at points outside of the causal horizon of any local agent who might have prepared
it at early times. This is equivalent to the existence of a particle horizon \cite{HawEll}, and is completely 
unlike that of flat Minkowski spacetime, where Cauchy data on a fixed time slice $t= const.$ may be 
prepared in principle by transmitting signals causally from a single point early enough in the past.
The existence of horizons and the absence of a global time coordinate connected with any symmetry
reveal the essential difficulties with defining a global vacuum state for QFT in eternal de Sitter space.

They also imply that the mathematical requirement of global de Sitter invariance cannot be realized 
by any local physics within de Sitter space itself, requiring instead an acausal fine tuning of initial data at 
spacelike past infinity $I_-$, with a view to the entire future manifold, which is presumed to be known in advance 
in order to specify a globally $O(4,1)$ invariant state. Although maximal $O(4,1)$ symmetry may seem natural 
mathematically, or by analytic continuation from the Euclidean ${\mathbb S}^4$ where no causal relations 
apply, it is quite unnatural with respect to the physical principles of locality and causality in real time, as well 
as lacking any de Sitter invariant Hamiltonian minimization principle.

For these reasons it is important to study the sensitivity of physical quantities in de Sitter space
to fluctuations and/or perturbations of states away from precisely the `right'  one for global $O(4,1)$
invariance, rather than simply assuming this symmetry. The calculation of the imaginary part of the
effective action of a simple scalar QFT in de Sitter space due to particle creation \cite{PartCreatdS,AndMot1}
already shows that de Sitter space is unstable to spontaneous creation of particle pairs from the
vacuum, just as is an `eternal' uniform electric field ${\bf E} = E\,{\bf \hat z}$ permeating all of space \cite{Schw}.
This electric field analogy and the close relation between fluctuations and dissipation in any causal
theory suggests that the `shorting of the vacuum' should result in the classical energy of de Sitter space
converting itself into standard matter and radiation, thus providing a route to a dynamical solution to the
cosmological `constant' problem  \cite{Fluc,NJP}.

As in electrodynamics, interactions in de Sitter space
are certainly relevant to understanding of the detailed evolution and final state, particularly
since spontaneous pair creation should be accompanied by induced emission processes which can create
an avalanche of particles that will inevitably interact and thermalize, leading to the final dissipation of vacuum
energy into matter and radiation. A fuller understanding of these non-equilibrium processes may well lead
to a satisfactory resolution of the cosmological `constant' problem, and be relevant to observational cosmology
through the residual dark energy in the present epoch \cite{Fluc,NJP}. However, this dynamics has not been
fully solved in in four dimensions even in flat space electrodynamics.  Moreover a number of 
questions persist about QFT in de Sitter space, even in the non-interacting case, and these should be settled 
definitively first, because they depend through the energy-momentum-stress tensor $T_{ab}$ only upon the 
universal coupling to the gravitational field, independently of any matter self-interactions.

In this paper we study the behavior of the renormalized energy-momentum tensor $\lag T_{ab}\rag$ of
QFT under perturbations of the $\vert \ups\rag$ state to nearby states of lower symmetry. In the expanding part
of de Sitter space $u>0$ of (\ref{hypermet}) or in the Poincar\'e coordinates of flat spatial sections (\ref{flatFRW}),
it has been shown that in a fixed de Sitter background, the expectation value $\lag T_{ab}\rag$
for a scalar field with effective mass $M^2 = m^2 + \xi R > 0$ approaches an $O(4,1)$ de Sitter invariant value 
at late times, for all {\it spatially homogeneous} UV allowed perturbations \cite{Attract}. Physically this 
result may seem intuitively obvious, since all deviations from the expectation value in de Sitter invariant state 
$\vert \ups\rag$ are redshifted in the de Sitter expansion and vanish in the $u\rightarrow \infty$ limit.  
Since global or `eternal' de Sitter space is time reversal invariant, the attractor behavior in the expanding phase
implies just the opposite behavior under time reversal in the contracting phase. That is, very small changes in the
initial state in the very distant past $u_0 \rightarrow -\infty$ of eternal de Sitter space with initially very
small $\lag T_{ab}\rag$ must necessarily produce larger and larger effects in $\lag T_{ab}\rag$ as the
contraction proceeds towards $u=0$. This is just the case where the aforementioned issues with causality
at spacelike $I_-$ arise, and this sensitivity to initial conditions at $I_-$ is the source of the instability.

By studying the general behavior of the renormalized 
$\lag T_{ab}\rag$ in states with lower symmetry, we show in this paper that the CTBD de Sitter 
invariant state $\vert \ups\rag$ is unstable, in the sense that there is a large class of initial state perturbations 
which have exponentially small energy density in the infinite past $u_0 \rightarrow - \infty$ but which grow 
large enough through exponential blueshifting proportional to $a^{-4}$, where $a= H^{-1} \cosh u$ is the 
scale factor in (\ref{hypermet}), to exceed the classical background energy $\Lambda/8\pi G$ and hence 
significantly disturb the de Sitter geometry at $u=0$. In fact, there are such states with $\lag T_{ab}\rag$ 
larger than {\it any} finite value at $u=0$.

This extreme sensitivity to initial conditions as $u_0 \rightarrow -\infty$ implies that $O(4,1)$ de Sitter
invariance is broken, and the spacetime will generally depart from de Sitter space when the backreaction
of $\lag T_{ab}\rag$  of any matter or radiation on the geometry is taken into account, through
the semiclassical Einstein eqs.,
\be
R^a_{\ b} - \frac{R}{2}\, \delta^a_{\ b} + \Lambda\, \delta^a_{\ b} = 8\pi G \,\lag T^a_{\ b}\rag_{_R}
\label{scE}
\ee
and quite apart from any matter self-interactions or higher loop effects. Although in a fixed de Sitter
background the energy density of spatially homogeneous perturbations will begin to decrease again 
for $u>0$, perturbations of the CTBD state and their backreaction through (\ref{scE}) will have already 
drastically altered the geometry in the contracting phase and broken the de Sitter symmetry by $u=0$,
rendering further evolution ignoring backreaction moot. This large backreaction of the energy-momentum 
tensor for perturbations of the CTBD state is independent of any definition of particles.

Although for definiteness we study this growth of  $\lag T_{ab}\rag$ explicitly in a scalar field theory,
the result is clearly much more general. A very useful tool for characterizing the behavior of the stress tensor
in any coordinates is the one-loop effective action of the trace anomaly and the stress tensor derived
from it \cite{MazMotWZ,MotVau,Zak}. The non-local form of this effective action, {\it c.f.} (\ref{Sanom})
already indicates infrared de Sitter breaking effects, and sensitivity to initial and/or boundary conditions
for conformal QFT's of any spin. The corresponding stress tensor may be found in closed form in de Sitter space in any
coordinates by solving a classical, linear eq. (\ref{phieq}) for a scalar condensate effective field, whose solutions
necessarily break de Sitter invariance, and allow wide classes of initial state perturbations for fields of any spin
to be surveyed at once. Because de Sitter space is conformally flat, this anomaly stress tensor is a complete
description of the full QFT stress tensor for conformal fields linearized around the CTBD state $|\ups\rag$ at
all length scales much larger than the Planck length $L_{Pl}$, where semiclassical methods should apply \cite{DSAnom}.

The $a^{-4}$ blueshifting of the energy density of even massive fields to eventually ultrarelativistic behavior
shows that the conformal anomaly stress tensor is relevant for long time evolution even if the underlying QFT
is not conformally invariant. When the scalar perturbations are spatially {\it in}homogeneous new effects may 
also become apparent. In Ref. \cite{DSAnom} we studied spatially inhomgeneous scalar perturbations in 
linear response of conformal QFT's around de Sitter space and found a class of gauge invariant perturbations, 
which do {\it not} redshift away but instead give diverging energy-momentum components at $r=H^{-1}$
in static coordinates (\ref{static}). These may be interpreted as fluctuations in the Hawking de Sitter temperature 
(\ref{HawkdS}) at the de Sitter horizon with respect to some arbitrary but fixed choice of origin, and clearly
respect only rotational $O(3)$ invariance around $r=0$ and static time $t$ translational invariance.
This result suggests that fluctuations on the horizon scale $H^{-1}$ may produce significant backreaction in
de Sitter space, and that the $O(4,1)$ symmetry is unstable to such spatially inhomogeneous scalar
fluctuations in the Hawking de Sitter temperature \cite{MottT}. Tensor perturbations have been 
studied recently in \cite{Fordetal}.

Using the anomaly form of $T_{ab}$ we shall show that there is even greater
sensitivity to spatially inhomogeneous non-$O(4)$ invariant initial data in the distant past $u_0 \rightarrow -\infty$
of global de Sitter space, so that $O(4)$ invariance is broken as well as the full $O(4,1)$ de Sitter invariance.
This strongly suggests that spatial inhomogeneities are more important in QFT in de Sitter space than previously
suspected, supporting the results of \cite{DSAnom}. Such spatially inhomogeneous perturbations clearly
are relevant even in the expanding Poincar\'e patch (\ref{flatFRW}). The interesting questions of the behavior 
of the stress tensor  in states of lower symmetry, such as the $O(3)$ symmetry evident in static coordinates 
(\ref{static}), and consequences for spatially inhomogeneous cosmologies will be taken up in future publications. 
An accompanying and closely related paper gives a fuller treatment of the instability of global de Sitter space 
to particle creation \cite{AndMot1}.

The paper is organized as follows. In the next section we construct the time symmetric invariant state analogous
to the CTBD state in de Sitter space, in the case of a uniform, constant electric field background ${\bf E} = E\, {\bf \hat z}$,
and show that it also is unstable to perturbations for which the mean current $\lag j_z\rag$ grows with time.
This growth of the current and breaking of background symmetries can be understood by consideration
of a quantum anomaly, in this case the chiral anomaly of massless fields in two spacetime dimensions.
The reader interested primarily in de Sitter space proper may skip this section upon first reading and
proceed directly to Sec. \ref{Sec:O4dS} where we begin discussion of the CTBD state and general states of
$O(4)$ symmetry in de Sitter space. In Sec. \ref{Sec:SET} we construct the renormalized expectation
value of the stress tensor of a massive scalar field with conformal coupling $\xi = \frac{1}{6}$ in general
$O(4)$ invariant states in the global hyperboloid coordinates (\ref{hypermet}) of de Sitter space,
and explicitly exhibit the class of states with large backreaction at $u=0$. In Sec. \ref{Sec:Anom} we
consider the effective action and stress tensor associated with the trace anomaly of conformal fields
in de Sitter space and show how the strong infrared effects, sensitivity to initial conditions, and  breaking
of de Sitter symmetry is inherent in the conformal anomaly for QFT's of any spin.
In Sec. \ref{Sec:LowerSym} we extend the analysis of the anomaly stress tensor to states of lower
than $O(4)$ symmetry, showing that these spatially inhomogeneous perturbations grow even more
rapidly to larger values at $u=0$ than $O(4)$ symmetric states. Sec. \ref{Sec:Concl} contains
our conclusions and a discussion of their possible consequences for cosmology and the 
problem of cosmological vacuum energy.

\section{Constant Uniform Electric Field: Invariant State and Instability}
\label{Sec:ConstantE}

\subsection{Time Symmetric Invariant State}

The example of a charged quantum field in the background of a constant uniform
electric field has many similarities with the de Sitter case. Although this problem has
been considered by many authors \cite{Schw,Nar,Nik,NarNik,FradGitShv,KESCM,GavGit,QVlas},
the existence of a time symmetric state analogous to the CTBD state in de Sitter space
does not appear to have received previous attention, and is particularly relevant to
our study of vacuum states in de Sitter space, so we consider this case first in some detail.

Analogous to choosing global time dependent coordinates (\ref{hypermet}) in de Sitter space,
one may choose the time dependent gauge
\be
A_z = -Et\,,\qquad A_t=A_x=A_y = 0
\label{Egauge}
\ee
in which to describe a fixed constant and uniform electric field ${\bf E} = E\, \bf \hat z$ in the $z$ direction.
Treating the electric field as a classical background field analogous to the classical gravitational field of
de Sitter space,  the wave equation of a non-self-interacting complex scalar field ${\bf \Phi}$ is
\be
\left[-(\partial_\mu -i eA_\mu) (\partial^\mu-i eA^\mu) + m^2\right]{\bf \Phi} = 0
\label{eomE}
\ee
in the classical electromagnetic potential (\ref{Egauge}).

The solutions of (\ref{eomE}) may be decomposed into Fourier modes
${\bf \Phi} \sim e^{i {\bf k\cdot x}} f_{\bf k}(t)$ with
\be
\left[ \frac{d^2}{dt^2} + \omega_{\bf k}^2(t)\right] f_{\bf k}(t) = 0
\label{Emodeq}
\ee
where the frequency function $\omega_{\bf k}(t)$ is defined by
\be
\omega_{\bf k}(t) \equiv \left[ (k_z+eEt)^2 + k_{\perp}^2 + m^2\right]^{\frac{1}{2}}
= \sqrt{2eE}\,\sqrt{\frac{u^2}{4\,} + \lam}\,.
\label{OmegaE}
\ee
We have defined the dimensionless variables
\be
u \equiv \sqrt{\frac{2}{eE}} \ (k_z + eEt)\,,\qquad \lam \equiv \frac{k_{\perp}^2 + m^2}{2\,eE} >  0
\label{ulamdef}
\ee
and chosen the sign of $eE$ to be positive without loss of generality. With $f_{\bf k}(t) \rightarrow  f_{\lam}(u)$,
the dimensionless mode eq. (\ref{Emodeq}) becomes 
\be
\left[ \frac{d^2}{du^2} + \frac{u^2}{4\,} + \lam \right] f_{\lam}(u) = 0
\label{yEmode}
\ee
the solutions of which may be expressed in terms of confluent hypergeometric functions $_1F_1(a;c;z)$
or parabolic cylinder functions $D_{\nu}$ \cite{Bate}.

Since (\ref{yEmode}) is real and symmetric under $u \rightarrow -u$, it is clear that its real solutions
can be classified into those which are either even and odd under this discrete reflection symmetry. Let us define
two fundamental real solutions of (\ref{yEmode}) $f_{\lam}^{(i)}(u), i=0,1$ by the conditions
\bes
\bea
&&[f_{\lam}^{(i)}(u)]^{\ast} = f_{\lam}^{(i)}(u)\,,\qquad i = 0,1 \hspace{1cm}  {\rm (real)}\\
&&f_{\lam}^{(0)} (u) = f_{\lam}^{(0)} (-u) \hspace{3.1cm}{\rm (even)}\\
&&f_{\lam}^{(1)} (u) = -f_{\lam}^{(1)} (-u) \hspace{2.85cm}{\rm (odd)}\,,
\eea
\ees
which are even or odd respectively, and which satisfy the initial data
\bes
\bea
&&f_{\lam}^{(0)} (0) = 1\,,\qquad f_{\lam}^{(0)\,\prime} (0) = 0\,,\\
&&f_{\lam}^{(1)} (0) = 0\,,\qquad f_{\lam}^{(1)\,\prime} (0) = 1\,
\eea
\label{uzero}\ees
at $u=0$, where the primes denote differentiation with respect to $u$. These fundamental real solutions
of (\ref{yEmode}) are most concisely expressed in terms of the confluent hypergeometric (Kummer) function
\be
\Phi(a,c;z) \equiv \,_1\!F_1(a;c;z) = \sum_{n=0}^{\infty} \frac{(a)_n}{(c)_n}\, \frac{z^n}{n!}\,,\qquad
(a)_n \equiv \frac{\Gamma(a+n)}{\Gamma(a)} \,,
\label{Phidef}
\ee
which has the integral representation \cite{Bate}
\be
\Phi(a,c;z) = \frac{\Gamma(c)}{\Gamma(a) \Gamma(c-a)} \int_0^1dx \, e^{xz}  \, x^{a-1} \, (1-x)^{c-a-1}
\,,\qquad {\rm Re}\, c > {\rm Re}\, a > 0
\ee
in the form
\bes
\bea
&&f_{\lam}^{(0)} (u) = e^{-\frac{iu^2\!\!}{4}}\, \Phi\left( \frac{1}{4} + \frac{i\lam}{2}, \frac{1}{2}; \frac{iu^2}{2}\right)
= e^{\frac{iu^2\!\!}{4}}\, \Phi\left( \frac{1}{4} - \frac{i\lam}{2}, \frac{1}{2}; -\frac{iu^2}{2}\right)\,,\\
&&f_{\lam}^{(1)} (u) = u\, e^{-\frac{iu^2\!\!}{4}}\, \Phi\left( \frac{3}{4} + \frac{i\lam}{2}, \frac{3}{2}; \frac{iu^2}{2}\right)
= u\, e^{\frac{iu^2\!\!}{4}}\, \Phi\left( \frac{3}{4} - \frac{i\lam}{2}, \frac{3}{2}; -\frac{iu^2}{2}\right)
\eea
\label{confly0y1}\ees
are clearly even or odd respectively, real by the Kummer transformation which yields the second
forms in (\ref{confly0y1}), and satisfy the initial data (\ref{uzero}).

It is also possible to express these fundamental real solutions $f^{(i)}_{\lam}$ as linear combinations
of parabolic cylinder functions $D_{\nu}$ in the forms \cite{Bate}
\bes
\bea
&&f_{\lam}^{(0)} (u) =2^{\frac{i\lam}{2} - \frac{3}{4}}\,
\frac{\Gamma\left(\frac{3}{4} + \frac{i\lam}{2}\right)}{\sqrt\pi}
\left[ D_{- \frac{1}{2} - i\lam}(e^{\frac{i\pi}{4}} u) +
D_{- \frac{1}{2} - i\lam}(-e^{\frac{i\pi}{4}} u)\right]\,,\\
&&f_{\lam}^{(1)} (u) = 2^{\frac{i\lam}{2} - \frac{5}{4}}\,
e^{-\frac{i\pi}{4}}\, \frac{\Gamma\left(\frac{1}{4} + \frac{i\lam}{2}\right)}{\sqrt\pi}
\left[ - D_{- \frac{1}{2} - i\lam}(e^{\frac{i\pi}{4}} u) +
D_{- \frac{1}{2} - i\lam}(-e^{\frac{i\pi}{4}} u)\right]  \;.
\eea
\label{y0y1}\ees
which representations are useful for identifying their relationship to the {\it in} and {\it out} positive
frequency scattering solutions defined as $u \rightarrow \mp \infty$ respectively in \cite{Nar,Nik,NarNik,AndMot1}.

From the fundamental real solutions (\ref{uzero})-(\ref{confly0y1}) one can construct the complex mode functions
\bea
\hspace{-1cm} \ups_{\lam}(u) &\equiv& (8eE\lam)^{-\frac{1}{4}} \left[ f_{\lam}^{(0)}(u)
- i \lam^{\frac{1}{2}} f_{\lam}^{(1)}(u)\right]\nn
&= &2^{-\frac{1}{2}}(k_{\perp}^2 + m^2)^{-\frac{1}{4}}\, e^{-\frac{iu^2\!\!}{4}}\,
\left[\Phi\left( \frac{1}{4} + \frac{i\lam}{2}, \frac{1}{2}; \frac{iu^2}{2}\right)
- i \lam^{\frac{1}{2}}\,u\, \Phi\left( \frac{3}{4} + \frac{i\lam}{2}, \frac{3}{2}; \frac{iu^2}{2}\right)\right]
\label{Esymodes}
\eea
which are normalized according to the Wronskian condition
\be
i\left(\ups_{\lam}^*\frac{d}{dt}\ups_{\lam} - \ups_{\lam} \frac{d}{dt} \ups_{\lam}^*\right) = 1
\label{WronE}
\ee
and which satisfy the time reversal conjugation property
\be
\ups_{\lam}^*(u) = \ups_{\lam}(-u)\,.
\label{timerev}
\ee
These $\ups_{\lam}$ mode functions satisfy the initial data
\be
\ups_{\lam} (0)= 2^{-\frac{1}{2}}(k_{\perp}^2 + m^2)^{-\frac{1}{4}} = \frac{1}{\sqrt{2 \omega_{\bf k}}}\Big\vert_{u=0} \,,\qquad
\frac{d\ups_{\lam} }{dt} \Big\vert_{u=0} = -\frac{i}{\sqrt 2}\, (k_\perp^2 + m^2)^{\frac{1}{4}} = -i \,\omega_{\bf k}\, \ups_{\lam}(u)
\label{upslaminit}
\ee
which coincides with the definition of the lowest order adiabatic frequency mode functions at the symmetric 
point $u=0$. The solution of (\ref{yEmode}) satisfying conditions (\ref{WronE})-(\ref{upslaminit}) 
is unique. Because of relations (\ref{y0y1}) and the simple asymptotic forms of the $D_{\nu}$ functions, the 
symmetric mode function $\ups_{\lam}$ is a coherent superposition of positive and negative frequency  
(particle and anti-particle) solutions as $u \rightarrow \pm \infty$, just as the CTBD mode
function is in de Sitter space \cite{AndMot1}.

The existence of such a time reversal invariant solution to (\ref{yEmode}) is related to the existence
of a maximally symmetric state constructed along the lines of the maximally $O(4,1)$ invariant
CTBD state in the de Sitter background. If the charged quantum field $\bf \Phi$ is
expanded in terms of these symmetric basis functions in a finite volume $V$
\be
{\bf \Phi} (t, {\bf x}) = \frac{1}{\sqrt{V}}\sum_{\bf k}
\left[ a_{\bf k}^{\ups}\,\ups_{\lam}(u)\, e^{i {\bf k} \cdot {\bf x}}+  b_{\bf k}^{\ups\,\dagger}\,\ups^*_{\lam}(u)\,
e^{-i {\bf k} \cdot {\bf x}}\right]\,,
\label{Phiopcharged}
\ee
with $u$ and $\lam$ defined by (\ref{ulamdef}), then a symmetric state
$\vert \ups \rag$ in a background constant uniform electric field may be defined by
\be
a_{\bf k}^{\ups} |\ups\rag = b_{\bf k}^{\ups}|\ups\rag = 0\,.
\label{Eannihops}
\ee
The symmetry in this case is isomorphic to the full Poincar\'e symmetry group of zero electric field in flat
Minkowski space. This is due to the remarkable fact that a canonical transformation exists that transforms the
algebra of position and momentum operators, $x^{\mu}$ and $p_{\nu} = -i \partial/\partial x^{\nu}$,
in a constant, uniform $\bf E$ field background to new position and momentum operators, $X^{\mu}$
and $P_{\nu} = - i\partial/\partial X^{\nu}$, such that the Klein-Gordon operator (\ref{eomE})
\bea
&&-(\partial_\mu -i eA_\mu) (\partial^\mu-i eA^\mu) + m^2
= -p_t^2 + (p_z +eEt)^2 + p_x^2 + p_y^2 +m^2 \nn
&& \hspace{2cm} = -P_T^2 + P_Z^2 + P_X^2 + P_Y^2 + m^2 =0
\label{KGeom}
\eea
(with $P_X= p_x, P_Y= p_y$ and $P_Z= p_z$) becomes that of flat space with zero field \cite{BeersNick}.
The existence of this transformation and symmetry may be less surprising when it is recognized that there
are two quantities
\bes
\bea
&&P_T = (p_t^2 -2eEt - e^2E^2t^2)^{\frac{1}{2}} = (p_z^2 + p_x^2 + p_y^2 +m^2)^{\frac{1}{2}}\\
&&\hspace{1cm}TP_Z + ZP_T = \frac{P_T}{eE} \ (eEz + p_t - P_T)
\eea
\label{symgen}\ees
that are conserved by virtue of the eq. of motion (\ref{KGeom}), and (together with $P_Z=p_z$ which generates
space $Z$ translations) they generate $T$ time translations and Lorentz boosts in the $Z$ direction in the
transformed $(T, Z)$ coordinates. This dynamical maximal Poincar\'e symmetry in the constant, uniform $\bf E$ field
is analogous to the maximal $O(4,1)$ point symmetry group of de Sitter space. In each case the existence of a  
maximally symmetric state $|\ups\rag$ which enjoys the full symmetries of the background follows.

The expectation value of the electric current operator is given in the symmetric state $|\ups\rag$ by
\be
\lag \ups |j_z |\ups\rag_{_R} = 2e \int \frac{d^3 \bf k}{(2\pi)^3} (k_z +eEt)
 \left[|\ups_{\lam}(u)|^2 - \frac{1}{2\omega_{\bf k}(t)}\right]
\label{jvac}
\ee
where the second term is the lowest order adiabatic vacuum subtraction sufficient for the constant $E$ field
background \cite{QVlas}. Actually by changing integration variables from $k_z$ to $u$ and using the fact that both
$|\ups_{\lam}(u)|^2$ and $\omega_{\bf k}(t)$ are even functions of $u$, it is clear that both terms in the integrand of
(\ref{jvac}) are odd under $u \rightarrow -u$ and thus give vanishing contributions if integrated symmetrically in $u$.
Hence as a consequence of time reversal invariance (or charge conjugation symmetry), the symmetric
state $\vert \ups\rag$ has exactly zero electric current expectation value
\vspace{-2mm}
\be
\lag \ups| \,j_z \,| \ups \rag_{_R} = \lag \ups|\, {\bf j}_{\perp}\, | \ups \rag_{_R} = 0
\label{vac0}
\ee
at all times, by the symmetry of this state.  Likewise the mean charge density 
$ \lag \ups|\, \rho \,| \ups \rag_{_R}$ vanishes in this charge symmetric state. Thus the state 
$|\ups\rag$ defined by (\ref{Esymodes})-(\ref{Eannihops}) in a constant,
uniform electric field background is an exact self-consistent solution of the semiclassical Maxwell eqs.
\bes
\begin{align}
&\hspace{7cm}\raisebox {3mm}{${\bf \nabla \cdot E} = \lag \ups| \,\rho\, | \ups \rag_{_R} = 0$}\rule{7cm}{0pt}\raisetag{11mm} \\
&\hspace{6.4cm}{\bf \nabla \times B} - \frac{\partial {\bf E}}{\partial t} =  \lag\ups|\, {\bf j}\, |\ups\rag_{_R} = 0 \label{Max}
\end{align}
\ees
with both sides vanishing identically. This is analogous to the maximally $O(4,1)$ symmetric and time reversal
invariant CTBD state $|\ups\rag$ which satisfies the semiclassical Einstein eqs. (\ref{scE}) in de Sitter space
with a simple redefinition of $\Lambda$, since $\lag \ups |T^a_{\ b}|\ups\rag_R = -\varepsilon_{\ups} \, \delta^a_{\ b}$\,,
{\it c.f.} Sec. \ref{Sec:SET}.

\subsection{Instability of the Maximally Symmetric State: Electric Current}

The existence of a state of maximal symmetry does not imply that it is the stable ground state of either
the de Sitter or electric field backgrounds. In the electric field case the imaginary part of the effective action
and spontaneous decay rate of the electric field into particle/anti-particle pairs was first calculated by Schwinger \cite{Schw}.
By time reversal invariance the imaginary part of the effective action (which changes sign under time reversal)
corresponding to the symmetric $|\ups\rag$ state vanishes, in disagreement with Schwinger's result. As
a precise coherent  superposition of particle and anti-particle pairs for all modes, the time symmetric state
defined by (\ref{Esymodes})-(\ref{Eannihops}) is a very curious state indeed, corresponding to the rather unphysical
boundary condition of each pair creation event being exactly balanced by its time reversed pair annihilation
event, these pairs having been arranged with precisely the right phase relations to come from great distances
at early times in order to effect just such a cancellation everywhere at all times.
While mathematically allowed in a time reversal invariant background, it would be difficult to arrange such an
artificial construction and fine tuning of initial and/or boundary conditions on the quantum state of the charged field
with any macroscopic physical apparatus, and certainly it would not be produced with a more realistic adiabatic
switching on and off of the electric field background in either finite time or over a finite region of space \cite{GavGit}.
Nor does the state $|\ups\rag$ minimize the Hamiltonian of the system which is time dependent in the gauge
(\ref{Egauge}), or unbounded from below in the static gauge $A_0 = Ez$.

The above physical considerations and Schwinger's earlier result suggest that there should be an instability
of the time symmetric state to nearby states in which the fine tuned cancellation between particle/anti-particle
creation and annihilation events is slightly perturbed. In order to probe these nearby states we return
to (\ref{Emodeq}), and express its general solution in the form
\be
f_{\bf k}(t) = A_{\bf k}\, \ups_{\lam}(u) + B_{\bf k}\,\ups_{\lam}^*(u)
\label{fdef}
\ee
with the (strictly time independent) Bogoliubov coefficients required to obey
\be
|A_{\bf k}|^2 - |B_{\bf k}|^2 = 1\quad {\rm for\ all} \quad {\bf k},
\label{ABnorm}
\ee
in order for the Wronskian condition
\be
i\left(f_{\bf k}^*\frac{d}{dt}f_{\bf k} - f_{\bf k} \frac{d}{dt} f_{\bf k}^*\right) = 1
\ee
to be satisfied. The Bogoliubov coefficients $(A_{\bf k}, B_{\bf k})$ may be regarded as specified by initial data
$f_{\bf k}(t_0)$ and $\dot f_{\bf k}(t_0)$ at $t=t_0$ according to
\bes
\bea
A_{\bf k}(t_0) &=& i\left(\ups_{\lam}^*\frac{d}{dt}f_{\bf k} - f_{\bf k} \frac{d}{dt} \ups_{\lam}^*\right)\Big\vert_{t=t_0}\\
B_{\bf k}(t_0) &=& i\left(f_{\bf k}\frac{d}{dt}\ups_{\lam} - \ups_{\lam}\frac{d}{dt}f_{\bf k}  \right)\Big\vert_{t=t_0}
\eea
\ees
The quantized charged scalar field operator (\ref{Phiopcharged}) may just as well be expressed in
terms of these general mode functions (\ref{fdef}) as
\be
{\bf \Phi} (t, {\bf x}) = \frac{1}{\sqrt{V}}\sum_{\bf k}
\left[ a_{\bf k}^{f}\, f_{\bf k}(t)\, e^{i {\bf k} \cdot {\bf x}}+  b_{\bf k}^{f\,\dagger}\, f^*_{-\bf k}(t)\,
e^{-i {\bf k} \cdot {\bf x}}\right]
\label{Phiopchargedf}
\ee
where upon setting the Fourier components of (\ref{Phiopcharged}) and (\ref{Phiopchargedf}) equal,
the corresponding Fock space operators $ a_{\bf k}^{f}, b_{\bf k}^{f\,\dagger}$ are related to the
previous ones by
\be
\left( \begin{array}{c} a_{\bf k}^{\ups\,} \\ b_{-\bf k}^{\ups\,\dag} \end{array}\right) =
\left( \begin{array}{cc} A_{\bf k}& B_{\bf k}^*\\
B_{\bf k} & A_{\bf k}^*\end{array}\right)\
\left( \begin{array}{c} a_{\bf k}^{f}\\  b^{f\,\dag}_{-\bf k}\end{array}\right)
\label{Bogfups}
\ee
or its inverse
\be
\left( \begin{array}{c} a_{\bf k}^{f\,} \\ b_{-\bf k}^{f\,\dag} \end{array}\right) =
\left( \begin{array}{cc} \ A_{\bf k}^*& -B_{\bf k}^*\\
-B_{\bf k} & \ A_{\bf k}\end{array}\right)\
\left( \begin{array}{c} a_{\bf k}^{\ups}\\  b^{\ups\,\dag}_{-\bf k}\end{array}\right)\,.
\label{Bogupsf}
\ee
Hence if we define the state $|f\rag$ by the condition
\be
a_{\bf k}^{f} |f\rag = b_{\bf k}^{f}|f\rag = 0\,.
\label{Eannihopsf}
\ee
this state contains a non-zero expectation value
\be
\lag f | a_{\bf k}^{\ups\,\dag}a_{\bf k}^{\ups}|f\rag = |B_{\bf k}|^2 = \lag f | b_{-\bf k}^{\ups\,\dag}b_{-\bf k}^{\ups}|f\rag
\ee
of $\ups$ quanta. Conversely the $|\ups\rag$ state contains a non-zero expectation value of $f$ quanta.
Since both the $|\ups\rag $ and general $|f\rag$ states are pure states, and each can be expressed as a coherent,
squeezed state with respect to the other, it is best not to use the term `particles' for either of these expectation values,
nor can one decide {\it a priori} which among them is the `correct' vacuum. This illustrates the fact that the
question of which vacuum state to choose is not limited to de Sitter space or gravitational backgrounds only, but is
characteristic of QFT in time dependent and persistent classical background fields more generally.

The most general state which is both spatially homogeneous and charge symmetric is the mixed state
with a density matrix $\rho_{f,N}$ and a finite expectation value of $f$ quanta \cite{QVlas}, which we denote by
\be
{\rm Tr} \left(a_{\bf k}^{f\,\dag}a_{\bf k}^{f}\rho_{f,N}  \right) = N_{\bf k}
= {\rm Tr} \left(b_{-\bf k}^{f\,\dag}b_{-\bf k}^{f}\, \rho_{f,N} \right)
\ee
Computing the renormalized mean value of the electric current in this charge symmetric state we find
\bea
&&{\rm Tr} \big(j_z \,\rho_{f,N}\big) = 2e \int \frac{d^3 \bf k}{(2\pi)^3} (k_z +eEt)
\left[ |f_{\bf k}(t)|^2(1 + 2 N_{\bf k}) - \frac{1}{2\omega_{\bf k}(t)}\right]\nn
&& =4e \int \frac{d^3 \bf k}{(2\pi)^3} (k_z +eEt) \Big\{ \big[N_{\bf k}  + |B_{\bf k}|^2(1 + 2 N_{\bf k}) \big] |\ups_{\lam}(u)|^2
+ (1 + 2 N_{\bf k}) \,{\rm Re} \left[A_{\bf k}\, B_{\bf k}^*\, \ups^2_{\lam}(u)\right]\Big\}
\label{jzR}
\eea
where we have used (\ref{vac0}) and (\ref{fdef})-(\ref{ABnorm}) in arriving at the second expression. Charge 
asymmetric states or spatially inhomogeneous states with lower symmetry could be considered as well. 
In a general state with $B_{\bf k} \neq 0$ or $N_{\bf k} \neq 0$, the charge conjugation and time reversal 
symmetry of the background is broken and the current Tr\,$ \big(j_z \,\rho_{f,N}\big) \neq 0$. Because such states
correspond to charged particle/anti-particle excitations that are rapidly accelerated to ultrarelativistic
energies by the background electric field, they lead to persistent currents that do not decay and which
destabilize the constant electric field background through the semiclassical Maxwell eq. (\ref{Max}).

\begin{figure}
\begin{center}
\vspace{-1cm}
\includegraphics[scale=0.3,angle=90,width=4.8in,clip]{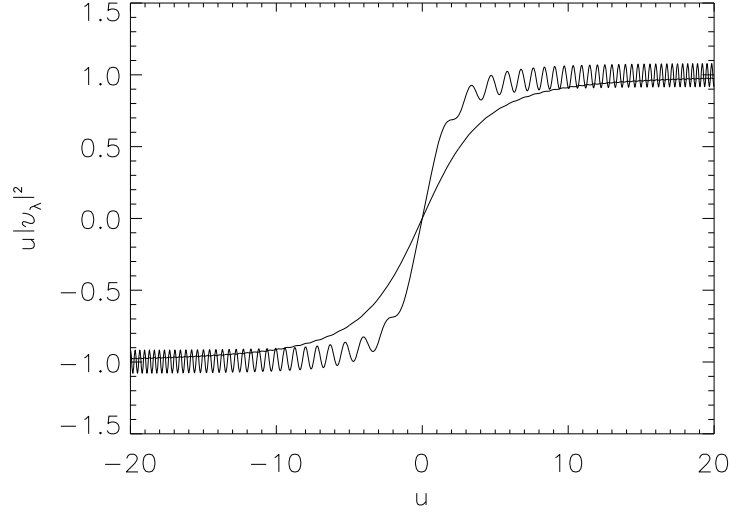}
\vspace{-1cm}
\caption{The $u|\ups_{\lam}|^2$ integrand in the current in (\ref{jvac}) or (\ref{jzR}) as a function of $u$
for two different values of $\lam$, {\it c.f} (\ref{ulamdef}). The curve with large oscillations
is for $\lam =1$ while the oscillations are much smaller in the curve for $\lam =5$.}
\label{Fig:uupsabs}
\end{center}
\vspace{-7mm}
\end{figure}

\begin{figure}[t]
\begin{center}
\vspace{-7mm}
\includegraphics[scale=0.3,angle=90,width=3.4in,clip]{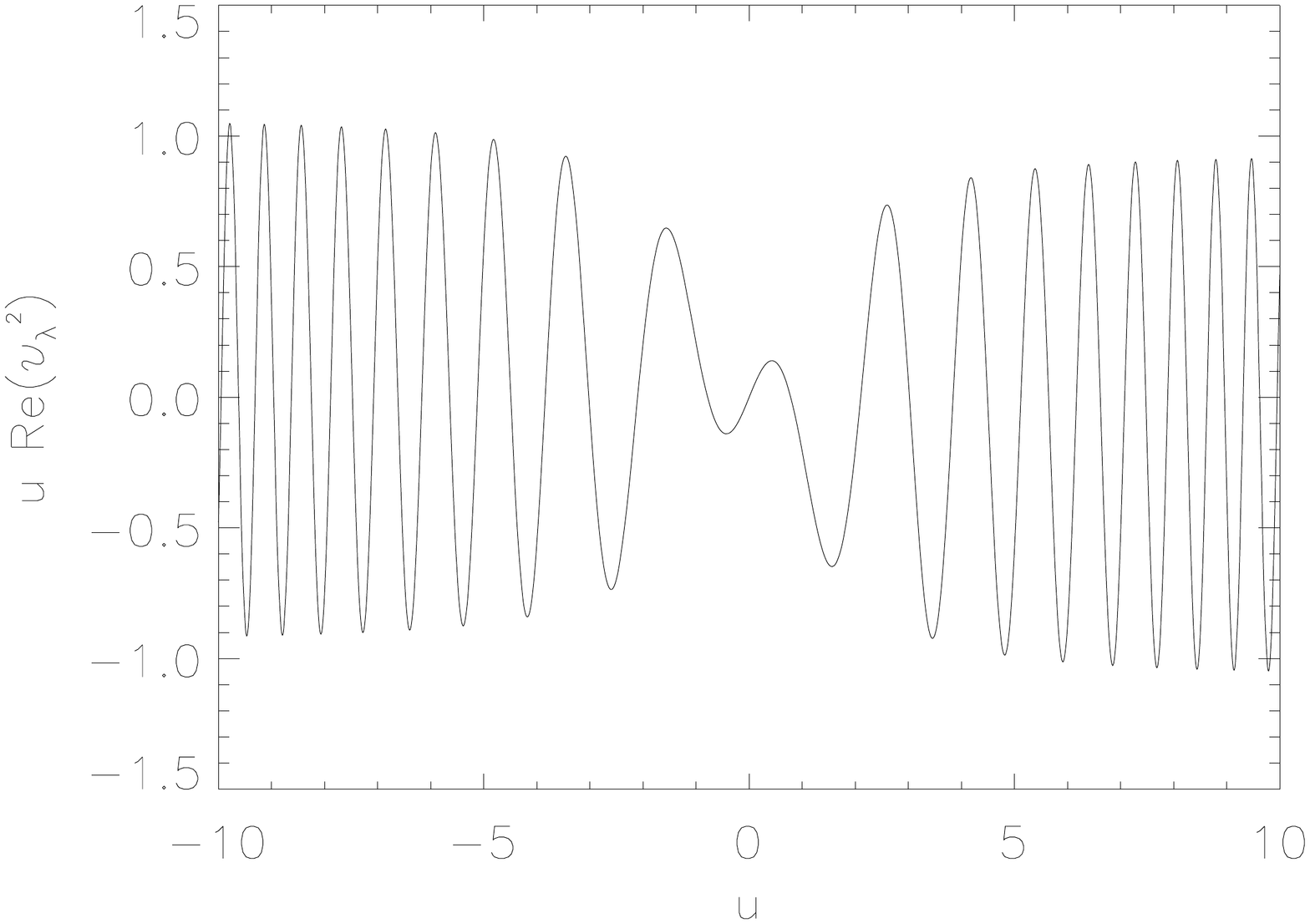}
\includegraphics[scale=0.3,angle=90,width=3.4in,clip]{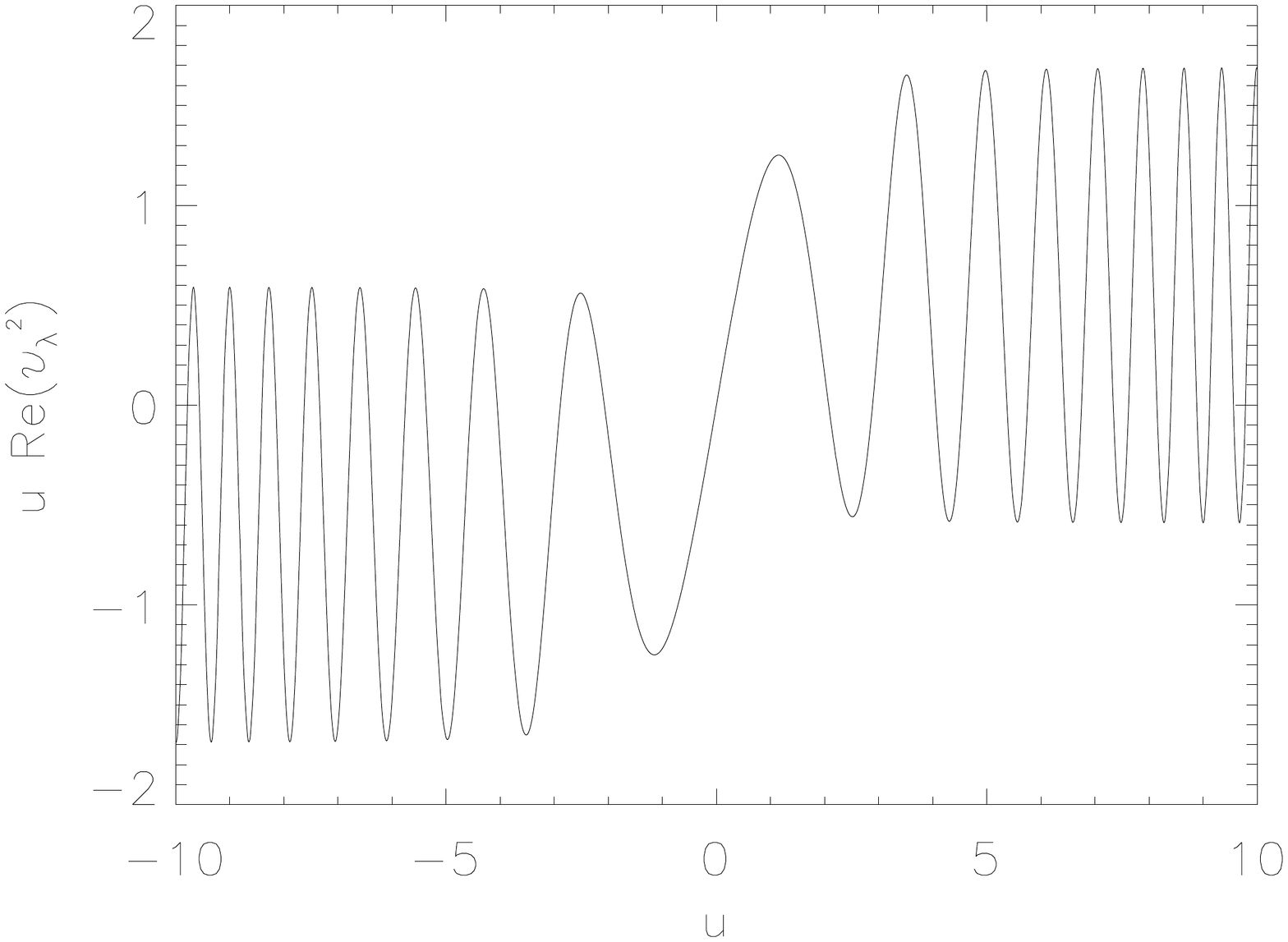}
\vspace{-1cm}
\caption{The $u\,{\rm Re}\,(\ups_{\lam}^{\,2})$ integrand in the current in (\ref{jzR}) as a function of $u$
for two different values of $\lam$. The left panel is for $\lam =1$ while the right panel
is for $\lam =0.1$ chosen to accentuate the asymmetry in $u\leftrightarrow -u$.}
\label{Fig:uupsre2}
\end{center}
\vspace{-8mm}
\end{figure}

\begin{figure}
\begin{center}
\includegraphics[scale=0.3,angle=90,width=3.4in,clip]{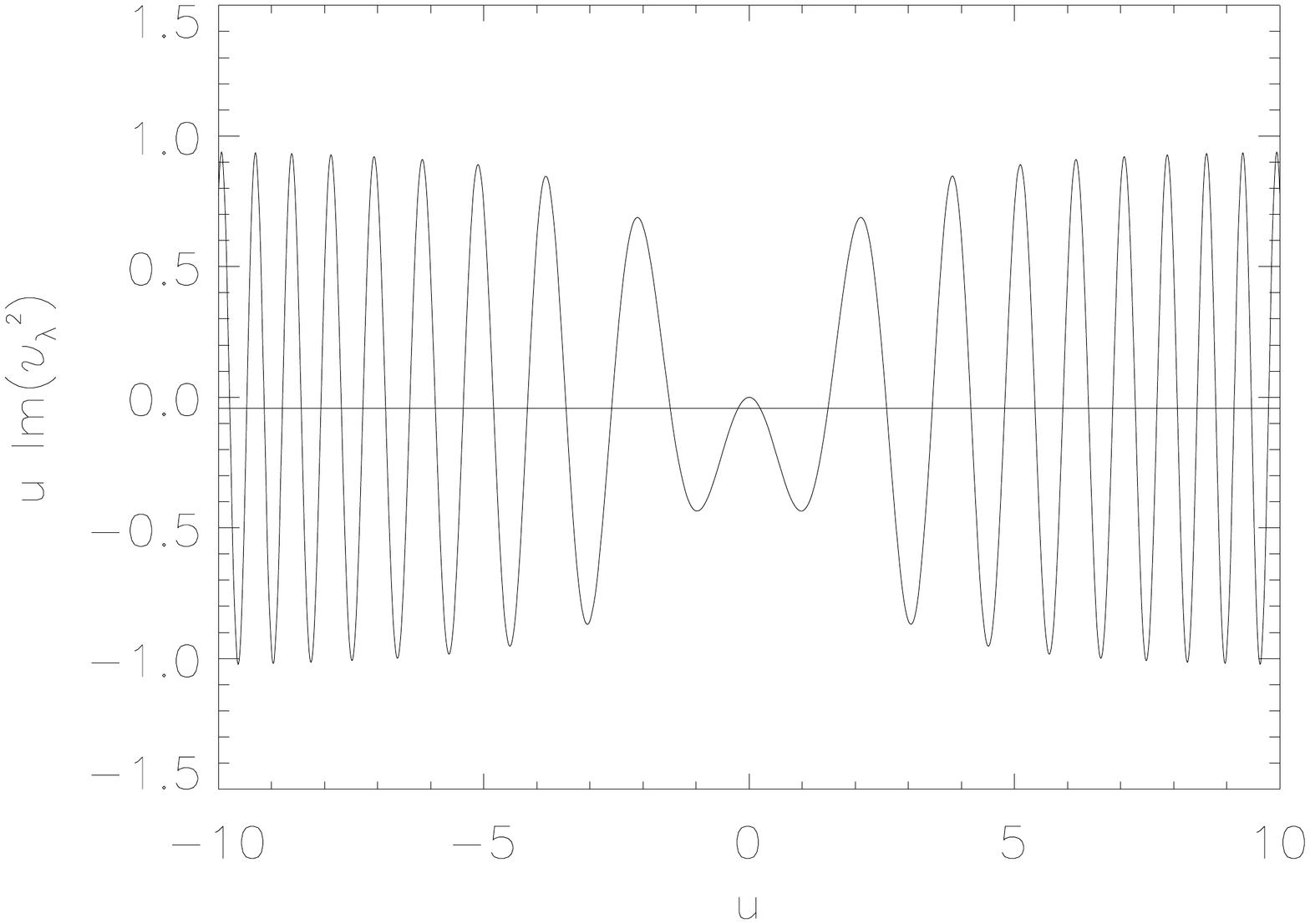}
\includegraphics[scale=0.3,angle=90,width=3.4in,clip]{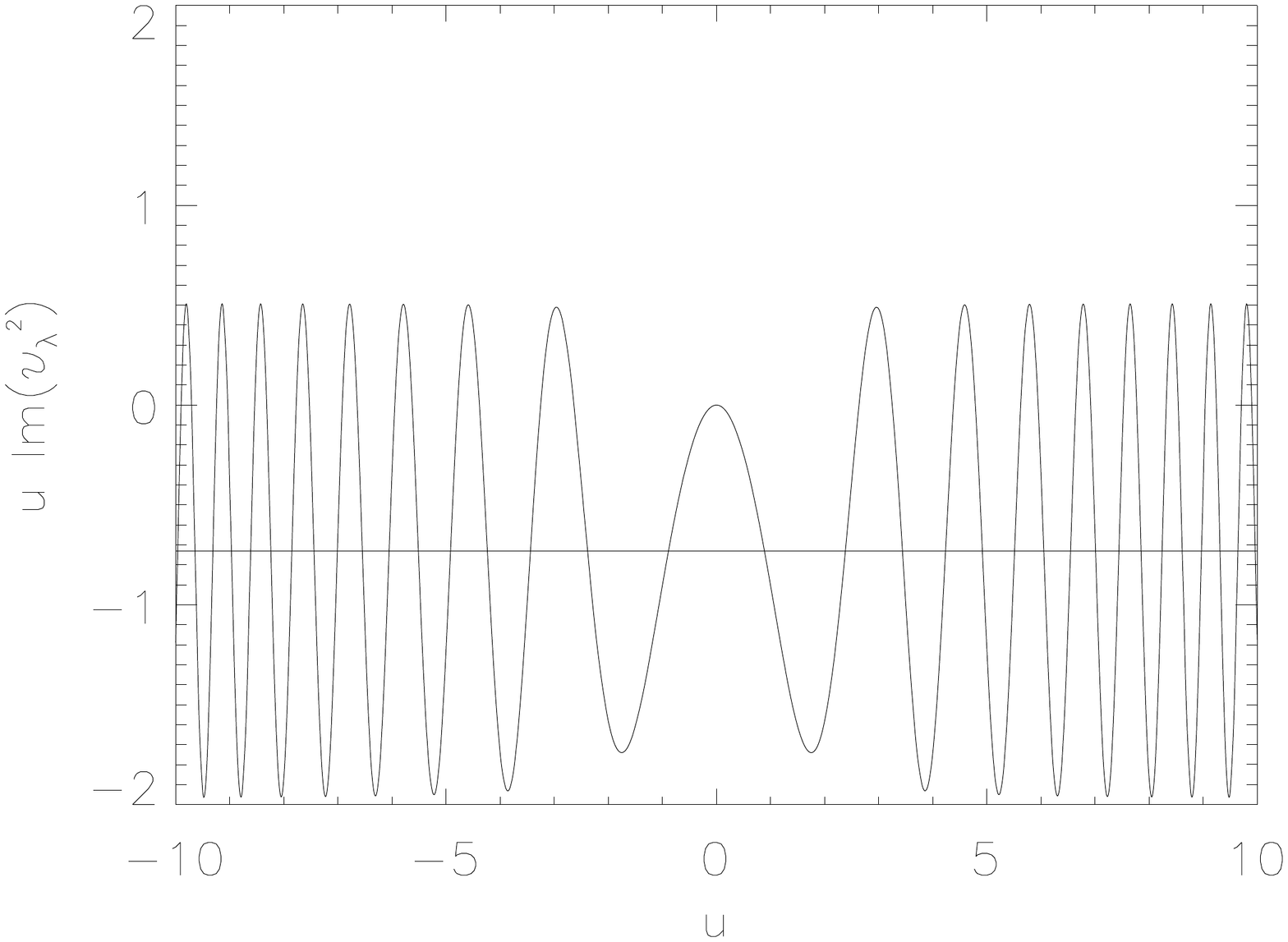}
\vspace{-1cm}
\caption{The $u\,{\rm Im}\,(\ups_{\lam}^{\,2})$ integrand in the current in (\ref{jzR}) as a function of $u$
for the same values of $\lam =1$ (left) and $\lam=0.1$ (right) as in Fig. \ref{Fig:uupsre2}. Shown also is
the horizontal line at $-\exp(-\pi \lam)$ around which the average amplitude of the oscillations are displaced.}
\label{Fig:uupsim2}
\end{center}
\vspace{-5mm}
\end{figure}

To see this requires only a qualitative understanding of the integrand in (\ref{jzR}). The three terms
$u|\ups_{\lam}|^2, u\,$Re\,$(\ups_{\lam}^{\,2}),$ and $u\,$Im\,$(\ups_{\lam}^{\,2})$ appearing in the
integrand of (\ref{jzR}) are shown as functions of $u$ for several values of $\lam$ in Figs.
\ref{Fig:uupsabs}-\ref{Fig:uupsim2}. In Fig. \ref{Fig:uupsabs} the saturation of the function
$u\,|\ups_{\lam}|^2$ at large $u$ is the result of acceleration of charged scalar particles to
ultrarelativistic energies by the electric field, where they make a constant contribution
to the current integrand. Hence if $N_{\bf k}$ and/or $|B_{\bf k}|^2$ in (\ref{jzR}) is non-zero for any range
of $\bf k$, such modes will make a contribution to the current proportional to the phase
space volume $\int d^3{\bf k}$ which can therefore give an arbitrarily large $\lag j_z\rag$
at late times, $u\gg 1$.

The oscillatory terms in the real and imaginary parts of $u\ups_{\lam}^{\, 2}$ are shown in Figs.
\ref{Fig:uupsre2}-\ref{Fig:uupsim2}. The envelope of the oscillations shows a saturation
behavior at large $|u|$ similar to Fig. \ref{Fig:uupsabs}. For smaller $\lam$ the oscillations
are significantly offset from the horizontal axis, by $\pm \exp(-\pi \lambda)$, showing that
there will also be a net contribution to the current from modes with $A_{\bf k}\,B^*_{\bf k} \neq 0$.
Hence these contributions to $\lag j_z\rag$ can also become arbitrarily large if the range of $\bf k$
for which $A_{\bf k}\,B^*_{\bf k}$ is non-zero is large.

An interesting special case in which to evaluate (\ref{jzR}) is the adiabatic vacuum state of initial data
\be
f_{\bf k} (t_0) = \frac{1}{\sqrt{2 \omega_{\bf k}(t_0)}} \,,\qquad
\dot f_{\bf k} (t_0) = \left(- i \omega_{\bf k} - \frac{\dot\omega_{\bf k}}{2 \omega_{\bf k}}\right) f_{\bf k}\Big\vert_{t=t_0}
\label{initadb}
\ee
and $N_{\bf k} = 0$. We denote this pure state which matches the lowest order adiabatic vacuum state at the
particular time $t=t_0$ by $|t_0\rag$. With these initial conditions it is shown in \cite{AndMot1}
that the Bogoliubov coefficients are given approximately by
\bes
\bea
A_{\bf k}(t_0) &\simeq & A_{\lam}^{in}\, \theta(-k_z-eEt_0) + A_{\lam}^{out}\, \theta(k_z+eEt_0)\\
B_{\bf k}(t_0) &\simeq & B_{\lam}^{in}\, \theta(-k_z-eEt_0) + B_{\lam}^{out}\, \theta(k_z+eEt_0)
\eea
\label{ABkt0}\ees
where
\bes
\bea
&&A^{in}_{\lam} = A_{\lam}^{out\,*} = \sqrt{\frac{\pi}{2}}\, \left(\frac{2}{\lambda}\right)^{\frac{i\lam}{2}}
\!\left[ \left(\frac{\lam}{2}\right)^{\frac{1}{4}}\! \frac{1}{\Gamma\left(\frac{3}{4} - \frac{i\lam}{2}\right)}
+ \left(\frac{2}{\lam}\right)^{\frac{1}{4}}\!\frac{ e^{\frac{i\pi}{4}}}{\Gamma\left(\frac{1}{4} - \frac{i\lam}{2}\right)}\right]
\exp \left( \tfrac{i\lam}{2}\!-\! \tfrac{i\pi}{8}\!-\! \tfrac{\pi \lam}{4}\right)\\
&&B^{in}_{\lam} =  B_{\lam}^{out\,*} = \sqrt{\frac{\pi}{2}}\, \left(\frac{2}{\lambda}\right)^{\frac{i\lam}{2}}
\!\left[ \left(\frac{\lam}{2}\right)^{\frac{1}{4}}\! \frac{1}{\Gamma\left(\frac{3}{4} - \frac{i\lam}{2}\right)}
- \left(\frac{2}{\lam}\right)^{\frac{1}{4}}\!\frac{ e^{\frac{i\pi}{4}}} {\Gamma\left(\frac{1}{4} - \frac{i\lam}{2}\right)}\right]
\exp \left( \tfrac{i\lam}{2} \!-\! \tfrac{i\pi}{8} \!-\! \tfrac{\pi \lam}{4}\right)\,.
\eea
\label{ABinE}\ees
In fact, the step functions in (\ref{ABkt0}) are smooth functions which interpolate between the two limits,
but this simple approximation is sufficient to illustrate the main features of the current expectation
value which is its linear growth in time from the initial time $t_0$.

Substituting (\ref{ABkt0}) into (\ref{jzR}) with $N_{\bf k}=0$, changing variables from $(k_z, k_{\perp})$ to $(u, \lam)$,
and making use of the fact that the functions $u\,|\ups_{\lam}(u)|^2$ and $u$ Re\,[$\ups_{\lam}^2(u)$] are odd functions
of $u$ ({\it c.f.} Figs. \ref{Fig:uupsabs} -\ref{Fig:uupsre2}), while $u$\,Im\,[$\ups_{\lam}^2(u)$] is even
({\it c.f.} Fig. \ref{Fig:uupsim2}), we obtain
\bea
&&\lag t_0 | j_z | t_0\rag_{_R} \simeq \frac{4e}{(2\pi)^2} \int _0^{\infty} k_\perp dk_{\perp}
\left\{\int_{-\infty}^{-eEt_0} dk_z (k_z+eEt) \Big[|B_{\lam}^{in}|^2 |\ups_{\lam}(u)|^2
+ {\rm Re} [A_{\lam}^{in} B_{\lam}^{in \,*} \ups_{\lam}^2(u)]\Big]\right.\nn
&&\hspace{2.8cm}+ \left.\int_{-eEt_0}^{\infty} dk_z (k_z+eEt) \Big[|B_{\lam}^{out}|^2 |\ups_{\lam}(u)|^2
+ {\rm Re} [A_{\lam}^{out} B_{\lam}^{out \,*} \ups_{\lam}^2(u)]\Big]\right\}\nn
&&\hspace{1.8cm}= -\frac{e^2E}{\pi^2} \int_{m^2/2eE}^{\infty} d\lam\, {\rm Im}[A_{\lam}^{in} B_{\lam}^{in\,*}]
\int_0^{\sqrt{2eE}\,(t-t_0)}du\, u
\, {\rm Im}[\ups_{\lam}^2(u)]\nn
&&\hspace{1.8cm}= \frac{e^3E^2}{2\pi^2}\, \frac{1}{\!\!\sqrt{2eE}}
\int_{m^2/2eE}^{\infty} d\lam \, e^{-\pi \lam}\,\int_0^{\sqrt{2eE}\,(t-t_0)}du
\, [uf_{\lam}^{(0)}(u) f_{\lam}^{(1)}(u)]
\label{currentint}
\eea
since from (\ref{ABinE}), or eq. (4.20b) of ref. \cite{AndMot1} and (\ref{Esymodes})
\bes
\bea
&&\hspace{2.5cm} |B_{\lam}^{in}|^2= |B_{\lam}^{out}|^2\\
&&\hspace{1.4cm}{\rm Re}[A_{\lam}^{in} B_{\lam}^{in\,*}] = {\rm Re}[A_{\lam}^{out} B_{\lam}^{out\,*}]\\
&&\hspace{-4mm}{\rm Im}[A_{\lam}^{in} B_{\lam}^{in\,*}] =- {\rm Im}[A_{\lam}^{out} B_{\lam}^{out\,*}]
= -\tfrac{1}{2}\, {\rm Im}\, B_{\lam}^{tot} = \tfrac{1}{2} e^{-\pi \lam}\\
&&\hspace{1cm}{\rm Im}[\ups_{\lam}^2(u)] = - \frac{1}{\sqrt{2eE}} \, f_{\lam}^{(0)}(u) f_{\lam}^{(1)}(u)\,.
\eea
\ees
Because of the offset from the $u$-axis of $ uf_{\lam}^{(0)}(u) f_{\lam}^{(1)}(u) = -u\, {\rm Im}[\ups_{\lam}^{\, 2}(u)]$
by $e^{-\pi\lam}$, around which the oscillations average to zero ({\it c.f.} Fig. \ref{Fig:uupsim2}), for large
$t-t_0 \rightarrow \infty$ the $u$ integral in (\ref{currentint}) is
\be
\int_0^{\sqrt{2eE}\,(t-t_0)}du\, [uf_{\lam}^{(0)}(u) f_{\lam}^{(1)}(u)]  \rightarrow \int_0^{\sqrt{2eE}\,(t-t_0)}du\, e^{-\pi\lam}
=\sqrt{2eE} \,(t-t_0)\, e^{-\pi\lam}
\ee
and hence (\ref{currentint}) gives for late times
\be
\lag t_0 | j_z | t_0\rag_{_R}  \rightarrow  \frac{e^3E^2}{4\pi^3} \,e^{-\pi m^2/eE}\,(t-t_0)
\label{jzlin}
\ee
which is the same result as that of Eq.\, (5.22) in Ref. \cite{AndMot1}, which was obtained 
much more naturally in the adiabatic particle basis by consideration of particle creation events. 
That treatment makes it clear that the growth of the current is a cumulative effect of particle creation 
from the quantum `vacuum' which continues unabated as long as the constant electric field is maintained.

Thus there are states for which the current grows linearly with time related to the steady rate of particle creation
in a constant electric field background. Moreover it is clear from the penultimate line of (\ref{currentint}) that any
perturbation of the symmetric $|\ups\rag$ state with Bogoliubov coefficients  $A_{\bf k}, B_{\bf k}$ of the form
(\ref{ABkt0}) obeying the conditions $A^{in}_{\lam} = A_{\lam}^{out\,*}$, $B^{in}_{\lam} = B_{\lam}^{out\,*}$
of (\ref{ABinE}), having constant but non-zero support for arbitrarily large and negative $k_z$ will produce a
cumulative effect on the current similar to (\ref{jzlin}), so that $\lag j_z\rag$ continues to grow linearly with time
for arbitrarily long times. This linear growth with time implies that however small the coupling $e$ and the
coefficient Im\,$[A_{\lam}^{in} B_{\lam}^{in\,*}]$ (which can be enhanced by taking $N_{\bf k} > 0$), the current
must eventually influence the background field through the semiclassical Maxwell eq. (\ref{Max}). Thus the
symmetric $|\ups\rag$ state in a fixed constant uniform electric field background is unstable to perturbations
of the kind (\ref{ABkt0}). If $B_{\bf k}$ has non-zero support up to some large but finite negative value
$(k_z)_{min}= -K_z$ the linear growth in (\ref{jzlin}) will be cut off at $t-t_0 = K_z/eE$ but still be large
and produce a large backreaction through (\ref{Max}).

If one goes beyond the simple mean field approximation considered here, it is also clear on physical
grounds that the introduction of a single electrically charged particle into the $|\ups\rag$ state will cause
it to be accelerated by the electric field to arbitrarily large energies, which would allow it to emit photons
and produce additional charged pairs resulting in an electromagnetic avalanche. Allowing these additional
channels opened up by self-interactions makes the physical instability of the symmetric $|\ups\rag$ state to small
perturbations more obvious, although that instability already exists even without self-interactions, in the mean
field approximation, as (\ref{jzlin}) and (\ref{Max}) show.

This example of the quantum states in a constant, uniform external electric field shows quite
clearly that the most symmetric state, with the full symmetry group of the background need not
be the stable ground state of the system. In this case it is well known that the background is
unstable to particle creation. In the accompanying paper \cite{AndMot1} we have shown how
the same conclusion follows in de Sitter space, for essentially the same reasons. The treatment
above shows that one need not be committed to any definition of particles to discover
the instability of the electric field background by perturbations of the symmetric $|\ups\rag$ state
which have support at large canonical momentum $|k_z|$. For $k_z < 0$ this may correspond to
small physical kinetic momentum $k_z +eEt_0$ at some early initial time $t_0$. The unlimited
growth of the physical momentum $k_z + eEt$ with time for fixed canonical momentum $k_z$ in terms
of which the initial state is specified is the essential feature, and this feature is found in
gravitational backgrounds such as de Sitter space as well.

\vspace{-2mm}
\subsection{Relation to Quantum Chiral Anomaly in Two Dimensions}
\label{Subsec:2D}

The linear secular growth of the current in a background constant electric field
can also be understood through the Schwinger anomaly in $1+1$ dimensions \cite{Schwmod}.
For that comparison we drop the $d^2 {\bf k}_{\perp}/(2\pi)^2$ integral in (\ref{jzR})
to reduce to $1+1$ dimensions, and further set the mass $m=0$. We have then
\be
\lag j_z\rag_{2d} \rightarrow \frac{e^2 E}{\pi}  \, (t-t_0)
\label{jz2d}
\ee
at late times. Since scalars are essentially the same as fermions in $1+1$ dimensions
one can use the bosonization results \cite{Cole} for fermionic QED to express the current
in the form
\be
\lag j^{\mu}\rag = \frac{e}{\!\sqrt{\pi}\ }\,\epsilon^{\mu\nu} \partial_{\nu} \chi
\label{curranom}
\ee
where $\epsilon^{\mu\nu}$ the antisymmetric symbol in two dimensions and $\chi$ is a pseudoscalar
field whose derivative is the chiral current
\be
\lag j^{\mu \,5}\rag = \frac{1}{\!\sqrt{\pi}\ }\partial^{\mu} \chi
\label{chicurr}
\ee
This current has the well-known chiral anomaly \cite{John}
\be
\partial_{\mu}\lag  j^{\mu\,5} \rag_{2d}= \frac{1}{\!\sqrt{\pi}\ }\sq \chi 
= \frac{e}{2\pi} \,\epsilon^{\mu\nu}F_{\mu \nu} = \frac{eE}{\pi}
\label{anom2d}
\ee
in a background electric field. The second order eq. (\ref{anom2d}) for $\chi$ with the anomaly source
in a constant, uniform field has solutions independent of $z$ of the form
\be
\frac{1}{\!\sqrt{\pi}\ }\chi = \frac{eE}{2\pi}\, \,(t-t_0)^2\,.
\label{chisoln}
\ee
Substituting this value of $\chi$ into the electric current (\ref{curranom}) gives
\be
\lag j_z\rag_{2d} =  \frac{e}{\!\sqrt{\pi}\ }\,\dot \chi =  \frac{e^2 E}{\pi}  \, (t-t_0)
\label{Schj}
\ee
which recovers (\ref{jz2d}). Thus the linear secular growth of the current with time in the massless limit
is related to the two-dimensional chiral anomaly and the particular $z$ independent solution (\ref{chisoln})
to the pseudoscalar field eq. (\ref{anom2d}). This particular solution to (\ref{anom2d}) is associated with the 
spatially homogeneous initial state condition (\ref{initadb}) and state specified by the mode functions (\ref{fdef}) and
(\ref{ABkt0}).

It is interesting to note that although the anomaly eq. (\ref{anom2d}) is Lorentz invariant,
because it is an inhomogeneous eq., {\it none} of its solutions are Lorentz invariant.
Thus the maximal Poincar\'e symmetry of the fixed electric field background is necessarily
broken by the solutions to the anomaly eq. (\ref{anom2d}), which leads to a spontaneous
breaking of symmetry of the background, at least in the semiclassical approximation
and neglecting backreaction. This may be seen also from the effective action corresponding to
the 2D chiral anomaly \cite{Schwmod,Jackiw}, {\it viz.}
\be
S_{anom}^{2D}[\chi] = \frac{e^2}{8\pi} \int d^2x \int d^2x'\, [\epsilon^{\mu\nu}F_{\mu\nu}]_x\,
\sq^{-1}(x,x')\, [\epsilon^{\alpha\beta}F_{\alpha\beta}]_{x'}
= \tfrac{1}{2}\int d^2 x \left[-\chi\sq \chi + \frac{e}{\!\sqrt\pi}\chi\, \epsilon^{\mu\nu}F_{\mu\nu}\right]
\label{2danomact}
\ee
where $\sq^{-1}(x,x')$ is the Green's function inverse of the scalar wave operator $\sq$ in two dimensions.
As is well-known, the usual construction of the Feynman Green's function for a massless scalar
in two dimensions is infrared divergent due to the constant $k=0$ mode, and consequently no
Lorentz invariant Feynman function exists in this case. Green's functions $\sq^{-1}(x,x')$ obeying
different boundary conditions exist, but these necessarily break some of the continuous or discrete
symmetries of the background. Thus the form of the effective action of the 2D chiral anomaly
(\ref{2danomact}), together with the absence of a Lorentz invariant Feynman Green's function
$\sq^{-1}(x,x')$ due to infrared divergences is sufficient to conclude that the maximally symmetric
state in a uniform constant background $\frac{1}{2}\epsilon^{\mu\nu}F_{\mu\nu} = E$ is sensitive to
non-invariant initial and/or boundary conditions which break that maximal symmetry. The linear
growth of the current found in (\ref{jzlin}) and reproduced by the solution (\ref{chisoln}) in (\ref{Schj})
is symptomatic of that necessary breaking of the maximal symmetry of the classical background
by the quantum chiral anomaly.

It is also interesting that this connection with the anomaly of massless fields in two dimensions
survives in four dimensions and even if the field has a non-zero mass $m$, whose main
effect is to suppress the coefficient of the linear growth by the Schwinger tunneling factor
$\exp(-\pi m^2/eE)$. We shall see there is also an interesting connection to a quantum
anomaly of massless fields in four dimensional de Sitter space,
a local condensate bilinear of the underlying quantum field(s) analogous to $\chi$, and
simple arguments analogous to (\ref{curranom})-(\ref{Schj}) which lead directly to the
analogous conclusion of instability of the symmetric state and breaking of maximal
de Sitter invariance in that case as well.

\section{$O(4)$ Invariant States in de Sitter Space}
\label{Sec:O4dS}

Turning to our primary topic of de Sitter space, we develop the quantization and discussion
of possible `vacuum' states in de Sitter space analogously to the electric field case of the previous
section. For an uncharged scalar field ${\bf \Phi}$ satisfying the free wave equation
\be
(-\sq + M^2) {\bf \Phi} \equiv \left[ - \frac{1}{\sqrt{-g}} \frac{\partial}{\partial x^a}
\left(\sqrt{-g}\,g^{ab}\frac{\partial}{\partial x^b}\right) + M^2\right] {\bf \Phi} = 0\,.
\label{waveq}
\ee
in a gravitational background, with $\xi$ the curvature coupling. The effective mass
\be
M^2 \equiv m^2 + \xi R = m^2 + 12\,\xi H^2
\ee
is a constant since the Ricci scalar $R=12H^2$ is a constant in de Sitter spacetime. In the geodesically
complete coordinates (\ref{hypermet}) the wave eq. (\ref{waveq}) may be separated into a complete basis
of functions of cosmological time $y_k(u)$ times $Y_{klm_l} (\hat N)$, the spherical harmonics on
$\mathbb{S}^3$. A unit vector on $\mathbb{S}^3$ is denoted by $\hat N$ with coordinates
\be
\hat N (\chi, \theta, \phi) = (\sin\chi \, {\bf \hat n}, \cos\chi) =
(\cos\chi ,\, \sin\chi\cos\theta ,\, \sin\chi\sin\theta\sin\phi ,\, \sin\chi\sin\theta\cos\phi ) \,.
\ee
The $Y_{klm}(\hat N)$  harmonics are eigenfunctions of the scalar Laplacian on the unit ${\mathbb S}^3$ satisfying
\be
-\Delta_3 \, Y_{klm_l} = -\frac{1}{\sin^2\chi}\left[\frac{\partial}{\partial \chi}\sin^2\chi \frac{\partial}{\partial \chi}
+ \frac{1}{\sin \theta}\frac{\partial}{\partial \theta}\sin\theta \frac{\partial}{\partial \theta}
+ \frac{1}{\sin^2\theta} \frac{\partial^2}{\partial \phi^2}\right] Y_{klm_l} = (k^2 -1)\, Y_{klm_l}
\ee
with the range of the integer $k = 1, 2,  \dots $ taken to be strictly positive, conforming to the notation
of \cite{Attract} and \cite{EMomTen}. These ${\mathbb S}^3$ harmonics are given in terms of Gegenbauer
functions $C^{l+1}_{k-l-1}(\cos\chi)$ and the familiar ${\mathbb S}^2$ spherical harmonics $Y_{lm_l}({\bf \hat n})$
in the form \cite{Bate}
\be
Y_{klm_l} (\hat N) = 2^l\,l! \sqrt{\frac{2k(k-l-1)!}{\pi\,(k+l)!}}\,(\sin\chi)^l \, C^{l+1}_{k-l-1}(\cos\chi)\, Y_{l m_l}({\bf \hat n})
\ee
with $l = 0, 1, \dots k-1$ and $m_l = -l, \dots , l$, normalized so that
\be
\int_{\mathbb{S}^3} d^3\Sigma \ Y_{k'l'm_l'}^*\, Y_{klm_l}
= \int_0^{\pi} d\chi \sin^2\chi \int_0^\pi d\theta \sin\theta \int_0^{2\pi} d\phi \,Y_{k'l'm_l'}^*\, Y_{klm_l}
=  \delta_{k'k}\delta_{l'l}\delta_{m_l'm_l}\,.
\label{Ynorm}
\ee
Note also that $Y_{klm_l}^*(\hat N) = Y_{kl\,-m_l}(\hat N)$.

The time dependent functions $y_k(u)$ satisfy
\be
\left[\frac{d^2}{du^2} + 3 \tanh u\, \frac{d}{du} + (k^2 -1)\, \sech^2 u
+ \left(\gamma^2 + \tfrac{9}{4}\right) \right]\, y_k = 0\,,
\label{modeq}
\ee
where the dimensionless parameter $\gamma$ is defined by
\be
\gamma \equiv \sqrt{\frac{M^2}{H^2}-\frac{9}{4}} \equiv i\nu\,.
\label{gamdef}
\ee
In the massive case $M^2 > \frac{9}{4}H^2$  (the {\it principal series}) $\gamma$ is real and positive.
With the  change of variables to $z = (1 -i \sinh u)/2$, the mode eq. (\ref{modeq}) can be recast in the form
of the hypergeometric equation. The fundamental complex solution $y_k \rightarrow \ups_{k\gamma}(u)$
may be taken to be
\be
\ups_{k\gamma}(u) \equiv H c_{k\gamma} \ (\sech u)^{k+1} \, (1 - i \sinh u)^k\,
F \left(\frac{1}{2} + i\gamma, \frac{1}{2} - i\gamma ; k+1 ; \frac{1 - i \sinh u}{2}\right)
\label{BDmode}
\ee
where $F \equiv \,_2F_1$ is the Gauss hypergeometric function and
\be
c_{k\gamma} \equiv \frac{1}{k!} \left[ \frac{ \Gamma\left(k + \frac{1}{2} + i\gamma\right)
\Gamma \left( k  + \frac{1}{2} - i\gamma\right)}{2}\right]^{\frac{1}{2}}
\label{Cdef}
\ee
is a real normalization constant, fixed so that $\ups_{k\gamma}$ satisfies the Wronskian condition
\be
iH a^3(u) \left[\ups_{k\gamma}^* \frac{d}{du}\ups_{k\gamma} - \ups_{k\gamma}  \frac{d}{du}\ups_{k\gamma}^*\right] = 1
\label{Wronups}
\ee
for all $k$, where $a(u) = H^{-1} \cosh u$ is the scale factor in coordinates (\ref{hypermet}). Note that under
time reversal  $u \rightarrow -u$ the mode function  (\ref{BDmode}) goes to its complex conjugate
\be
\ups_{k\gamma}(-u) = \ups_{k\gamma}^* (u)
\label{timerevBD}
\ee
for all $M^2 > 0$.

If $0< M^2 \le \frac{9}{4}H^2$, (\ref{gamdef})-(\ref{timerevBD}) continue to hold by analytic continuation to
pure imaginary $\gamma \equiv i \nu$, with $\ups_{k\gamma} \rightarrow \ups_{k, \nu}$. The mode functions
(\ref{BDmode}) reduce to elementary functions in the massless, conformally coupled case
\be
m=0\,, \,\xi = \tfrac{1}{6}\,,\, \nu = \tfrac{1}{2} : \quad \ups_{k\,,\frac{1}{2}} = \frac{H}{\sqrt{2k}}\,\sech u\,
(\sech u - i \tanh u)^k = \frac{H}{\sqrt{2k}}\,\cos \eta\, e^{-ik\eta} \,,
\label{m0conf}
\ee
and in the massless, minimally coupled case
\bea
&&\hspace{3cm}\raisebox {2mm}{$m=0\,, \,\xi = 0\,,\, \nu = \tfrac{3}{2}$:}\nn
&& \hspace{-1cm}\ups_{k\,, \frac{3}{2}} = \frac{H (k\, \sech u+ i\, {\rm tanh}\,u)}{\sqrt{2k(k^2-1)}}\, (\sech u - i \tanh u)^k
= \frac{H (k\cos\eta+ i\sin \eta)}{\sqrt{2k(k^2-1)}}\,e^{-ik\eta}\,,
\quad k=2,3,\dots\,,
\label{m0min}
\eea
where the conformal time variable $\eta$ is given by, {\it c.f.} Fig. \ref{Fig:dSCarPen},
\be
\eta \equiv \sin^{-1}({\rm tanh}\,u) \in \left(-\tfrac{\pi}{2}, \tfrac{\pi}{2}\right)\,,\qquad \cos\eta = \sech\,u\,,\quad \tan\eta =\sinh\,u\,.
\label{conftime}
\ee
The complex positive frequency modes $\ups_{k,\nu}$ of (\ref{BDmode}), (\ref{m0min}) are undefined for the case
$\nu = \frac{3}{2}$, $k=1$ since the solutions of (\ref{modeq}) are non-oscillatory in this case, and must be treated
separately \cite{Attract,AllFol}. This leads to the non-existence of a de Sitter invariant vacuum state or Feynman
Green's function $\sq^{-1}(x,x')$ for a massless, minimally coupled scalar in de Sitter space \cite{FordPark,AllFol}, that is
similar to that for a massless scalar in two dimensional flat space discussed in Sec. \ref{Subsec:2D}.

The scalar field operator ${\bf \Phi}$ can be expressed as a sum over the fundamental solutions
\be
{\bf \Phi} (u,\hat N) = \sum_{k=1}^{\infty}\sum_{l=0}^{k-1}\sum_{m_l=-l}^{l}
\left\{ a^{\ups}_{klm_l}\, \ups_{k\gamma}(u)\, Y_{klm_l} (\hat N) +  a_{klm_l}^{\ups\,\dag}\, \ups^*_{k\gamma}(u)\,
Y_{klm_l}^*\, (\hat N)\right\}
\label{Phiop}
\ee
with the Fock space operator coefficients $a^{\ups}_{klm_l}$ satisfying the commutation relations
\be
\Big[a^{\ups}_{klm_l}, a^{\ups\,\dag}_{k'l'm'_l}\Big] = \delta_{kk'}\delta_{ll'}\delta_{m_lm'_l}\,.
\label{commutator}
\ee
With (\ref{Ynorm}), (\ref{Wronups}), and (\ref{commutator}) the canonical equal time field commutation relation
\be
\Big[{\bf \Phi} (u, \hat N) , {\bf \Pi} (u, \hat N')\Big] = i\, \delta_{\Sigma}(\hat N, \hat N')
\label{cancom}
\ee
is satisfied, where ${\bf \Pi} = \sqrt{-g}\, \dot{\bf \Phi} = H a^3 \frac{\partial{\bf \Phi}}{\partial u}$ is the field
momentum operator conjugate to ${\bf \Phi}$, the overdot denotes the time derivative $H\, \partial/\partial u$
and $ \delta_{\Sigma}(\hat N, \hat N')$ denotes the delta function on the unit $\mathbb{S}^3$
with respect to the canonical round metric $d\Sigma^2$.

The Chernikov-Tagirov or Bunch-Davies (CTBD) state $\vert \ups\rag$ \cite{Nacht,CherTag,BunDav} is defined by
\be
a^{\ups}_{klm_l} \, \vert \ups\rag = 0 \qquad \forall\quad  k, l, m_l,
\label{dS}
\ee
and is invariant under the full $O(4,1)$ isometry group of the complete de Sitter manifold,
including under the discrete inversion symmetry of all coordinates in the embedding space,
$X^A \rightarrow - X^A$ ({\it c.f.} Fig. \ref{Fig:deShyper}), or  $(u,\hat N) \rightarrow (-u, -\hat N)$, which is not
continuously connected to the identity. The Feynman Green's function in this maximally symmetric state is invariant
under $O(4,1)$ and also coincides with that obtained by analytic continuation from the Euclidean ${\mathbb S}^4$
for $M^2 >0$ with full $O(5)$ symmetry \cite{DowCrit}. As in the electric field example of Sec. \ref{Sec:ConstantE}
the existence or construction of a maximally symmetric $O(4,1)$ invariant state does not imply that this state is
a stable vacuum.

Alternative Fock representations in real $u$ time are clearly possible. For example, since the general
solution of (\ref{modeq}) may be written as the linear combination
\be
y_{k}(u)= A_k\,\ups_{k\gamma}(u) + B_k\, \ups_{k\gamma}^*(u)\,,
\label{gensoln}
\ee
and normalized by (\ref{Wronups}) in the same way by requiring
\be
iH a^3(u) \left[y_{k}^* \frac{d}{du}y_{k} - y_{k}  \frac{d}{du}y_{k}^*\right] =
iH\left[f_{k}^* \frac{d}{du}f_{k} - f_{k}  \frac{d}{du}f_{k}^*\right] = |A_k|^2 - |B_k|^2 = 1\,.
\label{ABcond}
\ee
The general functions $y_k \equiv a^{-\frac{3}{2}}f_k$ may just as well be chosen as a basis of quantization
of the $\Phi$ field by
\be
\Phi (u,\hat N) = \sum_{k=1}^{\infty}\sum_{l=0}^{k-1}\sum_{m=-l}^{l}
\left[ a^{f}_{klm_l}\, y_{k}(u)\, Y_{klm_l} (\hat N) +  a^{f\,\dag}_{klm_l} y^*_{k}(u)\,
Y_{klm_l}^*\, (\hat N)\right]\,,
\label{Phigen}
\ee
with the Bogoliubov transformation between the corresponding Fock space operators 
\be
\left( \begin{array}{c} a_{klm_l}^{\ups\,} \\ a_{kl\,-m_l}^{\ups\,\dag} \end{array}\right) =
\left( \begin{array}{cc} A_k& B_k^*\\
B_k& A_k^*\end{array}\right)\
\left( \begin{array}{c} a_{klm_l}^{f}\\  a^{f\,\dag}_{kl\,-m_l}\end{array}\right)
\label{Bogupsy}
\ee
or its inverse
\be
\left( \begin{array}{c} a_{klm_l}^{f\,} \\ a_{kl\,-m_l}^{f\,\dag} \end{array}\right) =
\left( \begin{array}{cc} \ A_k^*& -B_k^*\\
-B_k & \ A_k\end{array}\right)\
\left( \begin{array}{c} a_{klm_l}^{\ups}\\  a^{\ups\,\dag}_{kl\,-m_l}\end{array}\right)
\label{Bogyups}
\ee
analogous to (\ref{fdef})-(\ref{Bogupsf}) of Sec. \ref{Sec:ConstantE}.

The mode function $f_k = a^{\frac{3}{2}}y_k$ also satisfies the eq. of an harmonic oscillator
\be
\frac{d^2f_k}{du^2} + \left[ \left(k^2 - \tfrac{1}{4}\right) \sech^2 u + \gamma^2\right] f_k = 0
\label{oscmode}
\ee
analogous to (\ref{yEmode}), and (\ref{oscmode}) which is the starting point for an adiabatic or WKB analysis
of particle creation in \cite{AndMot1}. Here we note that because of (\ref{ABcond}) the commutation
relations (\ref{commutator}) are also satisfied by $a^{f}_{klm_l}, a_{klm_l}^{f\,\dag}$, as is the
canonical field commutation relation (\ref{cancom}). Hence we may define a state $|f\rag$
corresponding to the general solution (\ref{gensoln}) of (\ref{modeq}) or (\ref{oscmode}) by
\be
a^{f}_{klm_l} \, \vert f\rag = 0 \qquad \forall\quad  k, l, m_l.
\label{fstate}
\ee
for any set of complex coefficients $\{A_k, B_k\}$ satisfying (\ref{ABcond}). Since the solutions $y_k(u)Y_{klm_l}(\hat N)$
at fixed $k$ form an irreducible representation of the group $O(4)$ for any $\{A_k, B_k\}$, these states are invariant under
$O(4)$ rotations of ${\mathbb S}^3$, but not the full $O(4,1)$ de Sitter group (unless $A_k =1, B_k = 0$ for all $k$).

The $O(4)$ invariant states are associated with a preferred $u$ time slicing which breaks the $O(4,1)$ symmetry.
The Bogoliubov coefficients $\{A_k, B_k\}$ and hence the particular $|f\rag$ state may be regarded as specified
by initial data $y_{k}(u_0)$ and $\dot y_{k}(u_0)$ on the $u=u_0$ Cauchy surface according to
\bes
\bea
A_k(u_0) &=& iHa^3\left(\ups_{k\gamma}^*\frac{d}{du}y_{k} - y_{k} \frac{d}{du} \ups_{k\gamma}^*\right)\Big\vert_{u=u_0}\\
B_k(u_0) &=& iHa^3\left(y_{k}\frac{d}{du}\ups_{k\gamma} - \ups_{k\gamma}\frac{d}{du}y_{k}  \right)\Big\vert_{u=u_0}
\eea
\label{ABinit}\ees
States with lower symmetry than $O(4)$ may be obtained by considering Bogoliubov transformations more general
than (\ref{Bogyups}), mixing $a^{\ups}$ and $a^{\ups\,\dag}$ of different $(klm_l)$. For example if the relation
(\ref{Bogyups}) is generalized to $a_{klm_l} = A_{kk'}^*\, a^{\ups}_{k'lm_l} - B_{kk'}^*\, a_{k'l\,-m_l}^{\ups\,\dag}$
so that the Bogoliubov coefficients are (non-diagonal) matrices in $k,k'$ (but still diagonal in $l, m_l$), the
corresponding states (\ref{fstate}) are $O(3)$ invariant only. These are appropriate for the static coordinates
of de Sitter space. All states related to $\vert \ups \rag$ by exact Bogoliubov transformations of this kind are {\it pure}
states and related to each other by a unitary transformation \cite{PartCreatdS}, whether they involve different
$(k l m_l)$ or not.

The expectation value of $a^{\ups\,\dag}_{klm_l}a^{\ups}_{klm_l}$ is non-vanishing in the general
$\vert f\rag$ state defined by (\ref{gensoln}), (\ref{Bogyups}) and (\ref{fstate}), {\it viz.}
\bes
\bea
&&\hspace{1cm}\lag f\vert a_{klm_l}^{\ups\,\dag} a^{\ups}_{klm_l}\vert f \rag = \vert B_k\vert^2 \\
&&\lag f\vert a^{\ups}_{klm_l}a^{\ups}_{kl\,-m_l}\vert f \rag = A_k B_k^*
= \lag f\vert a^{\ups\,\dag}_{klm_l}a^{\ups\,\dag}_{kl\,-m_l}\vert f \rag ^*  \;.
\eea
\label{ccdag}\ees
Hence the general $\vert f\rag$ `vacuum' state apparently contains `particles'  defined with respect to the
de Sitter invariant state $\vert \ups\rag$. However, the converse is also true as the expectation values
\bes
\bea
&&\hspace{2cm}\lag \ups\vert a^{f\,\dag}_{klm_l} a^{f}_{klm_l}\vert \ups \rag = \vert B_k\vert^2 \\
&&\lag \ups\vert a^{f}_{klm_l}a^{f}_{kl\,-m_l}\vert \ups\rag = - A_k^* B_k
= \lag \ups\vert a^{f\,\dag}_{klm_l}a^{f\,\dag}_{kl\,-m_l}\vert \ups \rag ^*
\eea
\label{aadag}\ees
are also non-zero, so that the de Sitter invariant `vacuum' may equally well be said to contain `particles'
with respect to the general $O(4)$ basis states $\vert f\rag$. Since each of the $|f\rag$ states in either case
is in fact a coherent, squeezed pure state with respect to the others, with all exact quantum phase correlations
maintained, it is better not to attach the label of `particles' to either set of expectation values (\ref{ccdag})
or (\ref{aadag}), or the term `vacuum' to any particular state in de Sitter space at this point.
As in the electric field case, mixed states which are $O(4)$ invariant can be defined through
a density matrix $\rho_{f,N}$ with \cite{ShortDistDecohere}
\be
{\rm Tr} \big(\rho_{f,N}\,a^{f\,\dag}_{klm_l} a^{f}_{klm_l}\big) = N_k
\label{Numk}
\ee
and $N_k=0$ reducing to the pure $|f\rag$ state defined in (\ref{fstate}). A non-zero $N_k$ above the
arbitrary $O(4)$ invariant $|f\rag$ `vacuum' is also best not identified with any physical particle number.
In previous works and in a companion paper to this one \cite{AndMot1,EMomTen}, we give a definition of
physical particle number in de Sitter space based on adiabatic or slowly varying positive frequency basis \cite{BirDav}.

\section{Energy-Momentum Tensor of O(4) Invariant States}
\label{Sec:SET}

The behavior of perturbations of the CTBD $O(4,1)$ symmetric state in de Sitter space may be
studied through the energy-momentum-stress tensor, and the potential backreaction effects on
the background geometry through the semiclassical Einstein eqs. (\ref{scE}), analogous to perturbations
of the symmetric $|\ups\rag$ state and backreaction effects of the electric current  through the
semiclassical Maxwell eq. (\ref{Max}).

The conserved energy-momentum-stress tensor of the free scalar field is
\be
T_{ab} = (\nabla_a \Phi) (\nabla_b \Phi) - \frac{g_{ab}}{2} \left[g^{cd} (\nabla_c \Phi) (\nabla_d \Phi)
+ m^2\Phi^2\right] + \xi \left[ R_{ab} - \frac{g_{ab}}{2} R -\nabla_a\nabla_b  +  g_{ab} \sq
\right] \Phi^2  \,.
\label{SET}
\ee
If the Heisenberg field operator in the general $O(4)$ basis (\ref{Phigen}) is substituted into this
expression, and (\ref{Numk}) is used, the expectation value of $T_{ab}$ in the general $\vert f\rag$
state may be expressed as a sum over modes. Since these states are spatially homogeneous and
isotropic, and $O(4)$ invariant, we find that
\bes
\bea
&&{\rm Tr} \big(\rho_{f,N}\, T^u_{\ u}\big)= -\varepsilon_{_{f,N}}\\
&&{\rm Tr} \big(\rho_{f,N}\,T^i_{\ j}\big)  = \delta^i_{\ j}\,p_{_{f, N}}\, \qquad i,j = \chi,\theta,\phi\,,
\eea
\label{genO4T}\ees
are the only non-vanishing components of the renormalized expectation value in coordinates ({\ref{hypermet}).
Since the renormalization counterterms are state independent, they may be subtracted from the mode sum for
the de Sitter invariant state with $A_k = 1, B_k =0$ once and for all. The renormalized expectation value
$\lag\ups|T_{ab}|\ups\rag_{_R}$ has been computed in the CTBD state \cite{BunDav}. Because
of its de Sitter invariance this expectation value satisfies (\ref{genO4T}) with $p_{\ups} =-\varepsilon_{\ups}$.
Collecting then the remaining finite terms which differ from this when $A_k \neq 1, B_k \neq 0$ in the general
$O(4)$ invariant mixed state, one obtains \cite{ShortDistDecohere}
\bes
\bea
&&\hspace{-1cm} \varepsilon_{_{f, N}} =
\varepsilon_{\ups} + \frac{1}{2 \pi^2} \sum_{k=1}^{\infty} k^2 \Big\{(1 + 2N_k)\,{\rm Re}\,[A_kB^*_k\, \varepsilon_k^A]
+ \big[N_k + \vert B_k\vert^2 (1 + 2N_k)\big] \,\varepsilon_k^B \Big\}\label{renorme}\\
&&\hspace{-1cm} p_{_{f, N}} = p_{\ups}
+ \frac{1}{2 \pi^2} \sum_{k=1}^{\infty} k^2  \Big\{(1 + 2 N_k) \,{\rm Re}\,[A_kB^*_k\, p_k^A]
+  \big[N_k + \vert B_k\vert^2 (1 + 2N_k)\big] \,p_k^B \Big\}\label{renormp}
\eea
\label{renormepsp}\ees
where we have defined
\bes
\bea
&&\varepsilon^A_k \equiv \dot \ups_k^2 + 2 h\,   \ups_k  \dot \ups_k  + (\omega_k^2 + h^2)\, \ups_k^2
\equiv 3p^A_k + 2m^2 \ups_k^2\,,\label{epspA}\\
&&\hspace{-1cm}\varepsilon^B_k \equiv \vert\dot\ups_k\vert^2 + 2 h\,  {\rm Re}\, [\ups^*_k  \dot \ups^{}_k]
+ (\omega_k^2 + h^2)\, \vert\ups_k\vert^2 \equiv 3p_k^B + 2m^2 \vert\ups_k\vert^2\,,\label{epspB}\\
&&\qquad\omega_k^2 \equiv \frac{k^2}{a^2} + m^2\,,\qquad h \equiv \frac{\dot a}{a} = H\,{\rm tanh}\, u\,,
\eea
\label{epspAB}\ees
an overdot denotes $H d/du$, $a = H^{-1} \cosh u$, we have suppressed the $\gamma$ subscript
and also set $\xi = \frac{1}{6}$ (but kept $m\neq0$) in order to simplify the expressions.
This is already sufficiently general for our purposes, as the general case $\xi \neq \frac{1}{6}$ adds
no essentially new features. By using the mode  eq. (\ref{modeq}) satisfied by $\ups_k$
one may readily check that the renormalized stress tensor is covariantly conserved,
\be
H \frac{d \varepsilon_{_{f,N}}}{du\ } + 3 h \, (\varepsilon_{f,N} + p_{f,N}) =
\frac{H}{a^3} \frac{d }{du} (a^3\varepsilon_{_{f,N}})+3 h\, p_{_{f,N}} = 0
\label{cons}
\ee
so that it is sufficient to focus attention on the energy density for the general $O(4)$ invariant state.

Since the renormalization subtractions have already been performed in defining the finite
$p_{\ups} =-\varepsilon_{\ups}$ in the $O(4,1)$ invariant state $\vert \ups\rag$, the additional state
dependent mode sums in (\ref{renormepsp}) must not give rise to any new UV divergences. This implies
that the Bogoliubov coefficients $B_k$ and numbers $N_k$ must satisfy
\be
\lim_{k\rightarrow \infty} \big[k^4 |B_k|\big] = \lim_{k\rightarrow \infty} \big[k^4 |B_k|^2\big]
= \lim_{k\rightarrow \infty} \big[k^4 N_k\big] = 0
\label{UVfinite}
\ee
so that all of the sums over $k$ for the remaining state dependent terms in (\ref{renormepsp}) converge.
States $\vert f\rag$ whose Bogoliubov coefficients satisfy (\ref{UVfinite}) in addition to (\ref{ABcond})
are {\it UV allowed} or {\it UV finite} $O(4)$ invariant states \cite{ShortDistDecohere}.  Finiteness
and conservation are clearly necessary conditions for the expectation value Tr\,$(\rho_{f,N} T^a_{\ b})$
to be used as a source for the semiclassical Einstein equations (\ref{scE}). These properties of $\lag T^a_{\ b} \rag_{_R}$
remain valid for all UV finite states, including those of lower symmetry, provided only that the Bogoliubov coefficients
fall off rapidly enough at large $k$, as in (\ref{UVfinite}).

\begin{figure}[t]
\vspace{-1.5cm}
\begin{center}
\includegraphics[scale=0.3,angle=90,width=3.7in,clip]{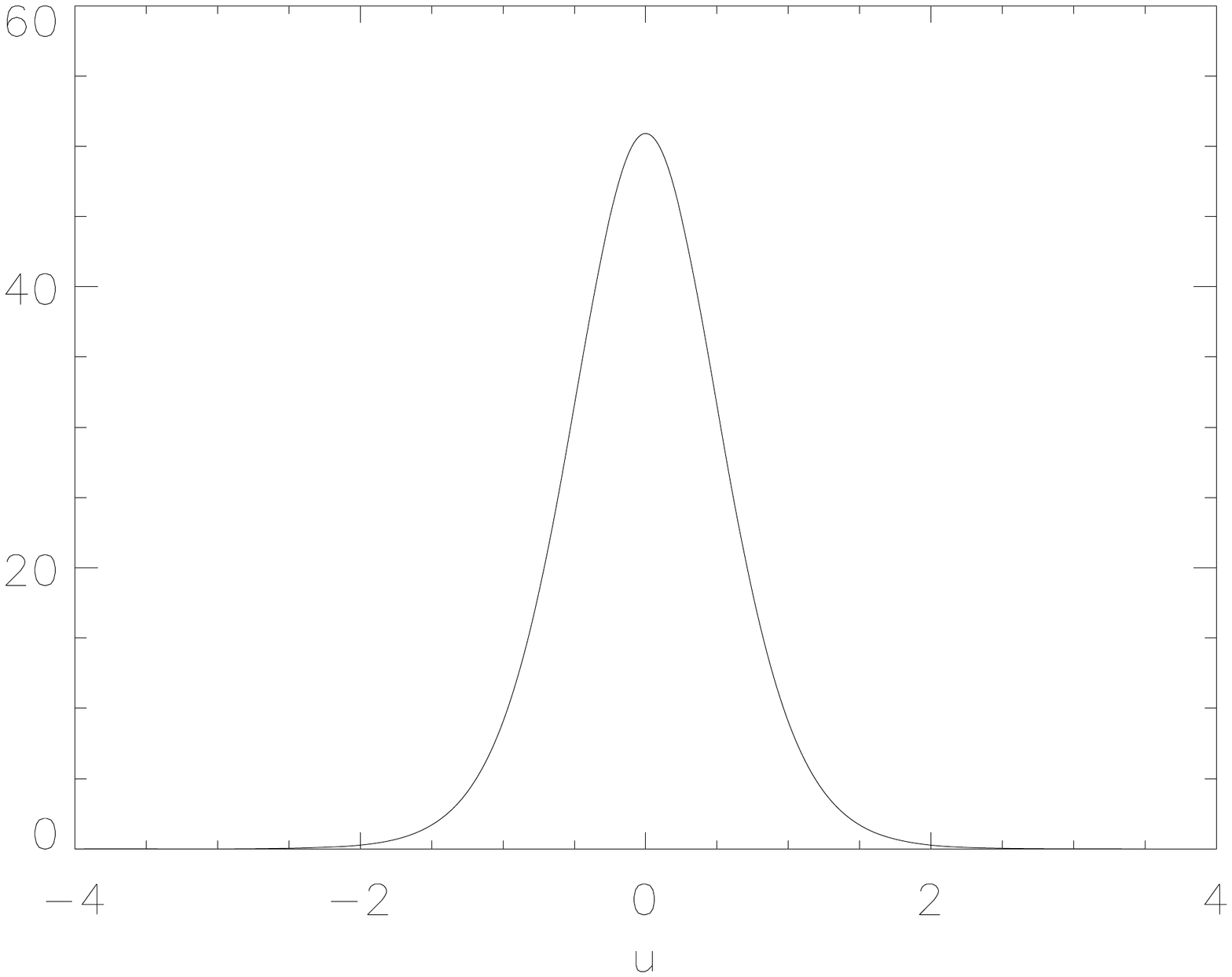}
\vspace{-1cm}
\includegraphics[scale=0.3,angle=90,width=3.2in,clip]{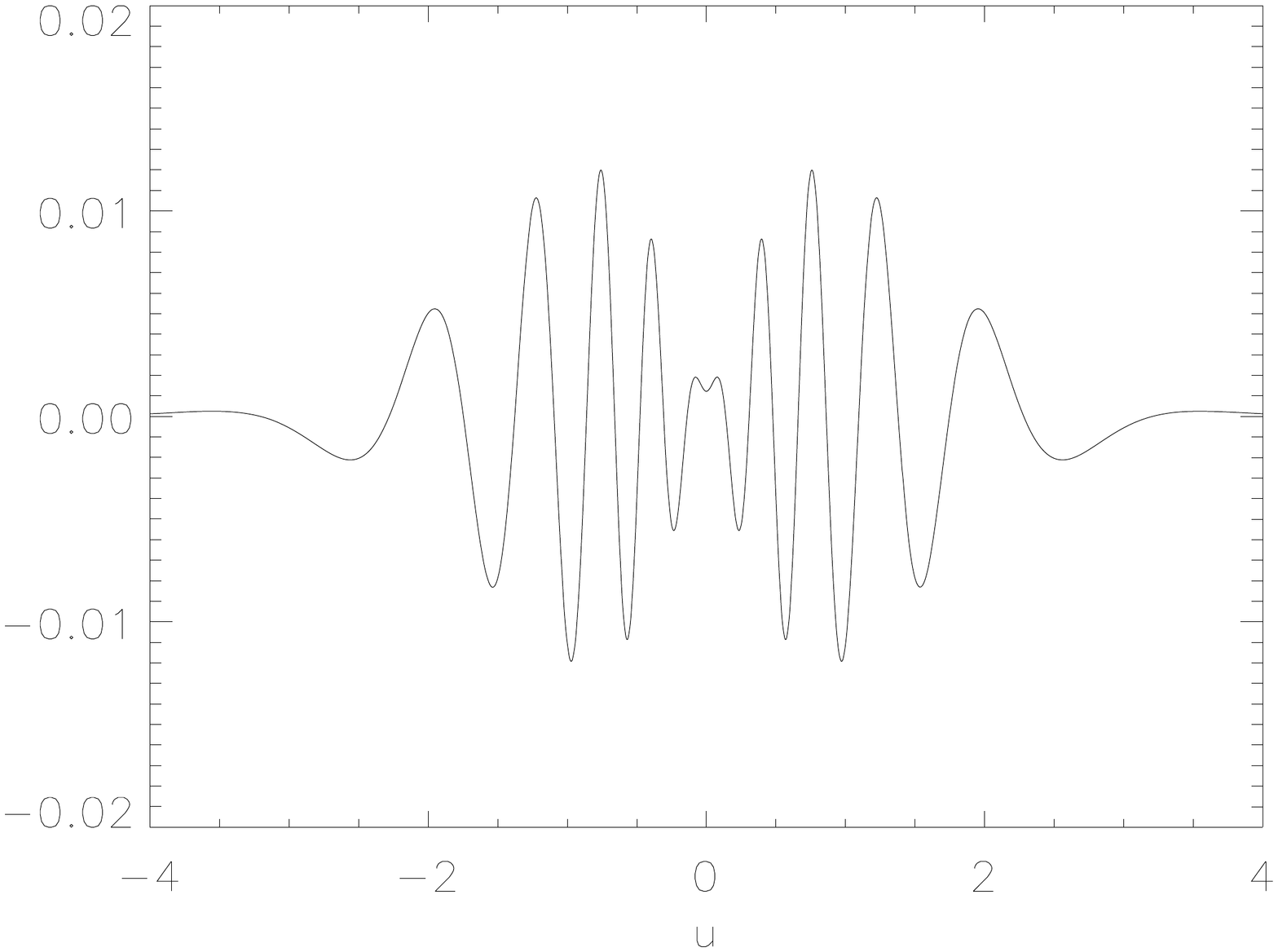}
\includegraphics[scale=0.3,angle=90,width=3.2in,clip]{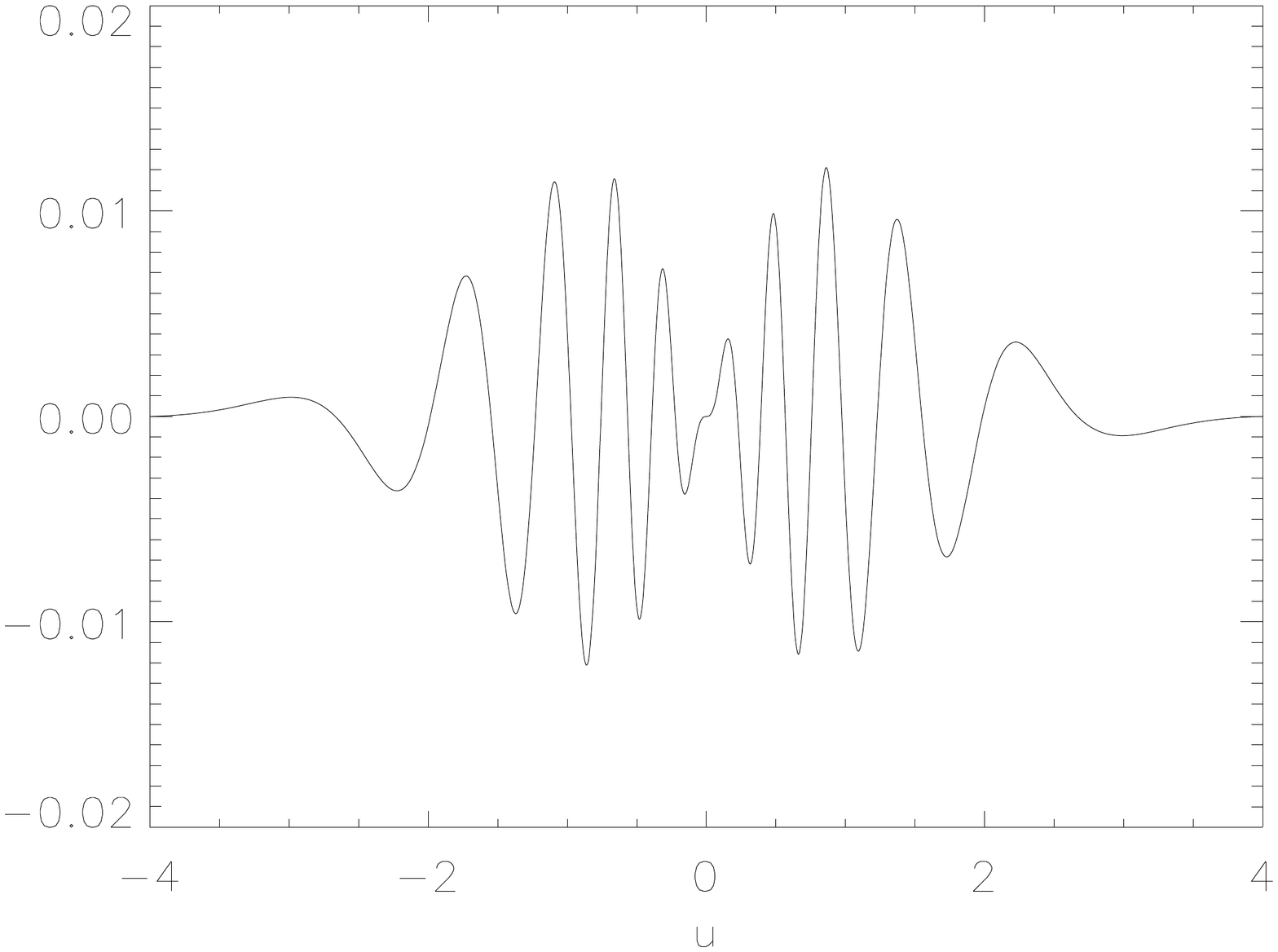}
\caption{The top panel shows the coefficient of the $|B_k|^2$ term in the energy density $k^2\varepsilon^B_k/(2\pi^2)$
of (\ref{renorme}), for $m=H$ and $k=10$ in units of $H^4$. The bottom left panel shows the real part and the bottom right
panel the imaginary part of the coefficient of the $A_k B^*_k$ term in the energy density, namely
$k^2\,{\rm Re}\,\varepsilon_k^A/(2 \pi^2)$ and $-k^2\,{\rm Im}\, \varepsilon_k^A/(2 \pi^2)$ 
respectively of (\ref{renorme}) and (\ref{epspA}) again for $m=H$ and $k=10$, in the same units.}
\vspace{-2mm}
\label{Fig:epsAM1K10}
\end{center}
\vspace{-1cm}
\end{figure}

\begin{figure}[t]
\vspace{-1.2cm}
\includegraphics[scale=0.3,angle=90,width=3.2in,clip]{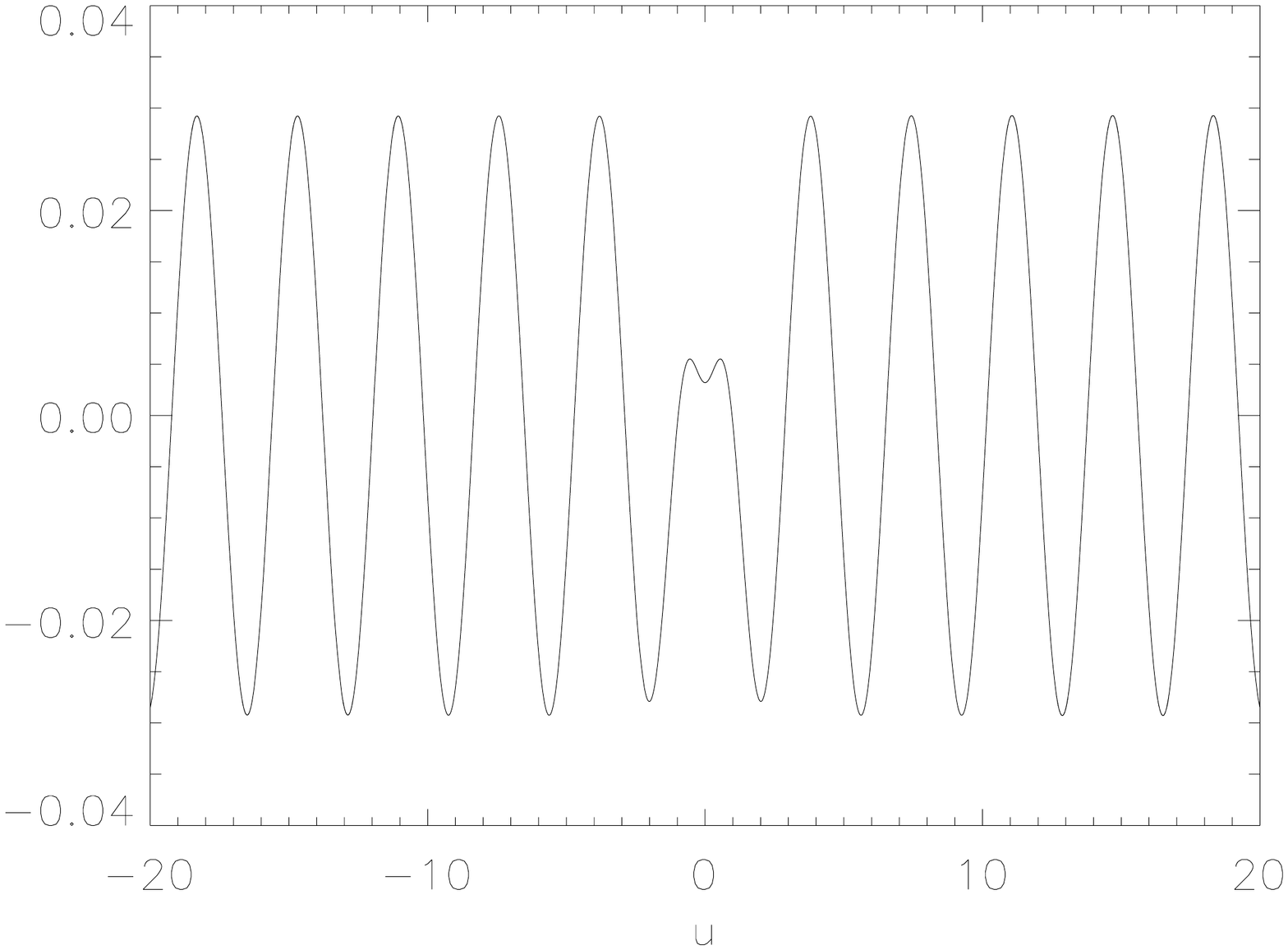}
\includegraphics[scale=0.3,angle=90,width=3.2in,clip]{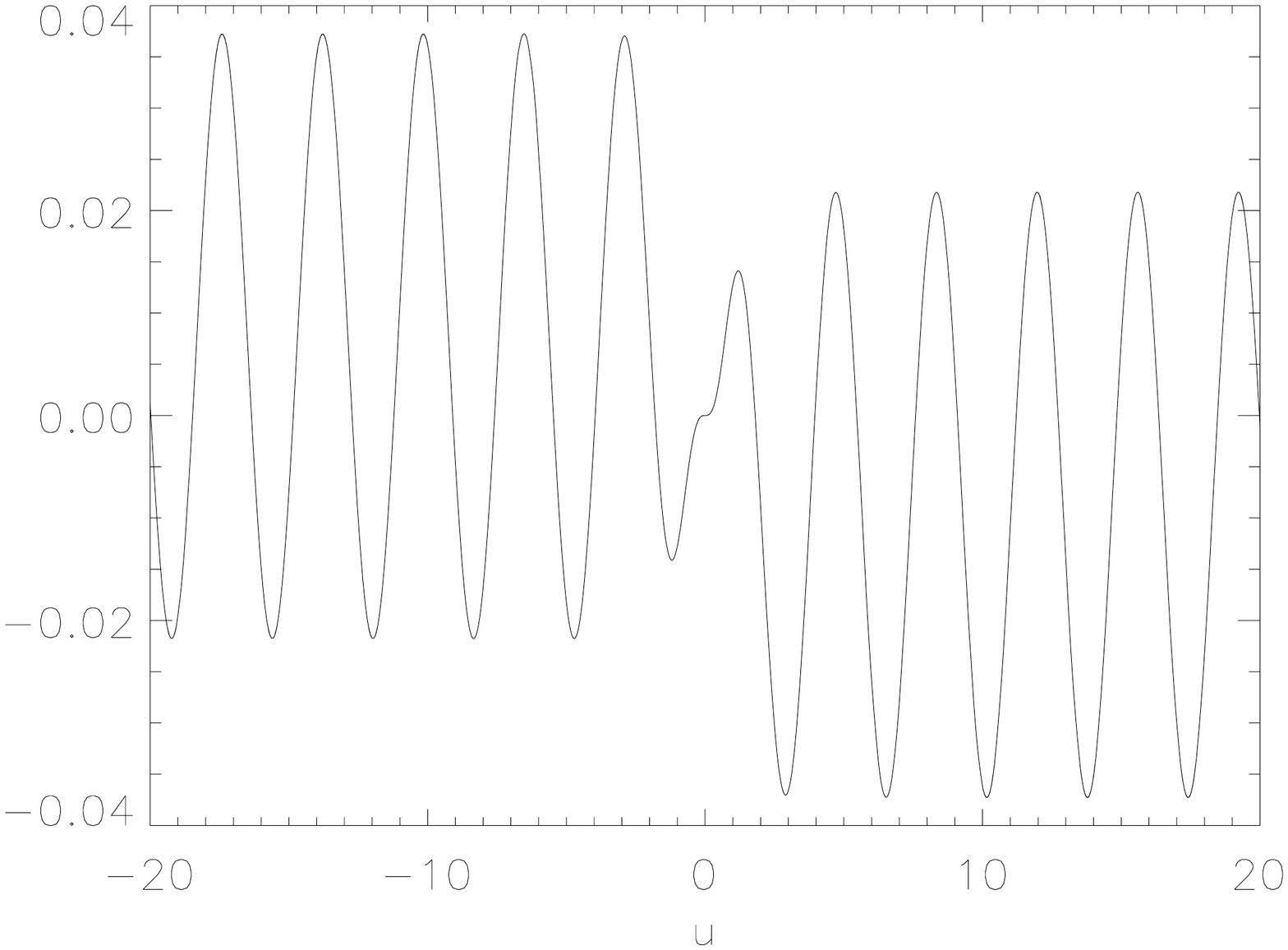}
\includegraphics[scale=0.3,angle=90,width=3.2in,clip]{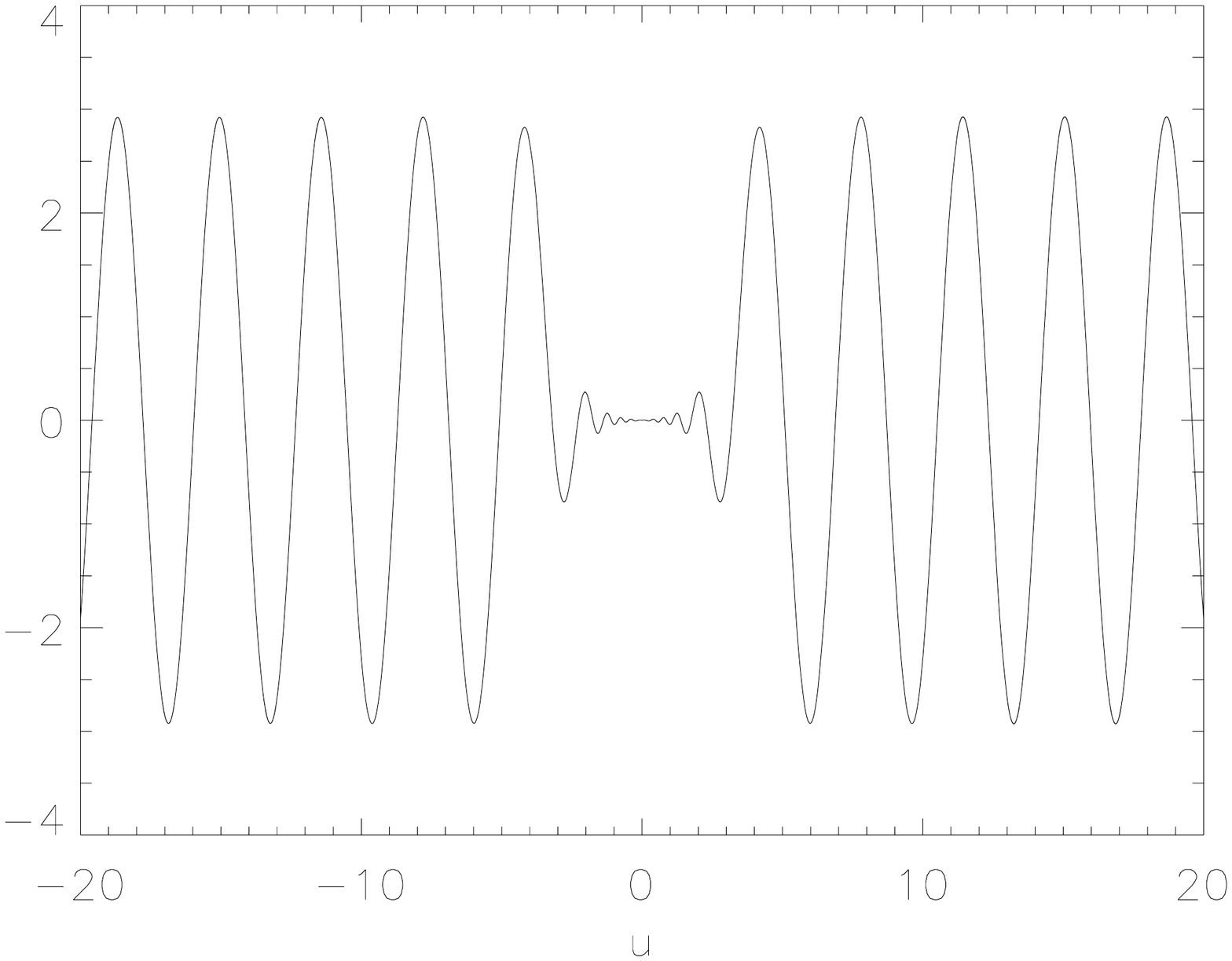}
\includegraphics[scale=0.3,angle=90,width=3.2in,clip]{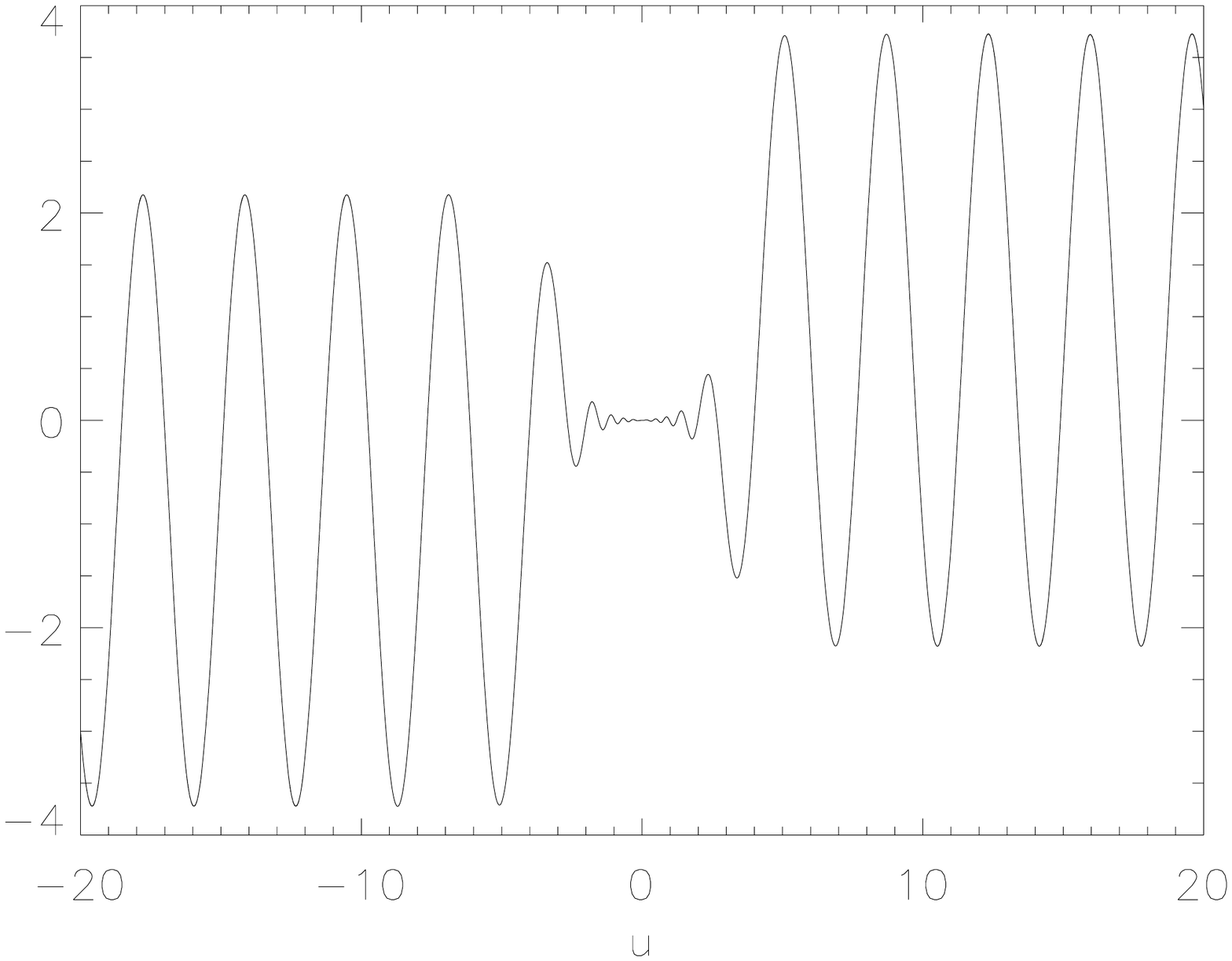}
\includegraphics[scale=0.3,angle=90,width=3.2in,clip]{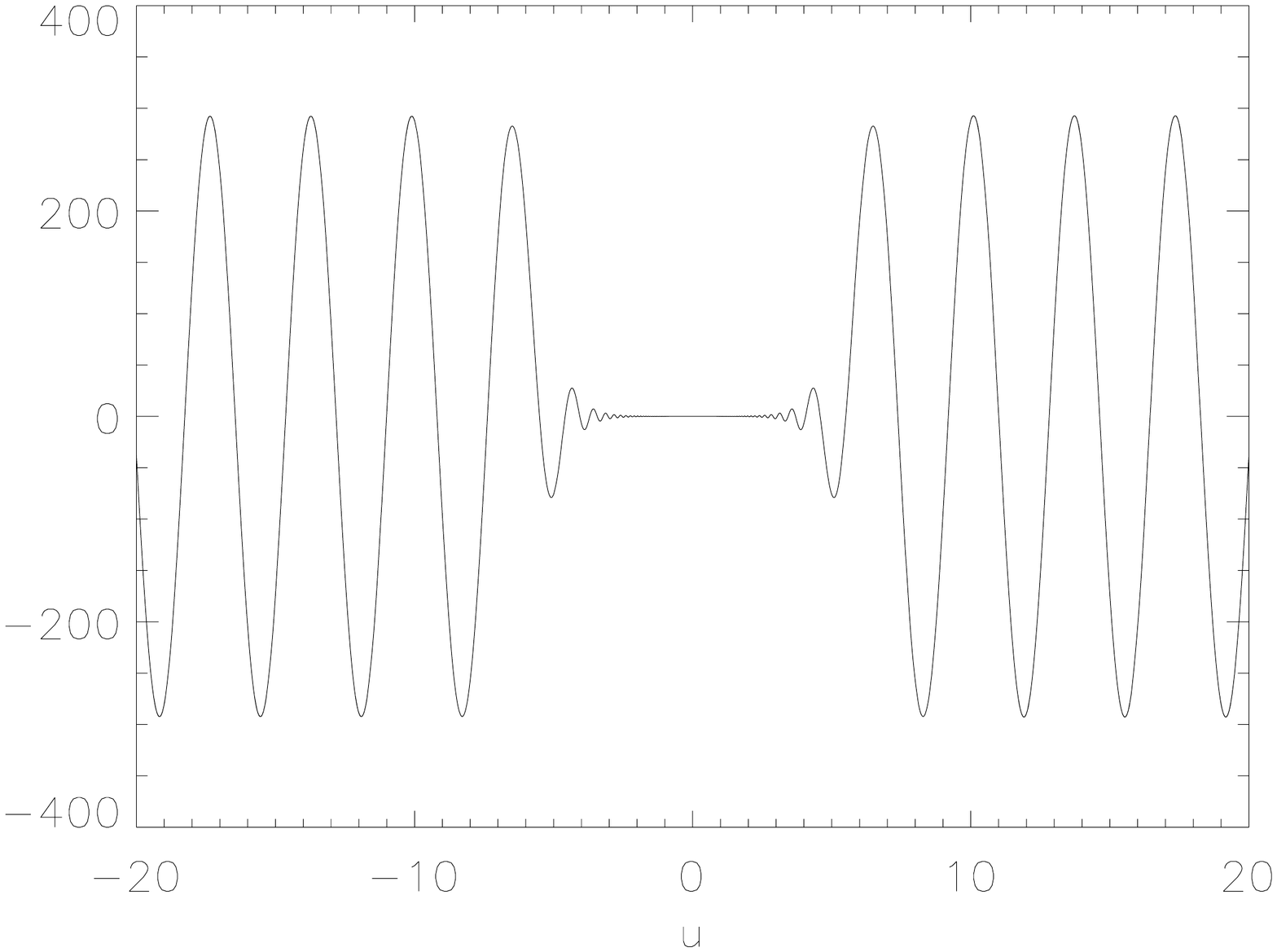}
\includegraphics[scale=0.3,angle=90,width=3.2in,clip]{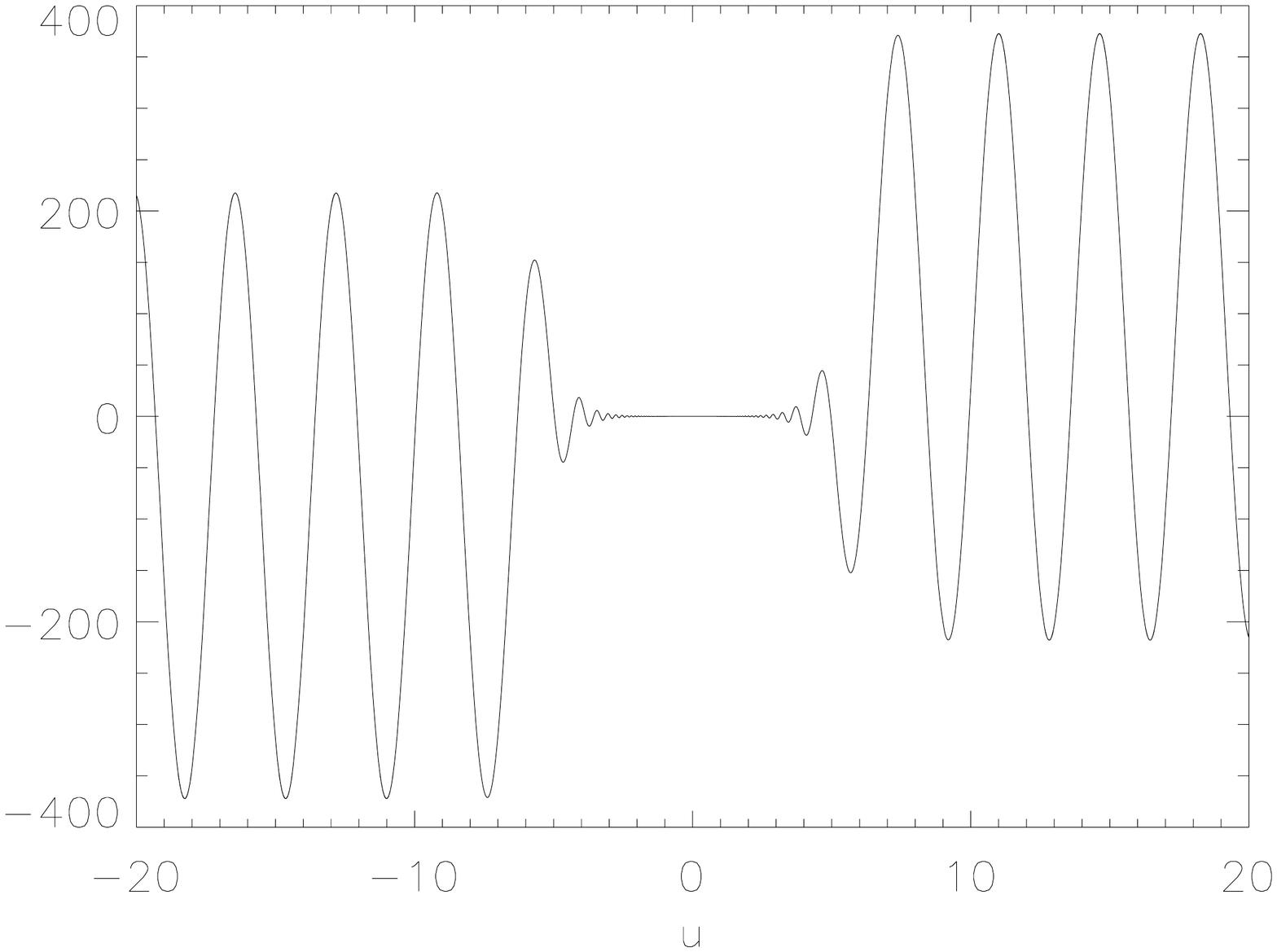}
\vspace{-8mm}
\caption{The panels at left show, for $m = H$, the summand $k^2\,{\rm Re}\,\varepsilon_k^A/(2 \pi^2)$
in the energy density of (\ref{renorme}) and (\ref{epspA}) multiplied by $a^3$, with
$a = H^{-1}\cosh u$ the scale factor.  The panels at right show $-a^3 k^2\,{\rm Im}\, \varepsilon_k^A/(2 \pi^2)$.
From top to bottom the values of $k$ are $k = 1$, $k = 10$, and $k = 100$.
The values of $u_{k\gamma}$ from (\ref{ukgam}) are $0.88, \,3.14$ and $5.44$ respectively. The plots show
that the envelope of the oscillations is proportional to $a^{-3}$ for $|u|> u_{k\gamma}$, but that the behavior
changes markedly for $|u| < u_{k\gamma}$.}
\label{Fig:epsAa3M1}
\vspace{-8mm}
\end{figure}

As in the current expectation value of Sec. \ref{Sec:ConstantE} we seek a qualitative understanding of the terms
contributing to the energy density in (\ref{renorme}) and (\ref{epspAB}). There are three kinds of terms
for a given $k$ in a general $O(4)$ invariant UV finite state, namely those multiplying
the factors $|B_k|^2$, ${\rm Re} (A_k B_k^*)$, and ${\rm Im} (A_k B_k^*)$ respectively. These are plotted in
Figs. \ref{Fig:epsAM1K10} through Fig. \ref{Fig:epsBa34M10}.  In Fig.\, \ref{Fig:epsAM1K10} the three summands in
(\ref{renormepsp}), namely $k^2\varepsilon_k^B/(2 \pi^2)$, $k^2\,{\rm Re}\,\varepsilon_k^A/(2 \pi^2)$ and
$-k^2\,{\rm Im}\, \varepsilon_k^A/(2 \pi^2)$ are shown in units of $H^4$ for the case $m = H$ and $k = 10$.
The $\varepsilon_k^A$ terms multiplying the complex $A_kB_k^*$ coefficient in (\ref{renorme}) are oscillatory,
while the $\varepsilon_k^B$ function multiplying the real coefficient $N_k + |B_k|^2(1+2N_k)$ is non-oscillatory.
The main difference between the coefficients of the real and the imaginary parts of $A_k B_k^*$ is that the former is symmetric
about $u = 0$  while the latter is antisymmetric. The plots also show that the maxima of the two oscillatory functions occur for
$|u|$ of order one, while the maximum of the third, non-oscillatory function is at the symmetric point $u=0$ and much larger
in magnitude. In all three cases the functions fall off for large values of the time $|u|$ where the scale factor $a(u)$ is large.

\begin{figure}[t]
\vspace{-1.2cm}
\includegraphics[scale=0.3,angle=90,width=3.2in,clip]{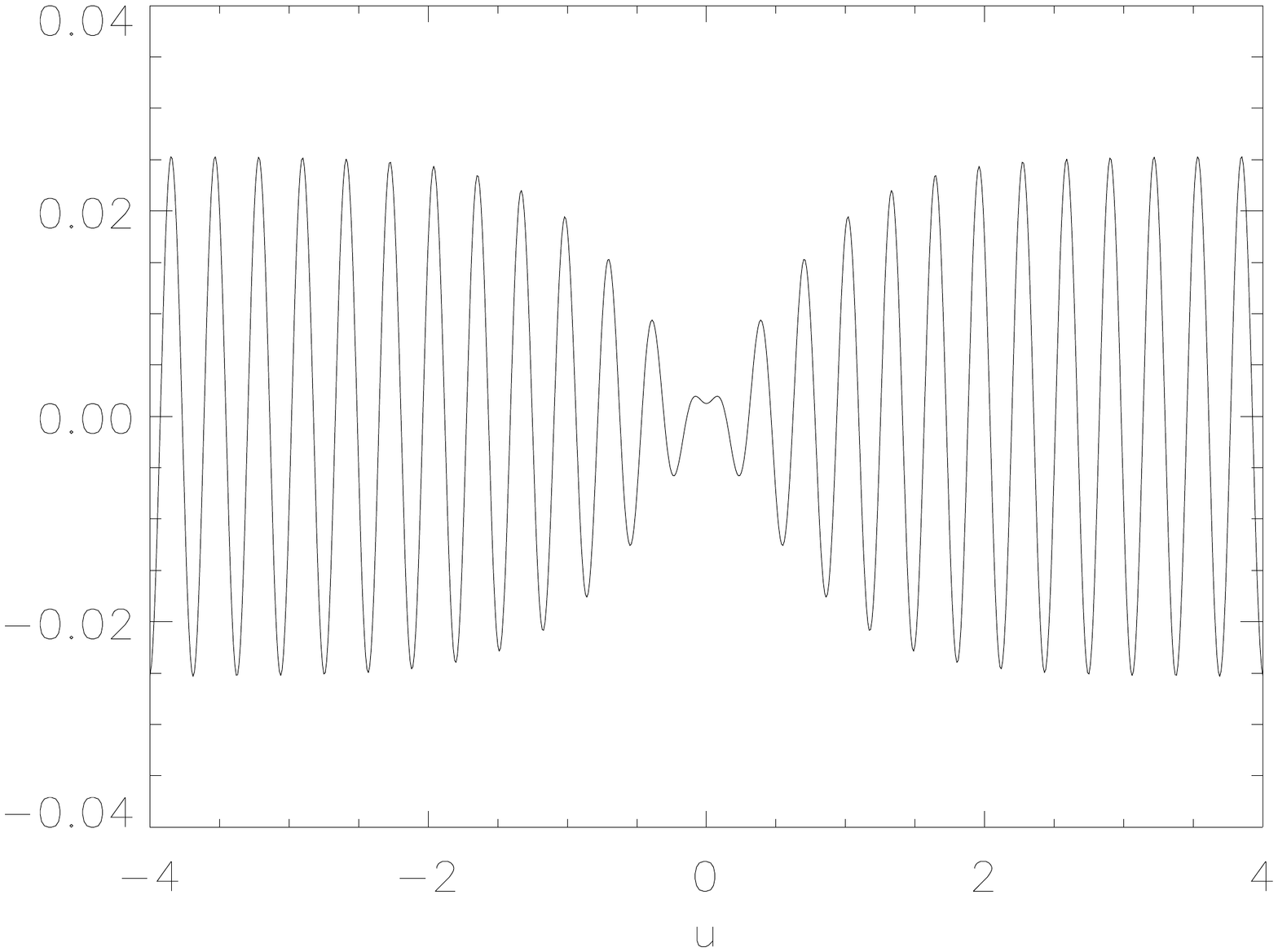}
\includegraphics[scale=0.3,angle=90,width=3.2in,clip]{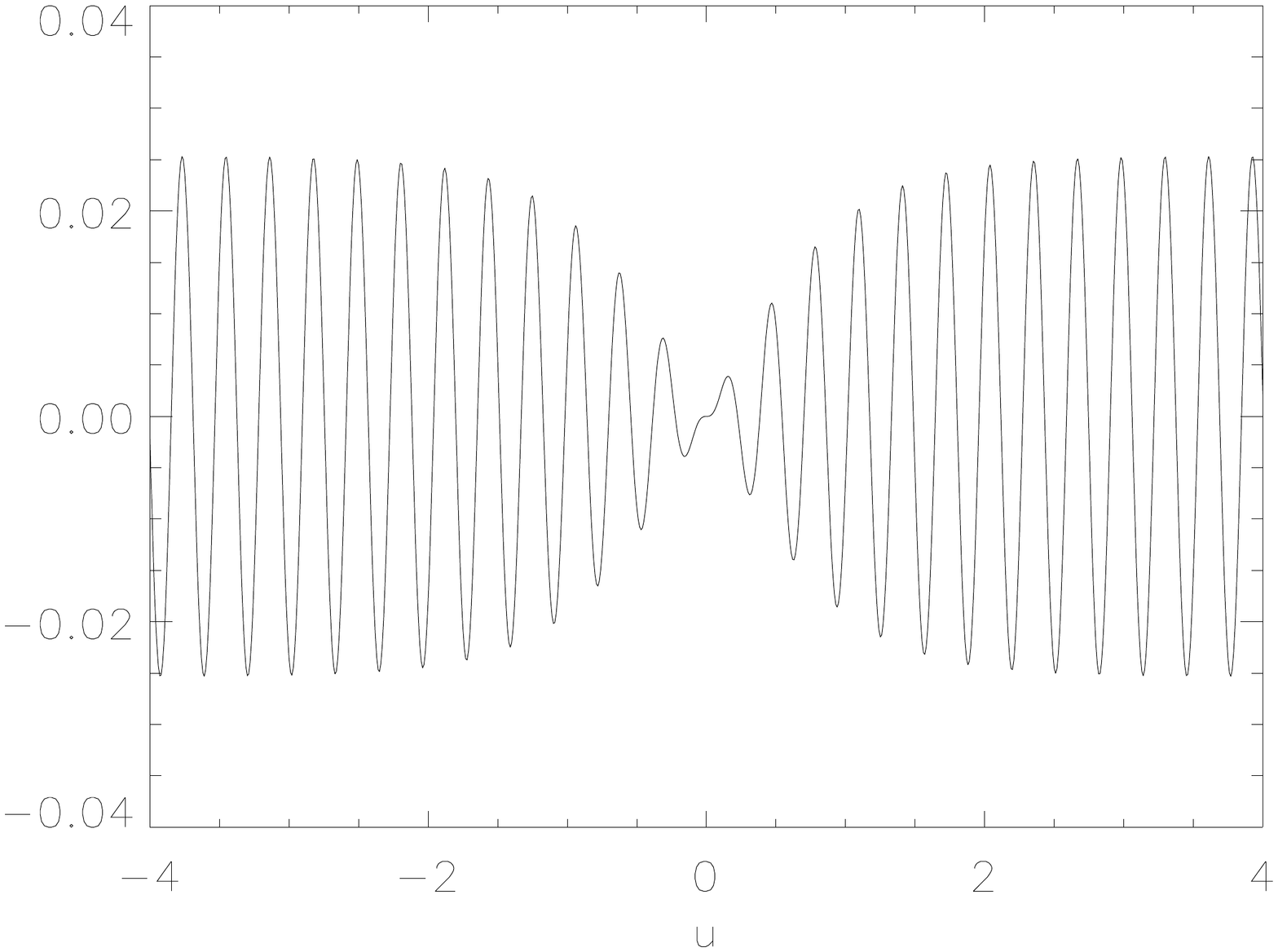}
\includegraphics[scale=0.3,angle=90,width=3.2in,clip]{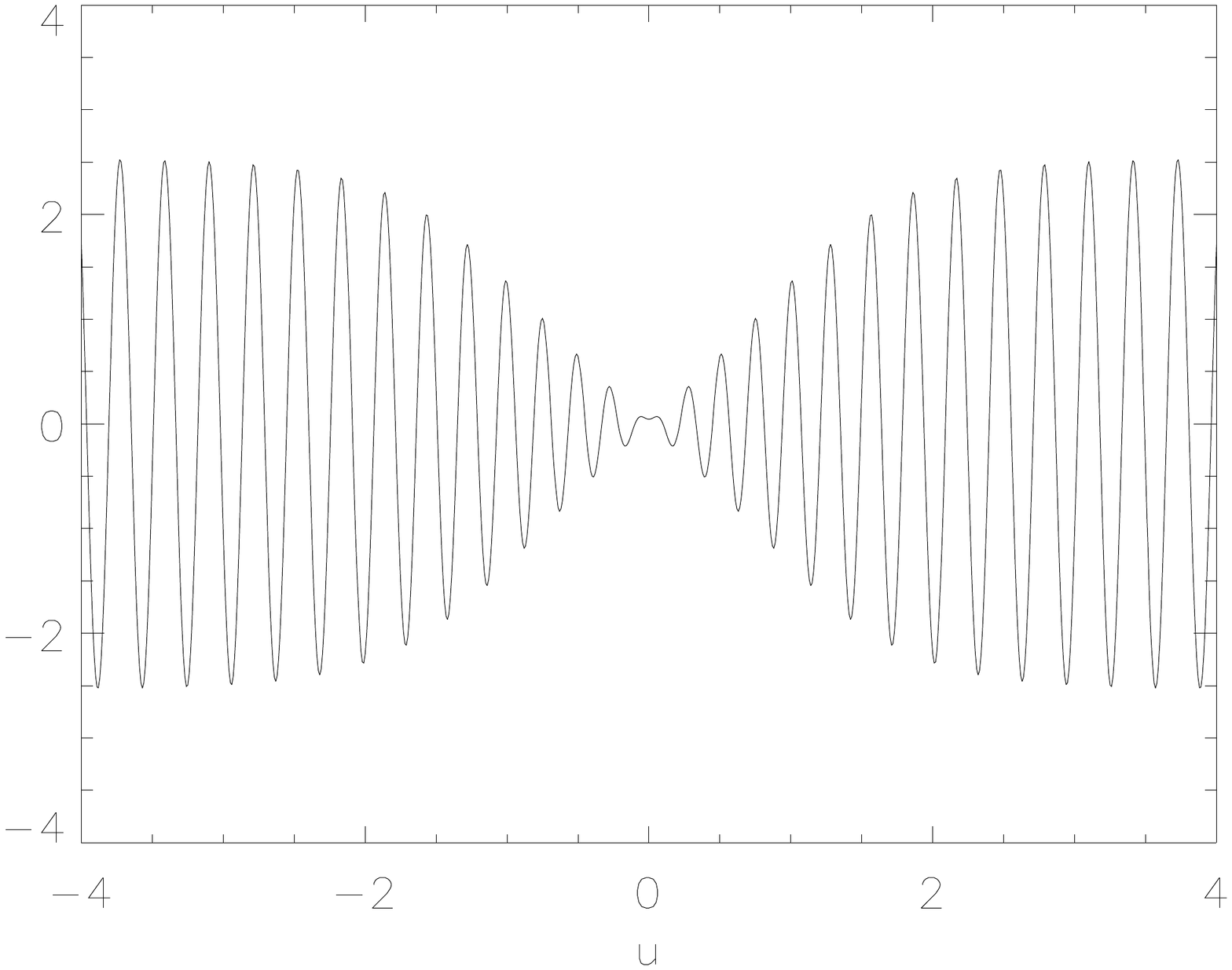}
\includegraphics[scale=0.3,angle=90,width=3.2in,clip]{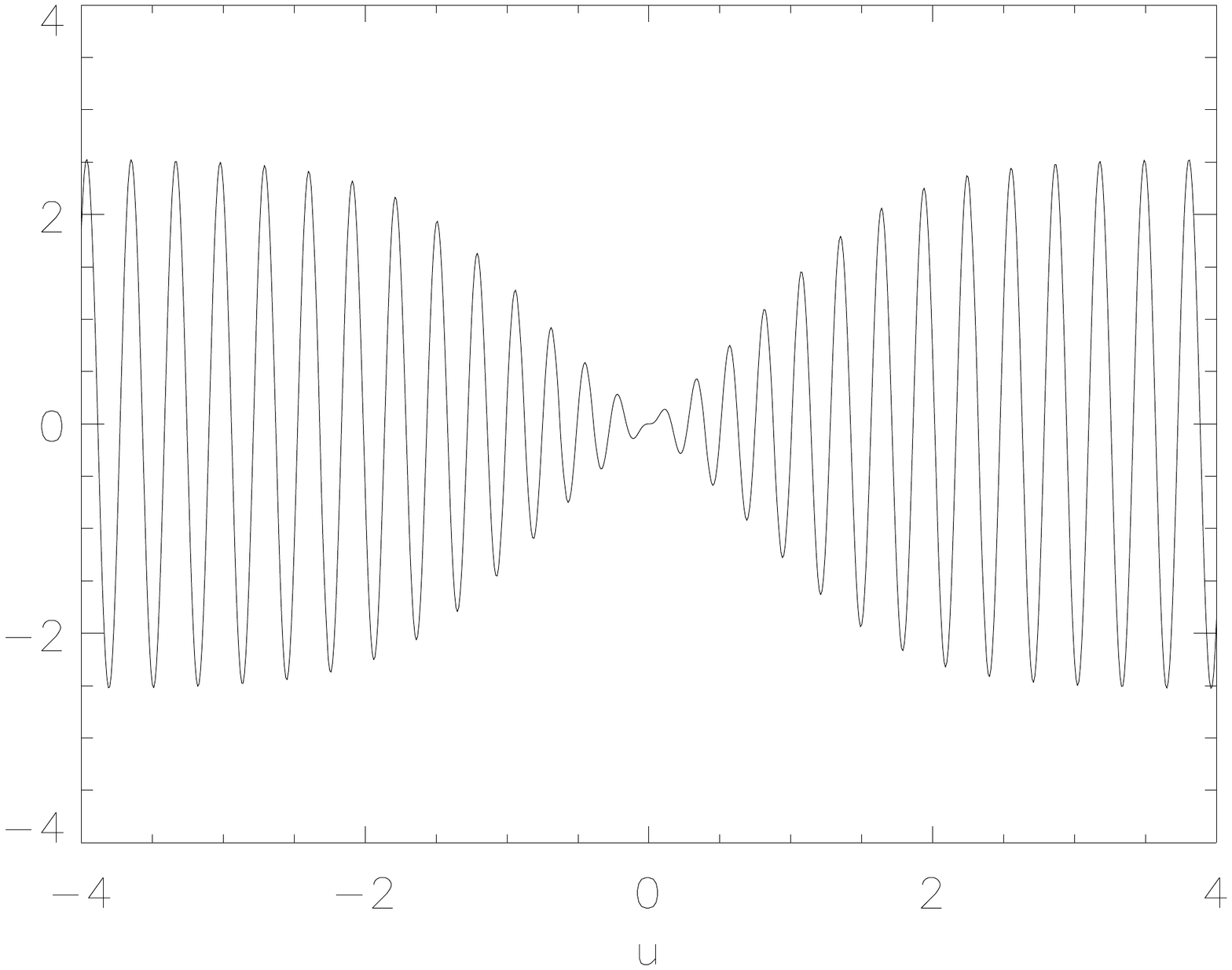}
\includegraphics[scale=0.3,angle=90,width=3.2in,clip]{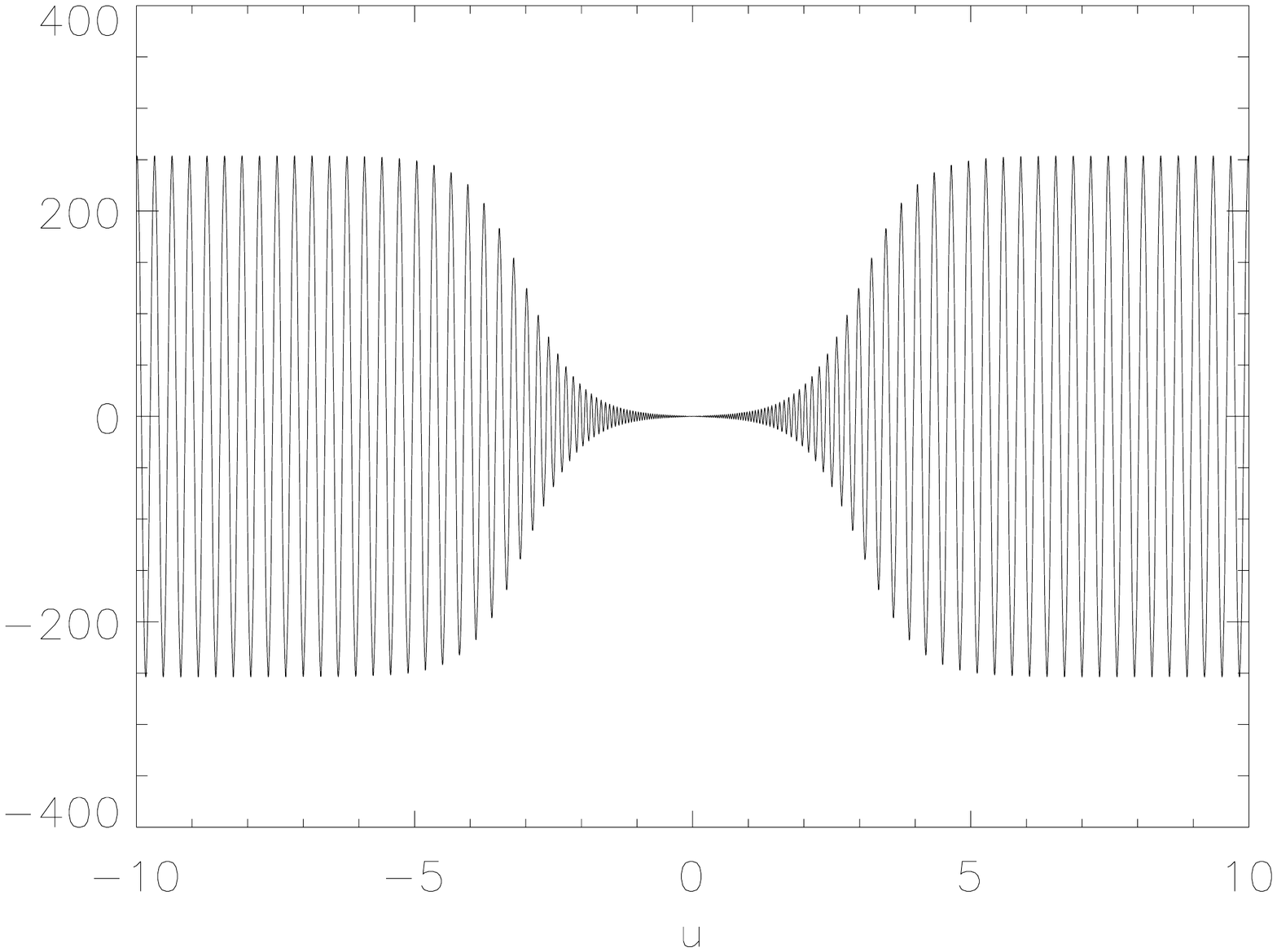}
\includegraphics[scale=0.3,angle=90,width=3.2in,clip]{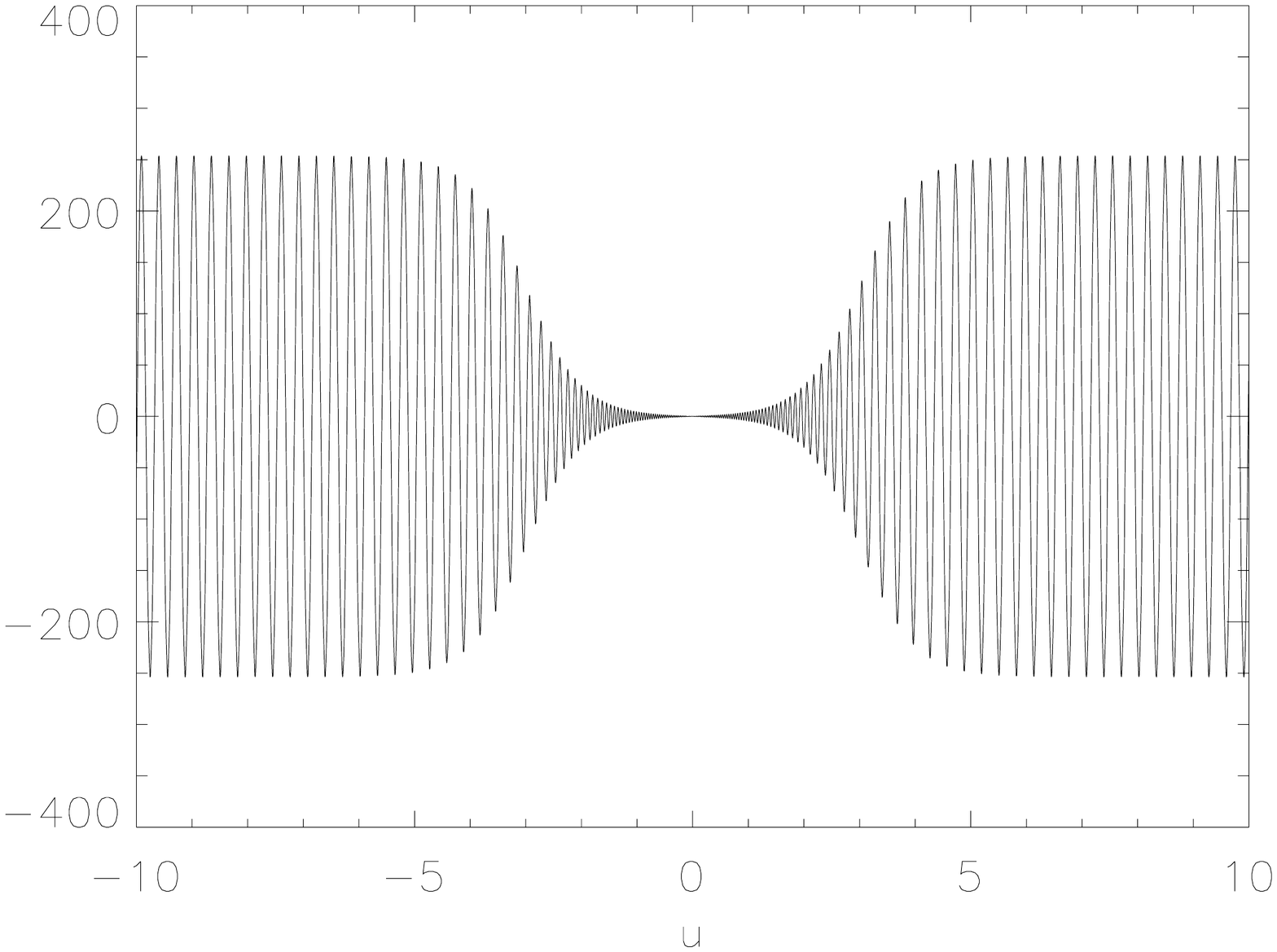}
\vspace{-1cm}
\caption{The panels at left show, for $m = 10H$, the summand $k^2\,{\rm Re}\,\varepsilon_k^A/(2 \pi^2)$
in the energy density of (\ref{renorme}) and (\ref{epspA}) multiplied by $a^3$, with
$a = H^{-1}\cosh u$ the scale factor.  The panels at right show $-a^3 k^2\,{\rm Im}\, \varepsilon_k^A/(2 \pi^2)$.  
From top to bottom the values of $k$ are $k = 1$, $k = 10$, and $k = 100$, with the corresponding values 
of $u_{k\gamma}$ from (\ref{ukgam}) $0.087,\, 0.88$ and $3.00$ respectively, where the behavior changes markedly.
It is also observed that the envelopes for $|u| > u_{k\gamma}$ also scale like $k^2$.}
\label{Fig:epsAa3M10}
\vspace{-7mm}
\end{figure}

Since the field is massive one might expect that at large values of the scale factor the contributions to the energy density
would scale like $a^{-3}$. To illustrate the power dependence on the scale factor we plot in Figs. \ref{Fig:epsAa3M1} and
\ref{Fig:epsAa3M10} the coefficients of the real and imaginary parts of $(1+ 2N_k) A_k B_k^*$ multiplied by $a^3$
for $k=1, 10, 100$ and $m=H$ and $m=10H$ respectively. We observe that the oscillations have an envelope
which does scale like $a^3$ for large $|u|$ and large $a(u)$. The envelope also scales like $k^2$ independently
of $m$, so that if we were to sum modes up to a large but finite $K$, we would expect a $K^3/a^3$ behavior
characteristic of a non-relativistic gas. However the rapid oscillations, particularly for larger values of $m$ and $k$,
highlight the fact that these are highly coherent quantum states, and the energy density is not that of
quasi-classical particles in any sense.

The $a^{-3}$ behavior of the envelope of the oscillations also does not hold for small $|u|$. As shown in detail
by a WKB analysis of the mode eq. (\ref{modeq}) in the accompanying paper \cite{AndMot1}, the mode
functions and adiabatic vacuum state change character around the times
$u =\mp u_{k\gamma}$ where
\be
u_{k\gamma} = \ln \left[ \frac{\sqrt{k^2 - \frac{1}{4}} + \sqrt{\gamma^2 + k^2 - \frac{1}{4}}}{\gamma}\right]\,.
\label{ukgam}
\ee
The modes are non-relativistic for $|u| > u_{k\gamma}$, but relativistic for $|u| < u_{k\gamma}$. 
For a conformal massless field $m=0$, with $\ups_k = \ups_{k,\frac{1}{2}}$ of (\ref{m0conf}),
$\varepsilon^A_k $ of (\ref{epspA}) vanishes identically. This accounts for the much smaller values of the energy densities
in Figs. \ref{Fig:epsAa3M1} and \ref{Fig:epsAa3M10} in the central regions where $-u_{k\gamma} < u < + u_{k\gamma}$,
where there is no simple behavior of the envelope of the quantum coherent oscillations.
The maximum of the oscillatory terms occurs for all values of $k$ and $m$ investigated at $|u| \sim 1$ in the central region.
This maximum saturates at a value of order one in $H^4$ units for large $k\gg 1$, as shown in Fig. \ref{Fig:epsmax}.

\begin{figure}[h]
\vspace{-8mm}
\includegraphics[scale=1,angle=90,width=3.8in,clip]{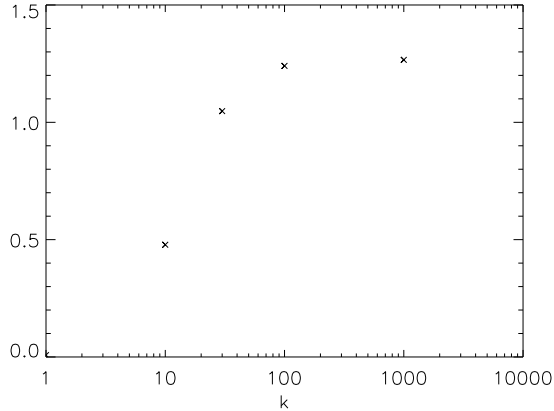}
\vspace{-1cm}
\caption{ The maxima of the oscillations of $k^2 Re \, \varepsilon_k^A/(2\pi^2)$ are plotted for several values of $k$.  The saturation of the value of maximum at large values of $k$ is apparent. }
\vspace{-5mm}
\label{Fig:epsmax}
\end{figure}

The non-oscillatory $k^2\varepsilon_k^B/(2 \pi^2)$ term in the energy density is shown in Figs. \ref{Fig:epsBa34M1}
and \ref{Fig:epsBa34M10} for $m=H$ and $m=10H$ respectively, for $k=1, 10$ and $100$. In the left panels
the term is multiplied by $a^3(u)$ and in the right panels the term is multiplied by $a^4(u)$. It is clear that in all 
cases the contribution from this $\varepsilon_k^B$ term is proportional to $a^{-3}$ at large values of the scale 
factor and is proportional to $a^{-4}$ near $u = 0$. In other words, it blueshifts  in the contracting phase of de Sitter 
space (and redshifts in the expanding phase) as a non-relativistic fluid for large $|u|$ but as a relativistic fluid for 
smaller $|u|$. At  $|u| = u_{k\gamma}$ given by (\ref{ukgam}), the energy density transitions from 
non-relativistic to relativistic behavior and for smaller $|u|$ the physical momentum $k/a$ dominates the mass term 
in the mode eq. (\ref{oscmode}). In the  non-relativistic region $u > |u_{k\gamma}|$ the $k$ dependence is $k^2$. 
However the maximum of the  $k^2\varepsilon_k^B$ term always occurs in the relativistic region 
$-u_{k\gamma} < u < + u_{k\gamma}$, where the $k$ dependence is $k^3$, so that this maximum value 
grows unbounded for $k\gg1$, in contrast to the oscillatory terms which are bounded for large $k$:  Fig.\, \ref{Fig:epsmax}.

\begin{figure}[t]
\vspace{-1cm}
\includegraphics[scale=0.3,angle=90,width=3.2in,clip]{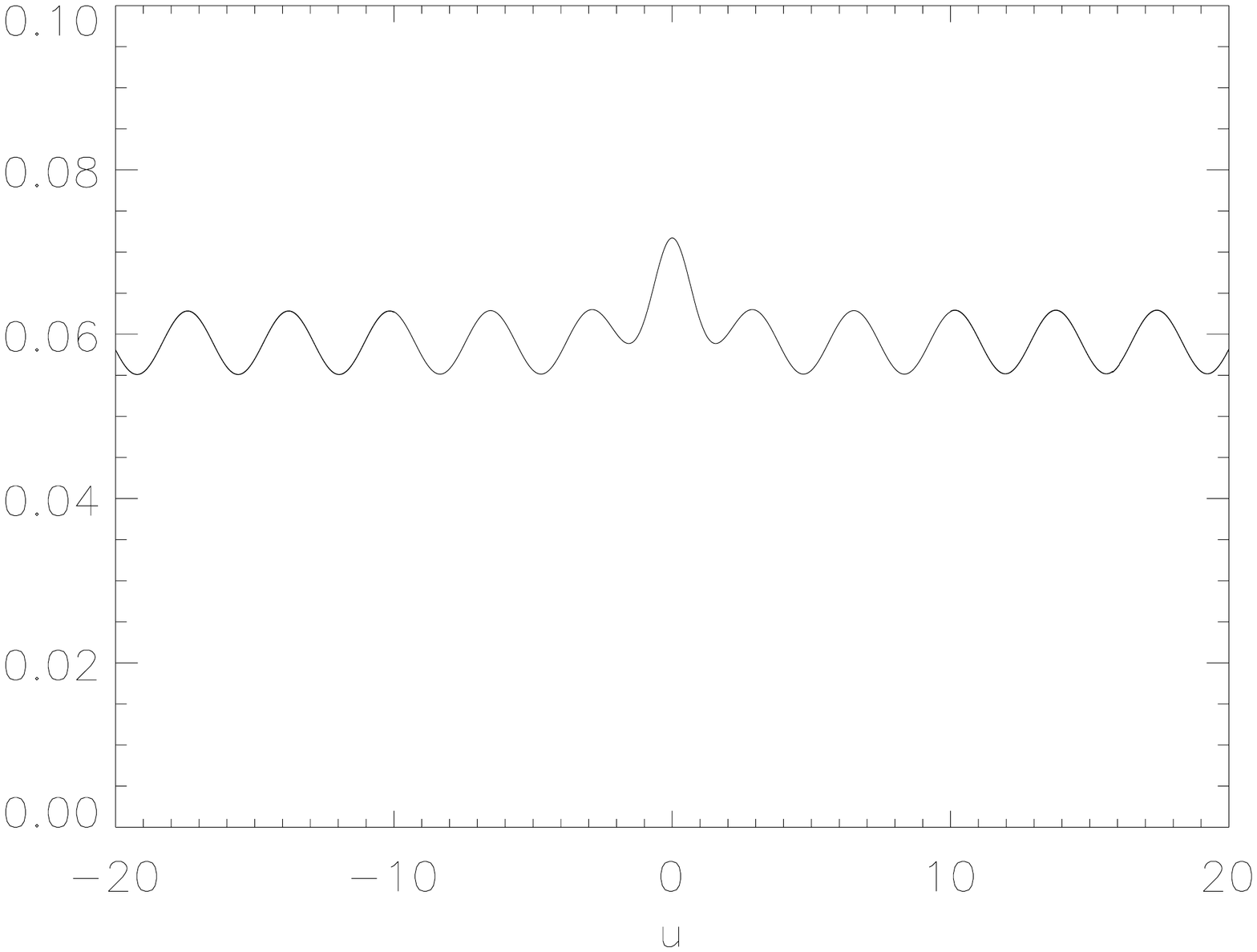}
\includegraphics[scale=0.3,angle=90,width=3.2in,clip]{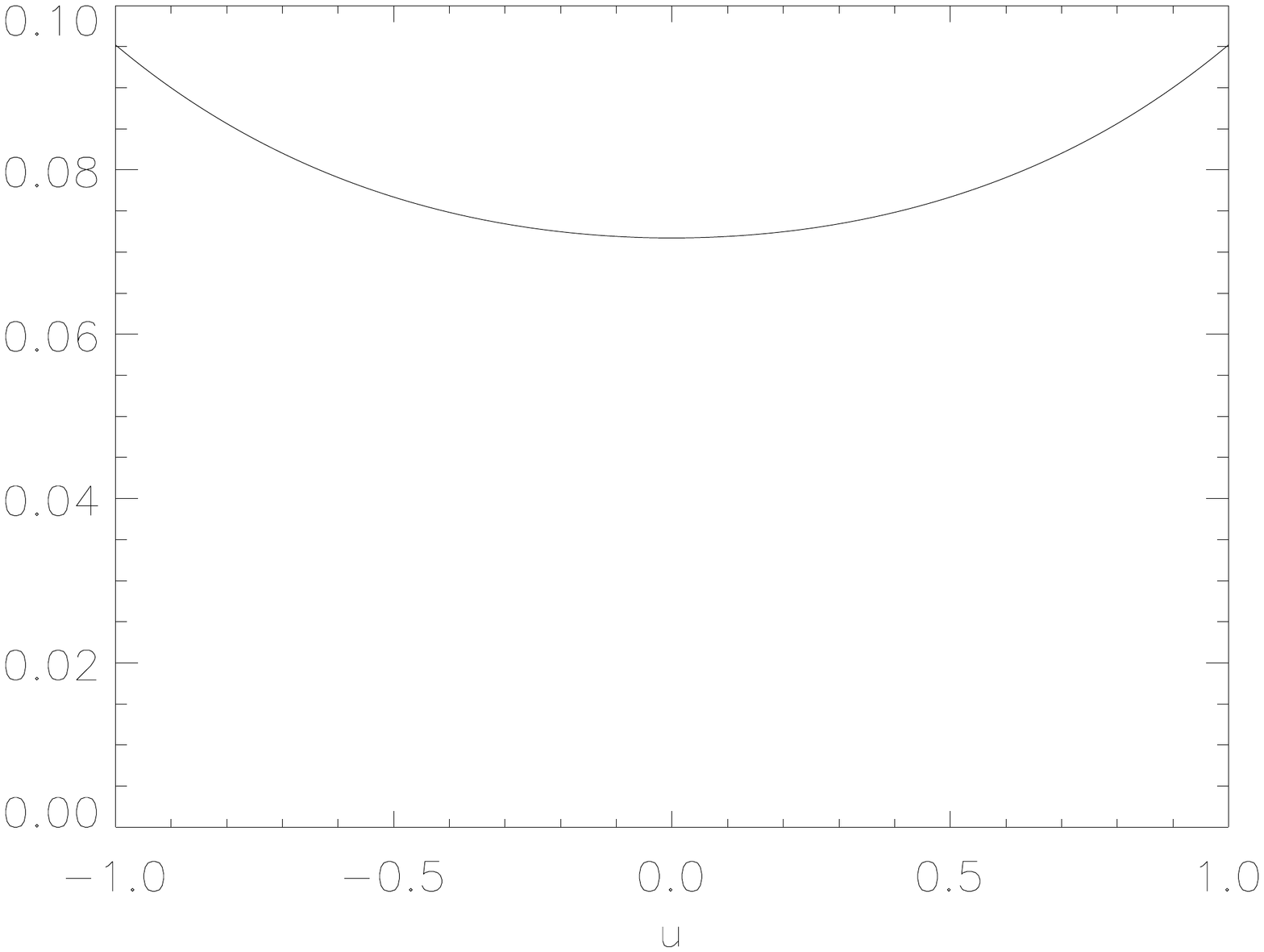}
\includegraphics[scale=0.3,angle=90,width=3.2in,clip]{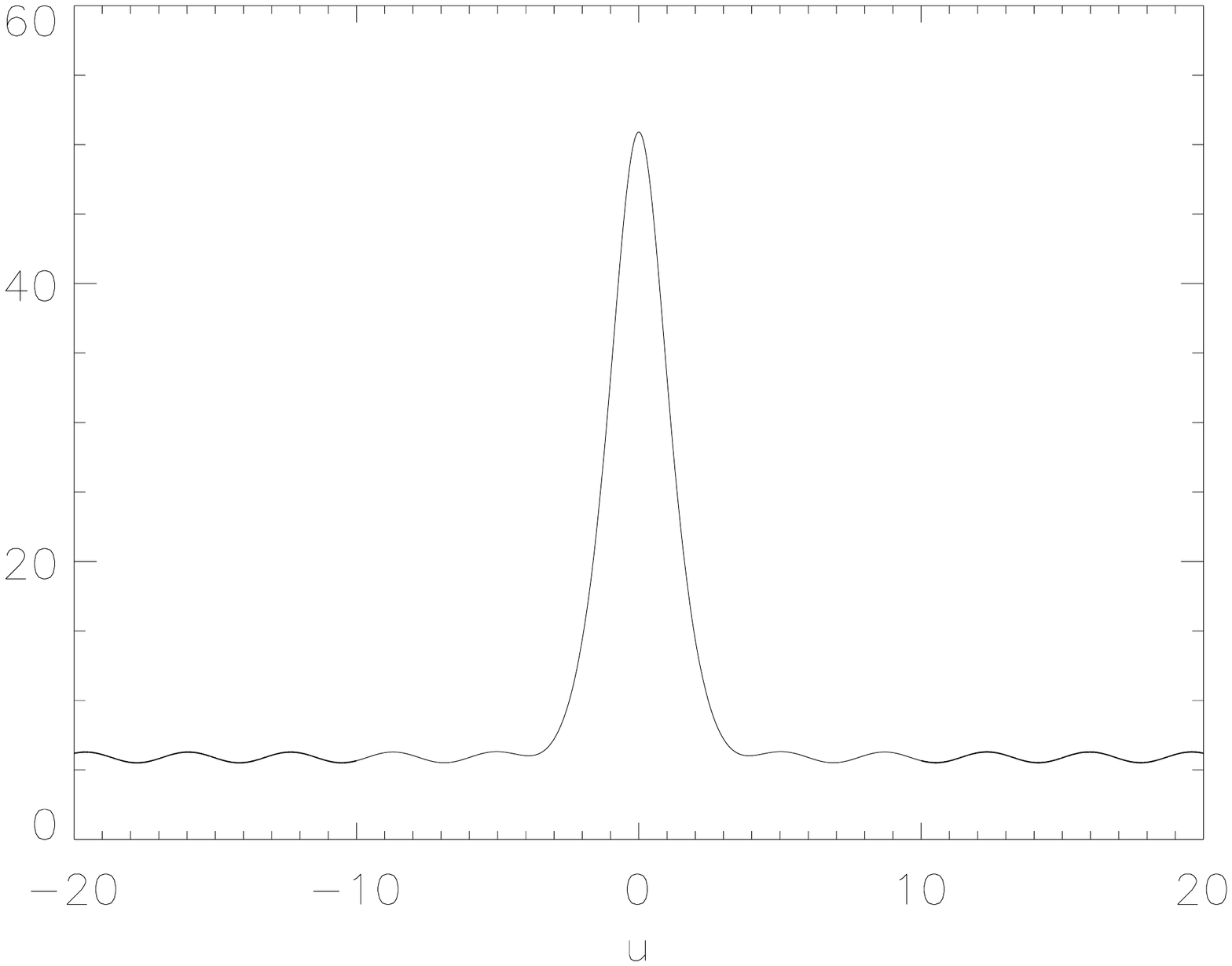}
\includegraphics[scale=0.3,angle=90,width=3.2in,clip]{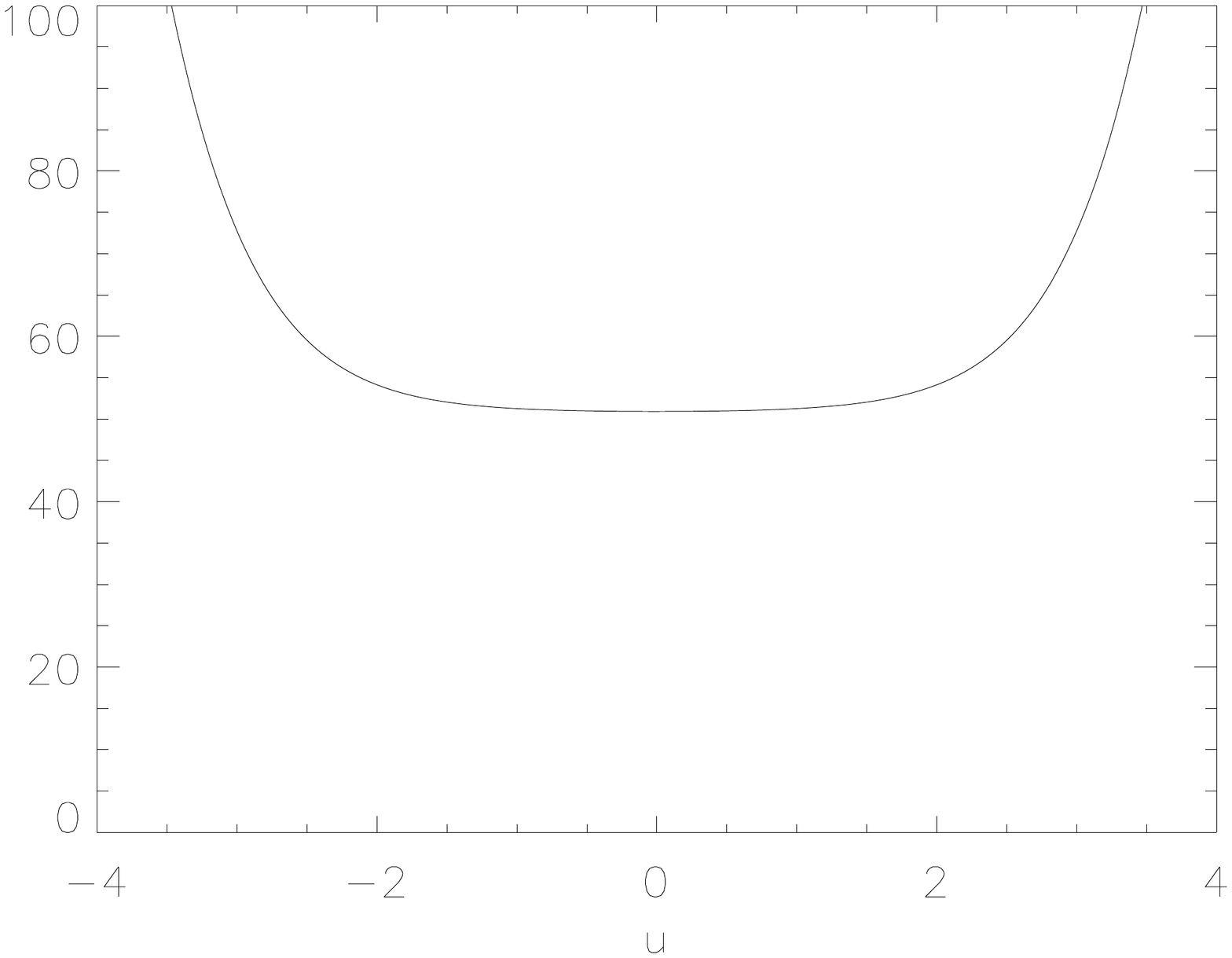}
\includegraphics[scale=0.3,angle=90,width=3.2in,clip]{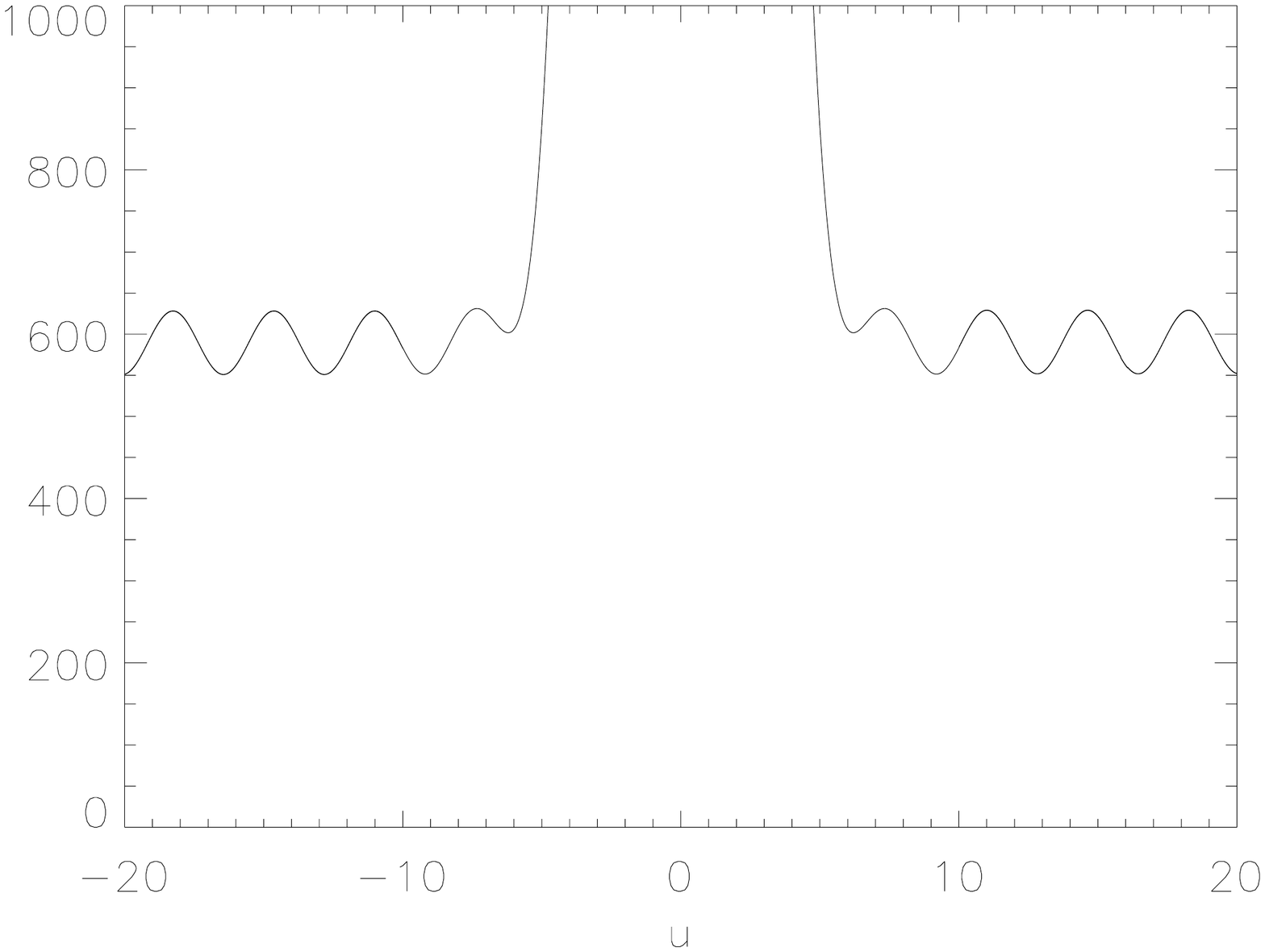}
\includegraphics[scale=0.3,angle=90,width=3.2in,clip]{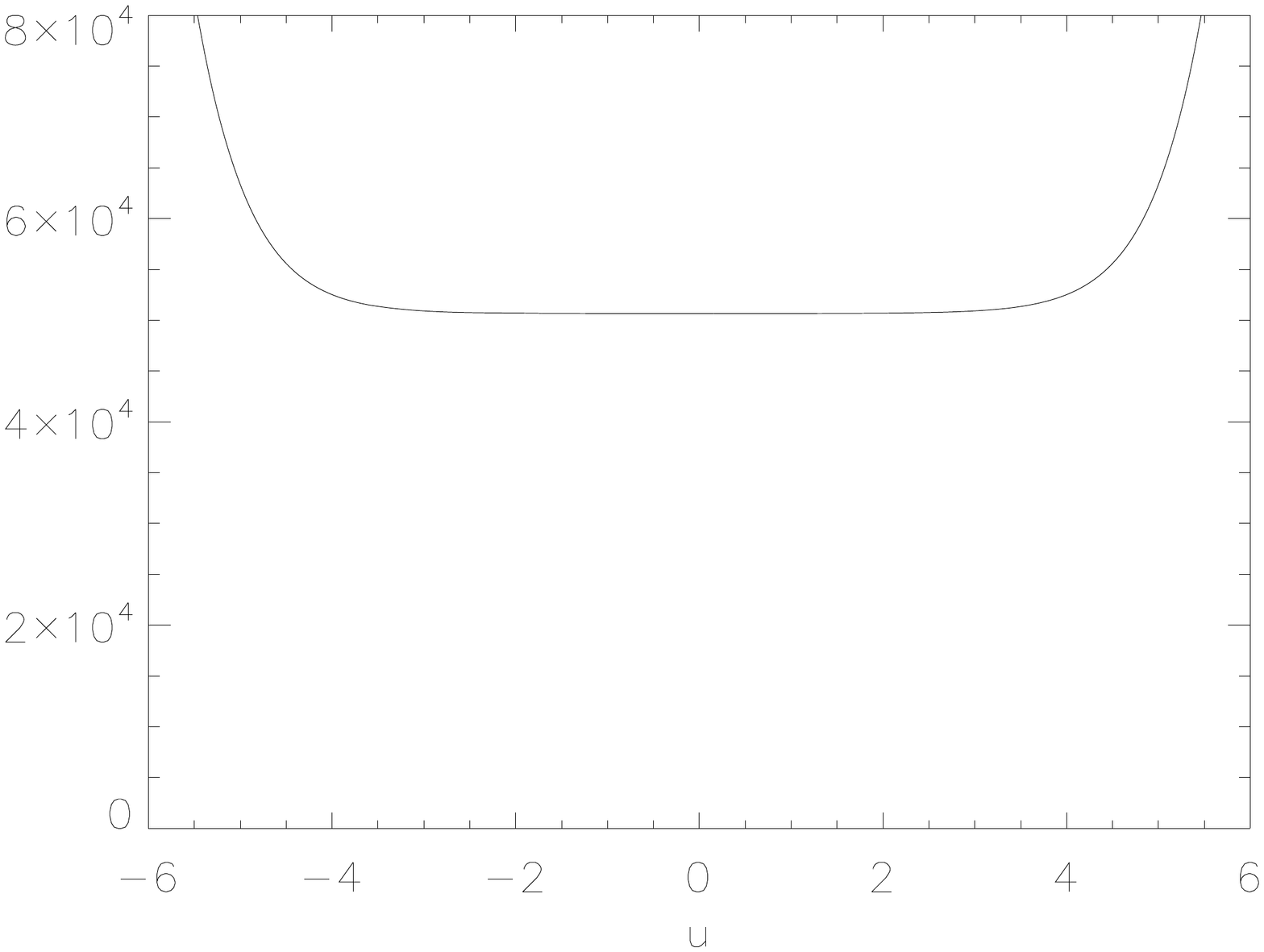}
\vspace{-1cm}
\caption{The panels on the left show for $m = H$, the $k^2\varepsilon^B_k/(2\pi^2)$ term in the energy density
of (\ref{renorme}) and (\ref{epspB}) in units of $H^4$ multiplied by a factor of $a^3$. The panels on the right show this
same term multiplied by a factor of $a^4$.  From top to bottom the value of $k$ which corresponds to each set of plots is
$k = 1$, $k = 10$, and $k = 100$, with values of $u_{k\gamma}$ of $0.88,\, 3.14$ and $5.44$ respectively,
where the behavior changes from non-relativistic to relativistic.}
\vspace{-5mm}
\label{Fig:epsBa34M1}
\end{figure}

\begin{figure}[t]
\vspace{-1cm}
\includegraphics[scale=0.3,angle=90,width=3.2in,clip]{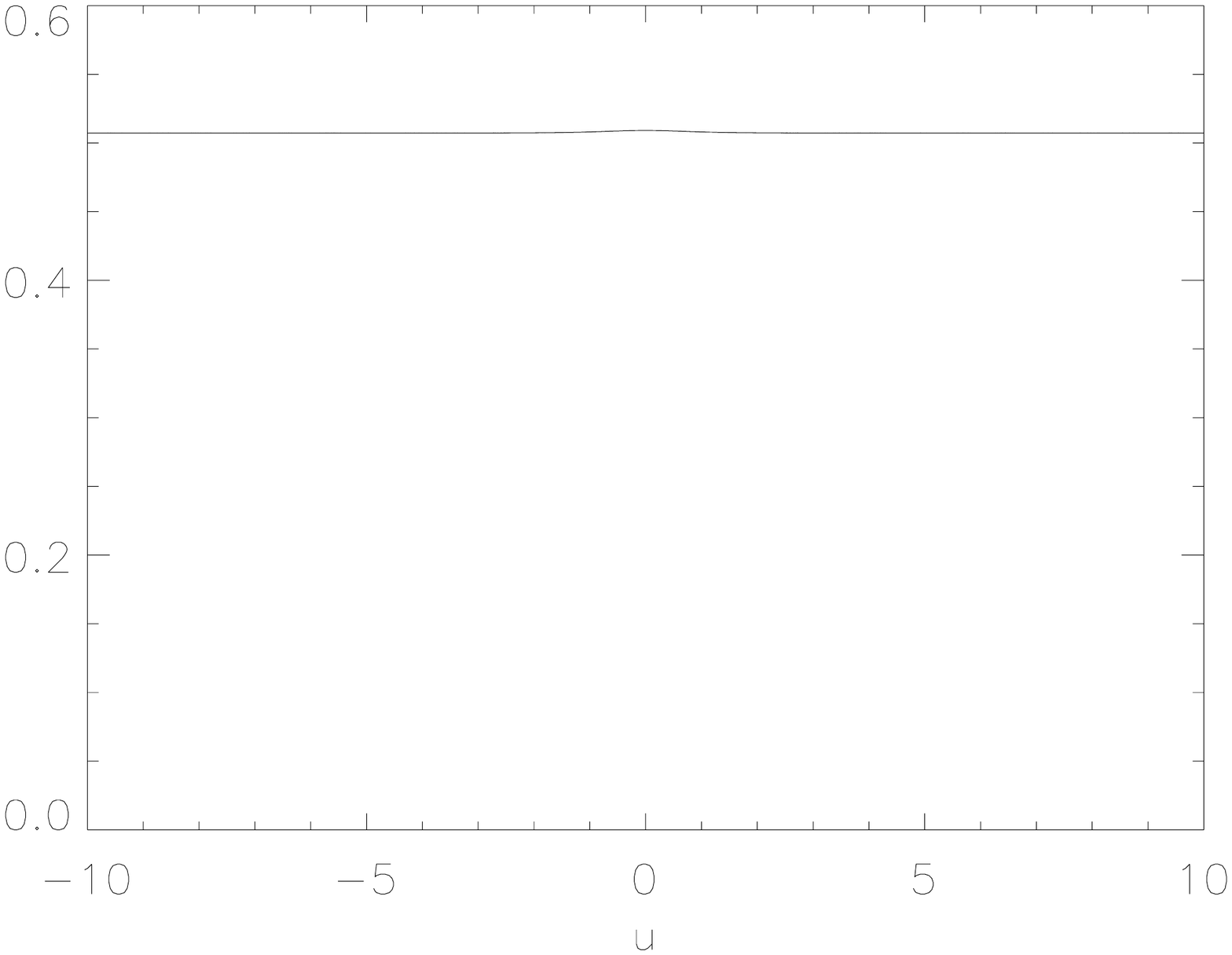}
\includegraphics[scale=0.3,angle=90,width=3.2in,clip]{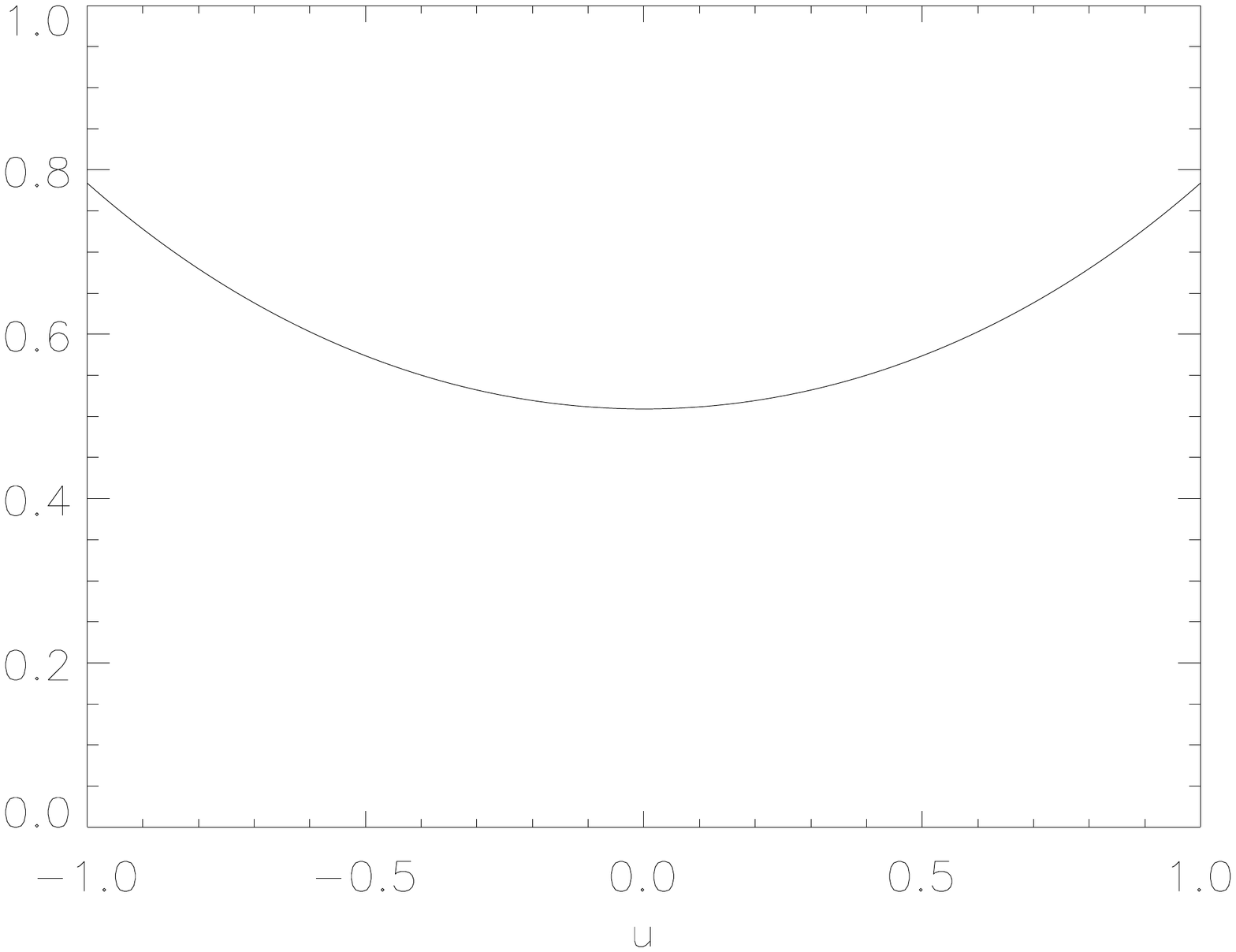}
\includegraphics[scale=0.3,angle=90,width=3.2in,clip]{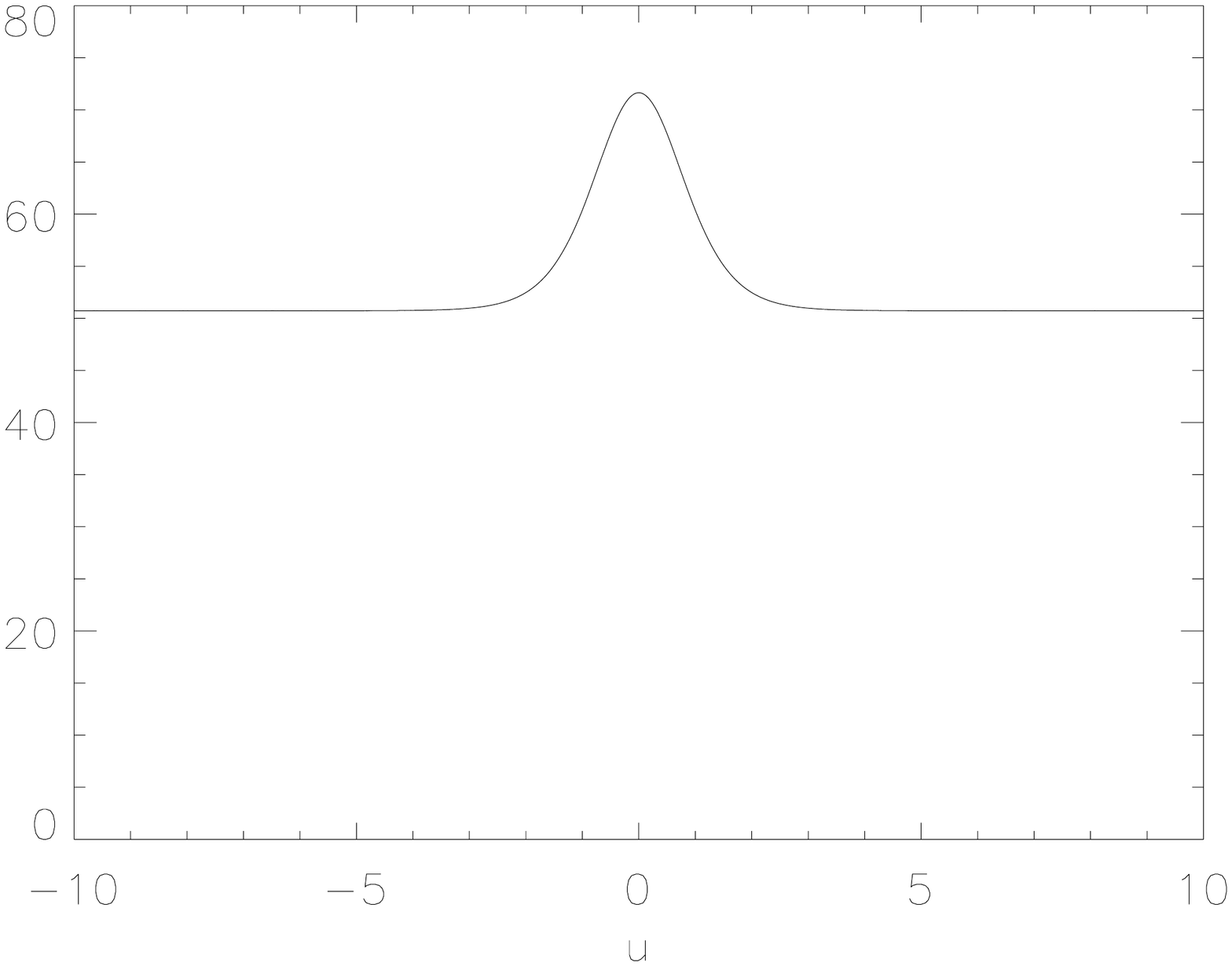}
\includegraphics[scale=0.3,angle=90,width=3.2in,clip]{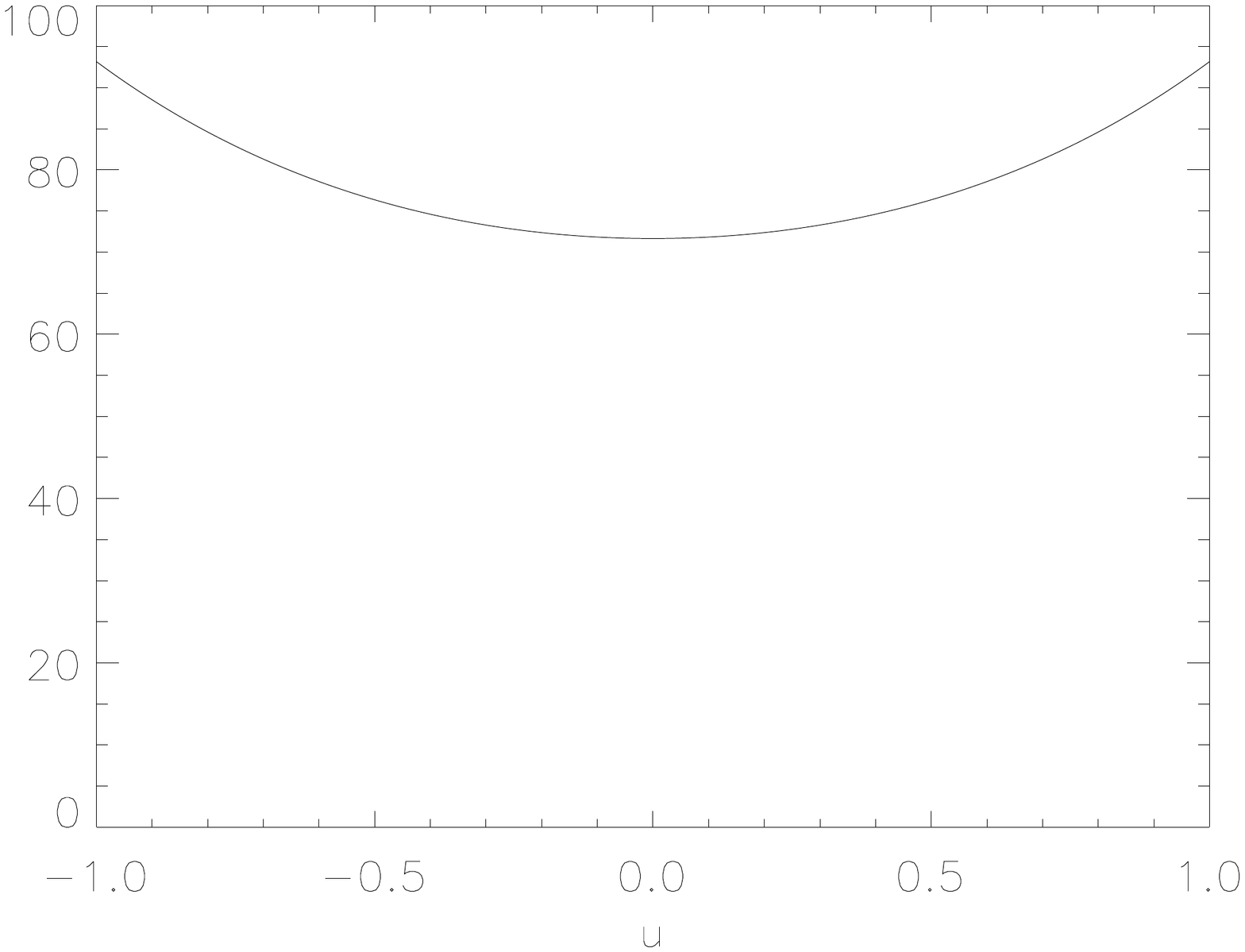}
\includegraphics[scale=0.3,angle=90,width=3.2in,clip]{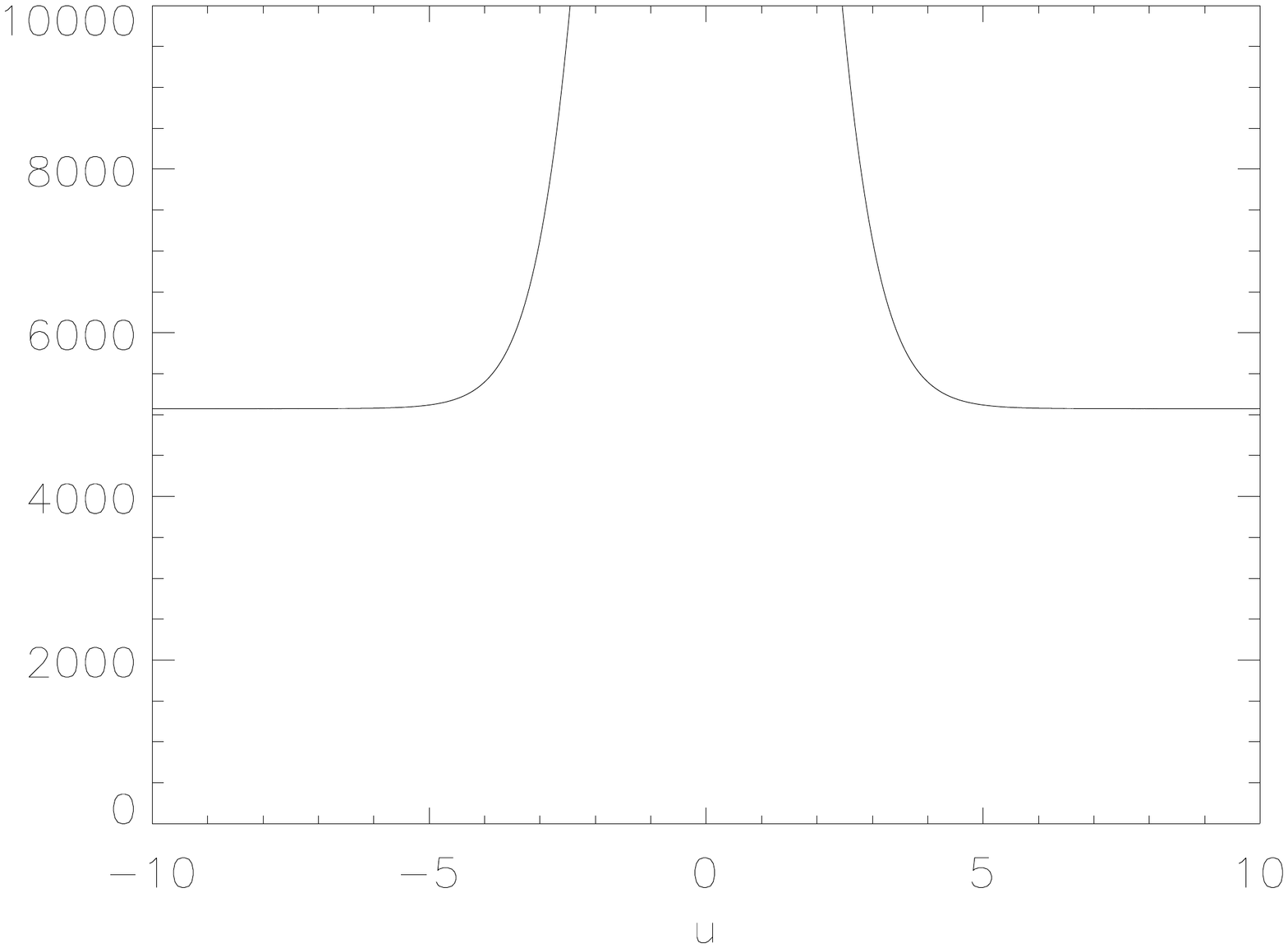}
\includegraphics[scale=0.3,angle=90,width=3.2in,clip]{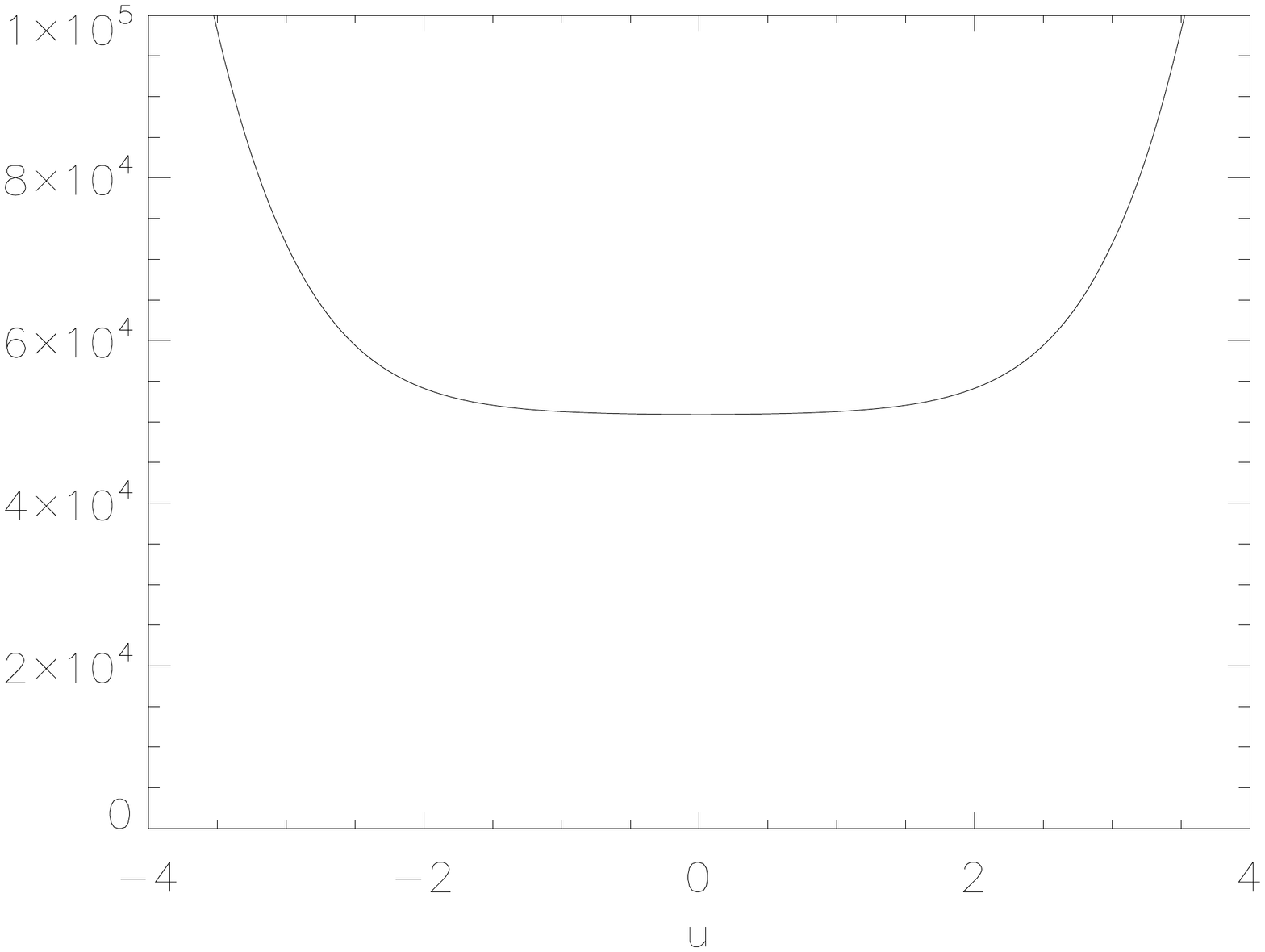}
\vspace{-1cm}
\caption{The panels on the left show for $m = 10H$, the $k^2\varepsilon^B_k/(2\pi^2)$ term in the energy density
of (\ref{renorme}) and (\ref{epspB}) in units of $H^4$ multiplied by a factor of $a^3$. The panels on the right show this
same term multiplied by a factor of $a^4$.  From top to bottom the value of $k$ which corresponds to each set of plots is
$k = 1$, $k = 10$, and $k = 100$. The values of $u_{k\gamma}$ where the behavior changes from non-relativistic 
to relativistic are $0.087,\, 0.88$ and $3.00$ respectively for this $m$. }
\vspace{-8mm}
\label{Fig:epsBa34M10}
\end{figure}

For the strictly massless conformally invariant scalar field, there are no oscillatory $\varepsilon_k^A$ terms
since $\varepsilon_k^A = 0$ identically for $\ups_k =\ups_{k, \frac{1}{2}}$, and hence there are no terms
linear in $B_k$ in the energy density or pressure of a general $O(4)$ invariant UV allowed state.
The only contributions come instead from the $\epsilon_k^B$ terms quadratic in the perturbation $B_k$
from the de Sitter invariant CTBD state $\vert \ups \rag$. Substituting $\ups_{k, \frac{1}{2}}$ into
(\ref{epspB}) with $m=0$ gives
\vspace{-6mm}
\be
\frac{k^2}{2\pi^2} \epsilon_k^B\Big\vert_{m=0} = \frac{k^3}{2 \pi^2 a^4}
\label{relB}
\ee
showing that the relativistic behavior observed in Figs. \ref{Fig:epsBa34M1}-\ref{Fig:epsBa34M10}
in the relativistic region $-u_{k\gamma} < u < u_{k\gamma}$ holds for all $u$ in the massless case.
This result can also be obtained by conformally transforming from flat space to de Sitter space the 
exact stress tensor for a conformal field in a state other than the Minkowski vacuum \cite{BirDav}. 
Although this is exactly the behavior one would expect for a gas of relativistic particles,
we emphasize that these are still coherent quantum excitations of the pure $|f\rag$ `vacuum' state, 
in which the exact phase relations of (\ref{ccdag})-(\ref{aadag}) are maintained.

In all cases the perturbations from the CTBD state $|\ups\rag$ fall off in the expanding
half $u>0$ of de Sitter space but grow in the contracting half $u<0$. The maximum value at
the symmetric point from (\ref{relB}) is given by
\be
\varepsilon_{_{f,N} max} = \frac{1}{2 \pi^2 a^4} \sum_{k=1}^{\infty} k^3 \big[N_k + \vert B_k\vert^2 (1 + 2N_k)\big]
\simeq  \frac{H^4}{8 \pi^2}\,  K^4\, \big[N_K + \vert B_K\vert^2 (1 + 2N_K)\big]
\label{epsfNmax}
\ee
where we have approximated the sum by an integral valid for large $k_{max} = K$, the maximum
value of $k$ for which $|B_k|^2$ and/or $N_k$  has support consistent with the UV finiteness conditions
(\ref{UVfinite}). For this to produce a significant backreaction on the classical de Sitter geometry through
the semiclassical Einstein eqs. it is necessary for this to be larger than the background
cosmological energy density, {\it i.e,}
\be
8 \pi G \varepsilon_{_{f,N}} \gtrsim  \Lambda = 3H^2\qquad {\rm or} \qquad
\frac{GH^2}{3 \pi}  \big[N_K + \vert B_K\vert^2 (1 + 2N_K)\big] \left(\frac{K}{\cosh u}\right)^4 \gtrsim 1\,.
\label{bound}
\ee
Clearly no matter how small $GH^2 \ll 1$, or the state dependent perturbation in brackets is,
as long as their product is non-zero there is always a large enough (but still finite $K$) for which
this inequality is satisfied at the maximum at $u=0$. Since all finite $k$ modes are redshifted in physical momentum
as $k/a \rightarrow 0$ for $a(u)\rightarrow \infty$, perturbations satisfying (\ref{bound}) at $u=0$ have
vanishingly small energy densities at early terms $u \rightarrow -\infty$. Hence for any finite $GH^2>0$
there is a large class of $O(4)$ invariant but de Sitter non-invariant states satisfying (\ref{bound}), which
give rise to energy densities that are large enough to produce significant de Sitter non-invariant
backreaction effects at the symmetric point $u=0$, all of which have exponentially vanishing
de Sitter non-invariant energy densities at times in the infinite past at $I_-$  of `eternal' de Sitter space.
The physical momentum $HK/a$ corresponding to the condition (\ref{bound}) for significant backreaction
is of order $\sqrt{HM_{Pl}} \ll M_{Pl}$, far less than the Planck scale $M_{Pl}$, so that the semiclassical approximation
is still reliable.

To see how the general condition (\ref{bound}) for large backreaction and de Sitter instability is realized
in a specific physical state, which is the one determined by adiabatically switching on of the background
analogous to switching on of the electric field in the infinite past \cite{AndMotSan3}, one can choose Bogoliubov
coefficients corresponding to the $O(4)$ invariant {\it in} state of \cite{AndMot1} prepared at the initial
time $u_0$. In the contracting phase of de Sitter space $u<0$ this corresponds to choosing the mode
functions according to the initial data at $u=u_0$
\vspace{-3mm}
\bes
\bea
&&A_k = A_{k\gamma}^{in}\, \theta(-u_{k\gamma} - u_0) + \theta (u_0+ u_{k\gamma})\\
&&B_k = B_{k\gamma}^{in}\, \theta(-u_{k\gamma} - u_0)
\eea
\ees
with $u_{k\gamma}$ defined in (\ref{ukgam}). The step functions are again simple approximations to
the actual smooth but rapid change at $-u_{k\gamma}$. Since $B_k =0$ for $u_{k\gamma} > -u_0$ the mode
sums in (\ref{renormepsp}) are cut off for $k > K_{\gamma}(u_0)$ where
\vspace{-3mm}
\be
K_{\gamma}(u_0) = \sqrt{\gamma^2 \sinh^2 u_0 + \tfrac{1}{4}} \simeq \frac{\gamma}{2} \, e^{|u_0|}
\label{largeu0}
\ee
for $|u_0| \gg 1$. Also
\be
A_{k\gamma}^{in} = \frac{i}{\sqrt{2 \sinh (\pi \gamma)}} \,e^{-\frac{ik\pi}{2}}\, e^{\frac{\pi \gamma}{2}}\,,\qquad
B_{k\gamma}^{in} = \frac{1}{\sqrt{2 \sinh (\pi \gamma)}}\,e^{\frac{ik\pi}{2}}\, e^{-\frac{\pi \gamma}{2}}
\label{ABinvalues}
\ee
so that
\be
A_{k\gamma}^{in} \,B_{k\gamma}^{in\,*} =  \frac{i\, (-)^k}{2 \sinh (\pi \gamma)}\ \theta(-u_{k\gamma} - u_0)
\label{ABin}
\ee
oscillates in $k$. Because the $\epsilon_k^A$ term is bounded in $k$ for any state as shown in
Fig. \ref{Fig:epsBa34M10} and its coefficient (\ref{ABin}) oscillates in $k$ for this {\it in} state, its
contributions tend to cancel when summed over $k$ in (\ref{renormepsp}), and are negligible
compared to the non-oscillatory $\epsilon_k^B$ term in the energy density for large $K_{\gamma}(u_0)$,
hence large $|u_0|$. Retaining only the latter we then have approximately
\bea
&&\varepsilon_{in} \simeq \varepsilon_{\ups}
+ \frac{1}{2\pi^2} \sum_{k=1}^{K_{\gamma}(u_0)} k^2 |B_{k\gamma}^{in}|^2 \varepsilon_k^B\nn
&&\simeq \frac{1}{4\pi^2}\frac{e^{-\pi\gamma}}{\sinh(\pi\gamma)} \int_1^{K_{\gamma}(u_0)}  dk\, \frac{k^3}{a^4(u)}  \nn
&&\simeq  \frac{H^4}{8\pi^2}\frac{1}{ e^{2\pi\gamma} - 1} \left(\frac{K_{\gamma}(u_0)}{\cosh u}\right)^4
\label{epsa4}
\eea
which agrees with eq. (8.17) of \cite{AndMot1} and (\ref{epsfNmax}) for the particular choice of $|B_K|^2$ from
(\ref{ABinvalues}) and $N_k =0$. This energy density, which has an arbitrarily small value for  $u \rightarrow -\infty$, is blueshifted as a relativistic fluid and by $u=0$ has grown large enough to comparable to the background
de Sitter energy density and hence significantly affect the background de Sitter geometry. 
It satisfies the inequality (\ref{bound}) for significant backreaction if
\be
K_{\gamma}(u_0) \gtrsim \left[ \frac{3 \pi}{GH^2} (e^{2 \pi \gamma} -1)\right]^{\frac{1}{4}}
\ee
or from (\ref{largeu0})
\be
|u_0| > \ln\left[\frac{2}{\gamma} \left(\frac{3 \pi}{GH^2}\right)^{\frac{1}{4}} (e^{2 \pi \gamma} -1)^{\frac{1}{4}} \right]
\ee
which can always be satisfied for early enough $u_0$, and non-zero $GH^2$ and $\gamma$ in eternal de Sitter space.

\section{\hspace{-2mm}Conformal Anomaly, Stress Tensor and de Sitter Symmetry Breaking}
\label{Sec:Anom}

As in the electric field example, the quantum vacuum instability in de Sitter space is illuminated
by consideration of a quantum anomaly, in this case the conformal trace anomaly
of the energy-momentum tensor \cite{BirDav,Duff}. The non-local covariant effective action that gives the
conformal anomaly in four dimensions is
\cite{Rie,MazMotWZ,MotVau,NJP,Zak}
\be
S_{anom}[g] = \frac {1}{2}\int\! d^4x\sqrt{-g}\int\! d^4x'\sqrt{-g'}\,
\left(\frac{E}{2} - \frac{\sq R}{3}\right)_{\!\!x} \Delta_4^{-1} (x,x')\left[ bF + b'
\left(\frac {E}{2} - \frac{\sq R}{3}\right)\right]_{\!x'}
\label{Sanom}
\ee
This non-local effective action (\ref{Sanom}) is the analog of (\ref{2danomact}) for the chiral anomaly in
two dimensions and the $\int d^2x\sqrt{-g} \int d^2x'\sqrt{-g'} R_x \sq^{-1}(x,x')R_x'$ effective action for the
conformal trace anomaly in two dimensions \cite{Polyanom}. In four dimensions there are two invariants
$E\equiv \, ^*\hskip-1mm R_{abcd}\,^*\hskip-1mm R^{abcd} = R_{abcd}R^{abcd} -4 R_{ab}R^{ab} + R^2$ and
$F\equiv C_{abcd}C^{abcd} =  R_{abcd}R^{abcd} -2 R_{ab}R^{ab} +\frac{1}{3} R^2$ contributing to the
non-local anomaly with corresponding dimensionless coefficients $b$ and $b'$ proportional to $\hbar$
in the notation of \cite{Duff}. Being non-local in terms of the curvature invariants $E$ and $F$,
the one-loop effective action (\ref{Sanom}) contains information about non-local and global quantum effects,
{\it i.e.} sensitivity to initial and/or boundary conditions, through the Green's function inverse $\Delta_4^{-1} (x,x')$
of the conformally covariant differential operator
\be
\Delta_4 \equiv \sq^2 + 2 R^{ab}\nabla_a\nabla_b - \frac{2}{3} R \sq +
\frac{1}{3} (\nabla^a R)\nabla_a = \nabla_a\left(\nabla^a\nabla^b + 2R^{ab} - \tfrac{2}{3} R g^{ab}\right) \nabla_b \,.
\label{Del4def}
\ee
To (\ref{Sanom}) it is possible to add any conformally invariant action (non-local or local) which does
not affect the anomaly. However, only the conformal breaking (\ref{Sanom}) term in the effective action
needs be retained in a low energy classification of operators in the effective action \cite{MazMotWZ},
and can have relevant infrared effects.

Moreover, the effective action of the anomaly (\ref{Sanom}) is distinguished by being responsible for
additional massless scalar degree(s) of freedom in low energy gravity, not present in the classical
theory \cite{GiaMot}, as seen also in two dimensions by the shifting of the central charge from $N-26$
to $N-25$ \cite{Polc}. In four dimensions this is made explicit by rewriting (\ref{Sanom}) in the local form
\be
S_{anom} = b' S^{(E)}_{anom} + b S^{(F)}_{anom}\,,
\label{allanom}
\ee
by the introduction of at least one additional scalar field, where for example
\be
S^{(E)}_{anom}[g;\varphi]\equiv \frac{1}{2} \int d^4x\sqrt{-g}\left\{
-\left(\sq \varphi\right)^2 + 2 \left(R^{ab}- \tfrac{1}{3} R g^{ab}\right)
(\nabla_a \varphi)\!(\nabla_b \varphi) + \left(E - \tfrac{2}{3}\sq R\right)
\varphi\right\}\label{SE}
\ee
is the term related to $E$ in terms of the additional scalar field $\varphi$. This scalar (analogous to $\chi$
of Sec. \ref{Subsec:2D}) is a new effective degree of freedom, not to be confused with the original scalar field $\Phi$,
which describes two-particle correlations or bilinears (relativistic `Cooper pairs') of the underlying scalar, fermion
or vector QFT \cite{Zak,GiaMot}. QFT's of different spin may all be studied via the effective action (\ref{SE}),
since the only dependence upon spin for free fields is through the trace anomaly coefficients $b'$ and $b$, where
\be
b'= -\frac{1}{360 (4 \pi)^2}\, (N_S + 11 N_F + 62 N_V)
\label{bprime}
\ee
is the coefficient for the $E$ term in the conformal anomaly for non-interacting scalar ($S$),
fermion ($F$), or vector ($V$) fields respectively. The $b$ term in the anomaly proportional to
$F$ gives rise to an effective action $S^{(F)}_{anom}$ similar to (\ref{SE}) but is less important
in de Sitter space where  $F= C_{abcd}C^{abcd} = 0$ \cite{MazMotWZ,MotVau,Zak,DSAnom}.

Formally solving (\ref{phieq}) for $\varphi$ in a general metric by inverting $\Delta_4$, {\it i.e.}
finding its Green's function $\Delta_4^{-1} (x,x')$, and substituting the solution into (\ref{SE}) returns the
$b'$ term of the non-local form (\ref{Sanom}). In de Sitter space we also have $E= 24H^4$, $\sq R= 0$,
and the operator $\Delta_4$ factorizes, so that the variation of (\ref{SE}) with respect  to $\varphi$ yields
the linear eq. of motion,
\be
\Delta_4\, \varphi\big\vert_{dS}= -\sq\big(-\sq + 2H^2\big)\varphi=  \frac{E}{2} - \frac{\sq R} {3}  = 12H^4
\label{phieq}
\ee
with a constant source. Because of this constant source, analogous to (\ref{anom2d}), and the fact
that the only invariant scalar in de Sitter space is a constant, it is clear that {\it no de Sitter invariant constant
solution to (\ref{phieq}) for $\varphi$ exists}. In the local form of the effective action (\ref{SE}), the freedom
to add homogeneous solutions to (\ref{phieq}) is equivalent to that of specifying the particular Green's
function inverse  $\Delta_4^{-1} (x,x')$ dependent upon initial/boundary conditions in the non-local
form (\ref{Sanom}). It is also clear from the factorized form (\ref{phieq}) of $\Delta_4$ in de Sitter space
that its inverse
\be
\Delta_4^{-1} \big\vert_{dS} = \frac{1}{2H^2} \left[ (-\sq)^{-1} - (-\sq + 2H^2)^{-1}\right]
\label{Del4invdS}
\ee
cannot be de Sitter invariant, since it is proportional to the difference of the inverses of a massless, minimally
coupled ($\xi = 0$) scalar and  a massless, conformally coupled ($\xi = \frac{1}{6}$) scalar, and no de Sitter
invariant form of the former exists \cite{AllFol}. Thus the breaking de Sitter invariance and infrared sensitivity
to initial/boundary conditions is already apparent from either the non-local one-loop effective action (\ref{Sanom})
and nonexistence of a de Sitter invariant Feynman Green's function (\ref{Del4invdS}), or equivalently from
the non-invariance of the solutions to (\ref{phieq}) and hence those of the local effective action (\ref{SE}).

The form of the breaking of de Sitter invariance may be studied through the stress tensor corresponding
to the local effective action (\ref{SE}), whose variation with respect to the metric gives the tensor
\bea
&&\hspace{-.5cm}E_{ab}[\varphi]\Big\vert_{dS} \equiv -\frac{2}{\sqrt{-g}} \frac{\delta S^{(E)}_{anom}}{\delta
g^{ab}} \bigg\vert_{dS}= -2\, (\nabla_{(a}\varphi) (\nabla_{b)} \sq \varphi)
+ 2\, (\nabla_c \varphi)(\nabla^c\nabla_a\nabla_b\varphi) +\, 2\, (\sq\varphi) (\nabla_a\nabla_b\varphi)\nn
&&- \frac{2}{3}\, \nabla_a\nabla_b\left[(\nabla \varphi)^2\right]
 - 4H^2 (\nabla_a\varphi)(\nabla_b\varphi) + \frac{1}{2}\, g_{ab} \left[- (\sq\varphi)^2
+ \frac{1}{3}\sq \left[(\nabla\varphi)^2 \right]
+ 2H^2 (\nabla\varphi)^2\right]\nn
&&\hspace{2cm} - \frac{2}{3}\, \nabla_a\nabla_b \sq \varphi
+ 4H^2 \nabla_a\nabla_b\varphi - \frac {2}{3}H^2\,g_{ab}\sq\varphi
+ 8H^4 g_{ab}
\label{Eab}
\eea
which is covariantly conserved by the use of (\ref{phieq}). In (\ref{Eab}) we have evaluated $E_{ab}$
in de Sitter space and used the notation $(\nabla \varphi)^2 \equiv g^{ab}(\nabla_a\varphi) (\nabla_b\varphi)$.
The stress tensor $T_{ab}^{(E)} = b' E_{ab}$ evaluated on solutions $\varphi$ satisfying the classical linear
eq. (\ref{phieq}) may be used to evaluate the renormalized expectation value $\lag T_{ab}\rag_{_R}$ of
the underlying QFT. This is exact up to state dependent (but curvature independent) terms if the spacetime is
conformally flat as is de Sitter space, and the QFT is classically conformally invariant \cite{BroCas}.
We show below in particular that (\ref{Eab}) reproduces the CTBD state value exactly for
classically conformally invariant fields of any spin with an appropriate choice of $\varphi$.

Since the fourth order linear operator $\Delta_4$ in (\ref{phieq}) factorizes into two second order wave
operators for a conformally coupled and minimally coupled massless scalar in de Sitter space,
the general homogeneous solution of (\ref{phieq}) in coordinates (\ref{hypermet}) is easily found in terms of
$\ups_{k,\frac{1}{2}}Y_{klm_l} $ and $\ups_{k,\frac{3}{2}}Y_{klm_l}$ and their complex conjugates.
Inspection of these solutions, (\ref{m0conf})-(\ref{conftime}) shows that the functions $\ups_{k,\frac{1}{2}}$
and $\ups_{k,\frac{3}{2}}$ may also be written as a linear combination of $\exp[-i(k\pm 1)\eta]$.
The reason this rearrangement of the solutions is possible is a consequence of the conformal
properties of the operator $\Delta_4$. In de Sitter spacetime (and in fact any conformally flat spacetime)
there is a second factorization of $\Delta_4$ into two second order operators, reflecting the fact that for a fixed $k$
(\ref{phieq}) may also be written \cite{AMMStates}
\be
\Delta_4\big\vert_{dS}\ \varphi_k(\eta) \,Y_{klm_l}(\hat N) = H^4 \cos^4\eta \left[\frac{d^2}{d\eta^2} + (k-1)^2\right]
\left[\frac{d^2}{d\eta^2} + (k+1)^2\right]\,\varphi_k(\eta) \,Y_{klm_l}(\hat N)
\label{del4conf}
\ee
in conformal time $\eta$, where $H^4 \cos^4\eta =a^{-4}$. Thus the homogeneous solutions of (\ref{phieq})
are clearly linear combinations of $\exp[-i(k\pm 1)\eta] \, Y_{klm_l}(\hat N)$ and their complex conjugates.
To these one must add a particular solution of the inhomogeneous eq. (\ref{phieq}), which
is easily found in coordinates (\ref{hypermet}) to be
\be
\varphi_0 \equiv 2 \ln (\cosh u) \,.
\label{phi0}
\ee
This particular solution is $O(4)$ invariant but not $O(4,1)$ invariant. Other choices correspond to states of
lower symmetry, but some choice must be made since the inhomogeneous term in (\ref{phieq}) disallows the
de Sitter invariant choice of constant $\varphi$.  Then we may express the general solution of (\ref{phieq}) in the form
\be
\varphi = \varphi_1(u) +  \frac{1}{2}\sum_{k=2}^{\infty}\sum_{l=0}^{k-1}\sum_{m_l = -l}^{l}
\left[\frac{a_{klm_l}}{\sqrt{2k(k-1)}}e^{-i(k-1)\eta}\ Y_{klm_l}+ \frac{b_{klm_l}}{\sqrt{2k(k+1)}}e^{-i(k+1)\eta}\ Y_{klm_l} + c.c.\right]
\label{genphi}
\ee
where $\varphi_1(u)$ is the general solution of (\ref{phieq}) for $k=1$, constant on ${\mathbb S}^3$, given by
\bea
\varphi_1(u) &=& 2 \ln (\cosh u) + c_0 + c_1 \sin^{-1}(\tanh\, u) + c_2\, \sech^2\,u + c_3\tanh u\,\sech\, u \nn
&=& 2 \ln (\sec \eta)  + c_0 +c_1\,\eta + c_2\, \cos^2\eta +c_3\,\sin\eta\,\cos\eta
\label{phi1}
\eea
with the $c_i$ arbitrary constants multiplying the $4$ homogeneous solutions which are functions only of $u$
or conformal time $\eta$ defined in (\ref{conftime}). The normalizations of the $k>1$ solutions in (\ref{genphi})
are chosen to correspond to a previous canonical analysis on the conformally related Einstein static cylinder
${\mathbb{R \otimes S}^3}$ where the $\varphi = 2 \sigma$ field was quantized and the $(a_{klm_l}, b_{klm_l})$
obey canonical commutation relations (the $b_{klm_l}, b^{\dag}_{klm_l}$ with negative metric) \cite{AMMStates}.
Here we treat all the expansion coefficients $(c_i, a_{klm_l}, b_{klm_l})$ of the general solution (\ref{genphi})
to (\ref{phieq}) for the effective action in de Sitter space as $c$-numbers.

For $O(4)$ invariant states the stress tensor can only be a function of $u$. Because of the terms linear in $\varphi$
in (\ref{Eab}) this corresponds to choosing all the coefficients $a_{klm_l}= b_{klm_l} = 0$ in (\ref{genphi}) for $k>1$.
With $\varphi = \varphi_1(u)$ substituted into (\ref{Eab}) we obtain the energy density
\be
-E^u_{\ u} [\varphi_1(u)]  =  \dddt \varphi_1\, \dt\varphi_1 - \frac{1}{2} \ddt \varphi_1^{\,2} +  2h\, \ddt \varphi_1\,\dt \varphi_1  +
3\Big(2 \dt h + \frac{3}{2}h^2-H^2\Big) \dt\varphi_1^2 - 2h\, \dddt\varphi_1 + 2\, (H^2 -3h^2)\, \ddt \varphi_1
-6 h\,(\dt h + H^2)\, \dt\varphi_1
\label{Ett}\ee
where a dot denotes the derivative $H^{-1}d/du$. Substituting (\ref{phi1}) into this expression gives
\be
\varepsilon = -b'E^u_{\ u}[\varphi_1(u)] =  -6b'H^4\,+\, \frac{2b'}{\,a^4}\, (c_1^2 - c_2^2 - c_3^2 + 4)\,.
\label{Euuanom}
\ee
The first term gives the constant value of the renormalized $\varepsilon_{\ups} = -6b'H^4$ for the de Sitter invariant state
of a free conformal field of any spin, with the corresponding pressure $p_{\ups} = - \varepsilon_{\ups}= 6b'H^4$.
The second $a^{-4}$ term shows that exactly the term corresponding to the relativistic limit obtained in Sec. \ref{Sec:SET}
from detailed analysis of the renormalized expectation value of the stress tensor of a quantum field in the general $O(4)$
invariant state is simply reproduced by the anomaly stress tensor (\ref{Eab}) with a classical effective field
$\varphi = \varphi_1(u)$. The spatial components
\be
E_{ij}[\varphi_1 (u)] =  6H^4g_{ij} + \frac{2}{3a^4}\,g_{ij}\, (c_1^2 - c_2^2 - c_3^2 + 4)
\label{Eijanom}
\ee
and eq. of state $p = \varepsilon /3$ for the second term are just that required by covariant conservation
(\ref{cons}) for this general $O(4)$ invariant state.

In (\ref{Euuanom}) the arbitrary coefficients $c_i$ of the homogenous solution in (\ref{phi1}) appear
and may be related to the sum over the state dependent coefficients $N_k, B_k$ and $K$ of (\ref{epsfNmax}).
The de Sitter invariant expectation value for $\lag T_{ab}\rag_{_R}$ is recovered iff
\be
c_1^2 + 4 = c_2^2 + c_3^2\qquad ({\rm de\  Sitter\  invariant \ } \lag T_{ab}\rag \,)
\label{deSci}
\ee
so that no relativistic radiation $a^{-4}$ term is present. Any solution of (\ref{phieq}) of the form (\ref{phi1})
with the condition (\ref{deSci}) on the coefficients $c_i$ may be taken as corresponding to the CTBD state
and de Sitter invariant stress tensor with $\varepsilon_{\ups} = - p_{\ups}$. It is interesting to note in passing
that for the particular values $c_1 = -2i$ and $c_0=c_2=c_3=0$, $\exp[\varphi_1(u)]$ is just the (complex)
conformal transformation that maps flat space and its Minkowski vacuum to de Sitter space and the CTBD
invariant state $|\ups\rag$. However, if we restrict $\varphi_1(u)$ of (\ref{phi1}) to be real, and invariant
under time reversal $u \leftrightarrow -u$, corresponding to the discrete inversion symmetry of the CTBD
state, then $c_1 = c_3 =0$, and from (\ref{deSci}) $c_2 = \pm 2$ so that
\be
\bar \varphi (u) = 2 \ln (\cosh u) + c_0  \pm 2\, \sech^2\,u = -2 \ln (\cos \eta) + c_0 \pm 2\cos^2\eta
\label{barphi}
\ee
is the background solution to (\ref{phieq}) with $\varepsilon_{\ups} = -6b'H^4 = -p_{\ups}$
most closely corresponding to $\vert \ups\rag$. Since the stress tensor (\ref{Eab})
depends only upon derivatives of $\varphi$, the constant $c_0$ is irrelevant and may be set to zero,
so that the choice of solution (\ref{barphi}) is determined up to the sign of the last term.

This $\bar\varphi(u)$ in (\ref{barphi}) is a kind of mean value condensate of the $\varphi$ effective field in
de Sitter. Although itself not de Sitter invariant, it gives a stress tensor corresponding to the
de Sitter invariant CTBD state of the underlying QFT. It seems that one has to consider more
complicated expectation values such as $\lag T_{ab}T_{cd}\rag$ in order to see directly the
de Sitter breaking effects of the inhomogeneous solution to (\ref{phieq}). This is similar to the
de Sitter invariant stress tensor $\lag T_{ab}\rag$ obtained for a massless, minimally coupled field in
de Sitter space despite the non-de Sitter invariant vacuum state \cite{AllFol}.

A small variation of $c_2$ away from $\pm 2$ produces a de Sitter non-invariant stress tensor of
the form (\ref{Euuanom})-(\ref{Eijanom}) which is infinitesimally small at asymptotic past infinity $I_-$
because of its $a^{-4}$ dependence upon the scale factor, but which grows to finite values at the
symmetric time $u=0$. The $c_i$ satisfying  (\ref{deSci}) are clearly a subset of a wider class of a
three parameter family corresponding to $O(4)$ invariant but non-$O(4,1)$ invariant states. In this
parameterization the condition (\ref{bound})  that the perturbations of the CTBD state produce a
large enough backreaction at $u=0$ to affect the classical geometry is
\be
16\pi GH^2\, \big\vert b'\,( c_1^2 - c_2^2 - c_3^2 + 4)\big\vert\, \gtrsim\, 1\,.
\label{ccond}
\ee
Clearly there are a large class of such states all which all have exponentially vanishing de Sitter
non-invariant energy densities at times $u \rightarrow -\infty$ in the infinite past. Since a perturbation
of the CTBD state with infinitesimally small energy density at $I_-$ with coefficients $c_i$ satisfying
(\ref{ccond}) produces a large backreaction on the geometry at $u=0$, we conclude that the de Sitter
invariant $|\ups\rag$ state is unstable to such state perturbations in the initial data of eternal de Sitter space.

Thus the anomaly effective action and stress tensor gives the same result of instability of the CTBD state
to perturbations, obtained previously for massive scalar fields, without any need of renormalization
subtractions or mode sums, although the connection to the large $K$ cutoff in (\ref{epsfNmax}) or (\ref{bound}),
or to particle creation in the {\it in} state of (\ref{epsa4}) or \cite{AndMot1} is no longer transparent in (\ref{ccond}).
The anomaly derivation of the instability condition (\ref{ccond}) emphasizes its generality, independent
of the particular case of a non-interacting scalar field, so that (\ref{ccond}) holds for fields of any spin simply
by changing $b'$ according to (\ref{bprime}), or more generally for interacting QFT's as well with the appropriate
$b'$. This result and the composite effective field $\varphi$ is similar to the generality of the axial anomaly
derivation of the linear growth of the current in a persistent electric field background in terms of the
bosonized effective field $\chi$ in (\ref{Schj}).

\section{States of Lower Symmetry: Spatially inhomogeneous Stress Tensor}
\label{Sec:LowerSym}

The expansion (\ref{genphi}) of the anomaly scalar $\varphi$ also enables a general study of states of lower
than $O(4)$ symmetry simply by allowing any of the parameters $a_{klm_l}$ or $b_{klm_l}$ in
the general solution (\ref{genphi}) to be different from zero. Substituting that general solution for
$\varphi$ in (\ref{Eab}) gives a $T_{ab}$ which is a function of directions $\hat N$ on ${\mathbb S}^3$
as well as $u$. In order to study the effect of these $O(4)$ breaking terms, we linearize the anomaly
stress tensor around the solution $\bar\varphi (u)$ of (\ref{barphi}) with a de Sitter invariant stress tensor by
\be
\varphi = \bar\varphi (u) + \phi(u, \hat N)
\ee
for $\phi$ a general solution of (\ref{phieq}) with $\Delta_4 \phi =0$, expressed as the sum of modes
(\ref{genphi}). To first order in $\phi$
\bea
&&E_{ab}^{(1)} = -2\, (\nabla_{(a}\bar\varphi) (\nabla_{b)} \sq \phi)  -2\,  (\nabla_{(a} \sq \bar\varphi) (\nabla_{b)}\phi)
+ 2\, (\nabla_c \bar\varphi)(\nabla^c\nabla_a\nabla_b\phi) + 2\, (\nabla^c\nabla_a\nabla_b\bar\varphi)(\nabla_c \phi)\nn
&&\hspace{0.7cm}+\, 2\, (\sq\bar\varphi)(\nabla_a\nabla_b\phi)  +\, 2\, (\nabla_a\nabla_b\bar\varphi) (\sq\phi)
- \tfrac{4}{3}\, \nabla_a\nabla_b\left[g^{cd}(\nabla_c \bar\varphi)(\nabla_d \phi)\right]
 - 8H^2 (\nabla_{(a}\bar\varphi)(\nabla_{b)}\phi)\nn
&&\hspace{2cm}  + g_{ab} \left\{- (\sq\bar\varphi)(\sq\phi)
+ \tfrac{1}{3}\sq \left[g^{cd}(\nabla_c\bar\varphi)(\nabla_d\phi)  \right]
+ 2H^2 g^{cd}(\nabla_c\bar\varphi)(\nabla_d\phi)\right\}\nn
&&\hspace{3cm}- \tfrac{2}{3}\, \nabla_a\nabla_b \sq \phi
+ 4H^2 \nabla_a\nabla_b\phi - \tfrac {2}{3}H^2\,g_{ab}\sq\phi
\label{delEab}
\eea
which is both covariantly conserved and traceless.  Using
\bes
\begingroup
\allowdisplaybreaks
\begin{align}
\nabla_{\tau} \nabla_{\tau}\, \varphi &= \ddt{\varphi}\\
\nabla_{\tau} \nabla_i\, \varphi &= \nabla_i\,(\dt{\varphi} - h \varphi)\\
\nabla_{\tau} \nabla_{\tau}\nabla_{\tau}\, \varphi &= \dddt{\varphi} \\
\nabla_{\tau} \nabla_{\tau}\nabla_i\, \varphi &= \nabla_i \,(\ddt{\varphi} - 2 h\dt{\varphi} - \dt{h}\varphi  + h^2\varphi)
\end{align}
\endgroup
\label{derivs}\ees
operating on any scalar function in the de Sitter metric in coordinates (\ref{hypermet}),
and the fact that $\bar\varphi(u)$ is a function only of $u$, we obtain
\bea
&&\hspace{-7mm}\delta E^{(1)\,u}_{\ \ u} = 2\,\dt{\bar\varphi} \,(\sq \phi)^{^{\dt{\, }}} + 2\, (\sq \bar\varphi)^{^{\dt{\, }}} \,\dt{\phi}
+ 2\,\dt{\bar\varphi}\,\dddt{\phi} + 2\, \dddt{\bar\varphi} \,\dt{\phi}
- 2\, (\sq\bar\varphi)\,\ddt{\phi}  - 2\, \ddt{\bar\varphi}\, (\sq\phi) - (\sq\bar\varphi)(\sq\phi) \nn
&&\hspace{-2mm} - \frac{4}{3} \left(\dddt{\bar\varphi}\,\dt{\phi} + 2 \ddt{\bar\varphi}\,\ddt{\phi}
+ \dt{\bar\varphi}\,\dddt{\phi}\right) + 6H^2 \dt{\bar\varphi}\,\dt{\phi}
- \frac{1}{3}\sq \left( \dt{\bar\varphi}\,\dt{\phi} \right)
+ \frac{2}{3}\, (\sq \phi)^{^{\ddt{\, }}}   - 4H^2 \ddt{\phi} - \frac {2}{3}H^2\,\sq\phi
\label{delEuu}
\eea
for the $a=b=u$ component of this linearized stress tensor. Since the off-diagonal metric components, $g_{u i}$, 
are zero, it is slightly easier to compute in this case
\be
E_{u i}^{(1)} \equiv \nabla_i V^{(1)} = \partial_i V^{(1)}
\label{Vdef}
\ee
where
\bea
&&\hspace{-7mm}V^{(1)} = -\dt{\bar\varphi} \,(\sq \phi) - (\sq \bar\varphi)^{^{\dt{\, }}} \,\phi
- 2\dt{\bar\varphi}\left[\ddt{\phi} - 2h\,\dt{\phi} + (h^2 - \dt{h})\,\phi\right] + 2\, (\sq \bar\varphi) \,(\dt{\phi} - h \phi) \nn
&&+ \frac{4}{3} \left(\ddt{\bar\varphi}\,\dt{\phi}  + \dt{\bar\varphi}\,\ddt{\phi} - h\, \dt{\bar\varphi}\, \dt{\phi} \right)
  - 4 H^2\, \dt{\bar\varphi}\, \phi - \frac{2}{3} \big[(\sq \phi)^{^{\dt{\, }}} - h\,\sq\phi\big] + 4H^2\, (\dt{\phi} - h\, \phi)
\label{Vterm}
\eea
The energy density can be obtained from this component by using the conservation equation
$\nabla_a T^{(1)\,au} = 0$ with $T^{(1)\,ab} = b' E^{(1)\,ab}$, or
\be
H\, \frac{\partial \varepsilon^{(1)}\!\!\!}{\partial u\,}\  +\, 3h \left(\epsilon^{(1)} + p^{(1)}\right) = b'\, \frac{\Delta_3}{a^2} \,V^{(1)}
\label{gencons}
\ee
together with tracelessness
\be
p^{(1)} = \tfrac{1}{3}\,\varepsilon^{(1)}
\label{notrace}
\ee
so that
\be
\varepsilon^{(1)}_{klm_l} = \frac{b'}{Ha^4} \int^u_{-\infty} du\, a^2 \Delta_3\, V_{klm_l}^{(1)}
= - b' \,\frac{(k^2 -1)}{Ha^4} \int^u_{-\infty} du\, a^2\,  V_{klm_l}^{(1)}
\label{Vk}
\ee
for the linearized energy density perturbation in a given $(k,l,m_l)$ mode.

Substituting (\ref{genphi}) and (\ref{barphi}) into (\ref{Vterm}), we obtain in particular the contributions
\bea
&&\hspace{-2cm}-\frac{2}{3}  \dt{\bar\varphi}\,\ddt{\phi} = \frac{2H^3}{3} k^2\, \sech^2u\, \tanh u\, (1 \mp 2\,\sech^2u) \times\nn
&&\left[\frac{a_{klm_l}}{\sqrt{2}\,k}\,e^{-i(k-1)\eta}\ Y_{klm_l}+ \frac{b_{klm_l}}{\sqrt{2}\,k}\,e^{-i(k+1)\eta}\ Y_{klm_l} + c.c.\right] + \dots
\label{phiddot}
\eea
and
\be
-\frac{2}{3} (\sq \phi)^{^{\dt{\, }}} = \frac{2H^3}{3} k^2\, \sech^2 u\, \left[-i\, \frac{a_{klm_l}}{\sqrt{2}\,k}\,e^{-ik \eta}\ Y_{klm_l}
+ i\,\frac{b_{klm_l}}{\sqrt{2}\,k}\,e^{-ik\eta}\ Y_{klm_l} + c.c.\right] + \dots
\label{boxphidot}
\ee
both of which are leading in $k$, where the ellipsis and all other terms in (\ref{Vterm}) are subleading in $k$ for $k\gg1$.
Since these terms are linear in $k$ for large $k$, and because of the additional factor of $k^2$
from (\ref{Vk}), these leading terms in $k$ in $\varepsilon_k^{(1)}$ are proportional
to $k^3$. Taking into account the time dependence next, we observe that since $e^{\pm i\eta} = \sech u \pm i \tanh u
\rightarrow \mp i$ as $u\rightarrow -\infty$, the leading $\sech^2 u$ behavior of $V^{(1)}_{klm_l}$ cancels in the sum
of (\ref{phiddot}) and (\ref{boxphidot}), so that the integrand of (\ref{Vk}) vanishes at its lower limit, making the
integral convergent. The surviving subleading term then gives a contribution to (\ref{Vk}) of
\bea
\varepsilon_{klm_l}^{(1)} &\simeq &- \frac{b'k^3}{a^4}\frac{\sqrt{2}}{3}\int^u_{-\infty} du\, \sech u\tanh u
\left[ (a_{klm_l} + b_{klm_l}) \,e^{-ik \eta}\ Y_{klm_l}  + c.c.\right]\nn
&\simeq & \frac{b'\sqrt{2}}{3}\, H^4\, k^3\,\sech^5 u \left[ (a_{klm_l} + b_{klm_l}) \,i^k\ Y_{klm_l}  + c.c.\right]
\label{eps1asymp}
\eea
as $u \rightarrow -\infty$. The integral in (\ref{eps1asymp}) can be computed exactly for $u=0$ 
with the result
\be
\varepsilon_{klm_l}^{(1)}\Big\vert_{u=0} \rightarrow - b' H^4\frac{\sqrt{2}}{3} \, k^2
\left[ (a_{klm_l} + b_{klm_l}) \,i^{k+1}\ Y_{klm_l}  + c.c.\right]
\label{eps1at0}
\ee
for $k \gg 1$. Since the contribution of this $O(4)$ breaking leading term in $k$ to the total linearized energy density is
\vspace{-5mm}
\be
\varepsilon^{(1)} = \sum_{k=1}^{\infty}\sum_{l=0}^{k-1} \sum_{m_l=-l}^l \, \varepsilon_{klm_l}^{(1)}
\label{ksum}
\ee
it falls off proportional to $a^{-5}$ from (\ref{eps1asymp}) and hence a factor of $a^{-1}$ faster than the
$O(4)$ symmetric terms in (\ref{Euuanom}) as $u\rightarrow -\infty$. From (\ref{eps1at0}) at $u=0$ its maximum
value grows with the maximum momentum $K$ for which the coefficients $a_{klm_l}$ or $b_{klm_l}$ are non-zero proportionally to
\be
\varepsilon^{(1)}\Big\vert_{u=0} \sim -b' H^4\sum_{k=1}^K\sum_{l=0}^{k-1} \sum_{m_l=-l}^l  k^2
\sim -b' H^4\int_{k=1}^K dk\, k^4 \simeq  -b' H^4 \frac{K^5}{5}
\label{sumapprox}
\ee
and hence can easily exceed (\ref{Euuanom}) at the symmetric point $u=0$ if $K \gg 1$.
The backreaction of these $O(4)$ breaking terms in the stress tensor becomes significant when
\be
8 \pi GH^2 \,\vert b'\vert\, \frac{K^5}{5} \, \big\vert a_{Klm_l}\vert
\sim 8 \pi GH^2\,\vert b' \vert\, \frac{K^5}{5} \, \big\vert b_{Klm_l}\vert  \gtrsim\, 1
\label{abcond}
\ee
which is ever easier to satisfy for a larger range of state coefficients as $K$ is increased.
Thus these $k$ dependent $O(4)$ breaking terms begin smaller and for large enough $K$
grow larger to dominate the $a^{-4}$ de Sitter breaking $O(4)$ symmetric terms (\ref{Euuanom})
in the stress tensor as $a$ decreases from infinity at $I_-$.

We conclude that the general $O(4)$ invariant `vacuum'  states $|f\rag$ defined by (\ref{gensoln})-(\ref{fstate})
are dynamically unstable to producing large deviations in the stress tensor, even more so than
the $O(4,1)$ invariant $\vert \ups\rag$, the larger the $k$ of the solution of (\ref{genphi}) considered.
Hence the $O(4)$ symmetry subgroup and spatial homogeneity is also spontaneously broken in `eternal'
de Sitter space. This conclusion which follows from the stress tensor of the anomaly could also be obtained by
calculating the expectation value $\lag T_{ab}\rag_R$ of the underlying QFT in $O(4)$ non-invariant
states in de Sitter space.

\section{Conclusions}
\label{Sec:Concl}

The main conclusion of our analysis of possible states in both de Sitter space and the
example of a constant, uniform electric field is that the most symmetric state in such persistent
background fields is {\it not} the stable vacuum state. Unlike flat Minkowski space
where the Poincar\'e invariant vacuum is determined by a physical minimization of
energy, no such conserved Hamiltonian bounded from below exists in either de Sitter
space or in a constant, uniform electric field. Instead both of these systems are characterized
by a mixing of particle and anti-particle modes with respect to any proposed Hamiltonian
generator, and are therefore unstable to spontaneous particle pair creation from the
vacuum \cite{PartCreatdS,AndMot1}. In each case the persistent or `eternal' background
classical field provides an inexhaustible supply of energy to create pairs at a finite
rate and subsequently accelerate them to ultrarelativistic particle energies. In this situation
one should expect the symmetric state to be unstable to perturbations and capable of
generating large backreaction effects, even in a semiclassical mean field approximation
in which particle self-interactions are neglected. The study of particle creation in real time
and the resulting vacuum decay rate given in an accompanying publication \cite{AndMot1}
is perhaps the clearest path to the instability of `eternal'  de Sitter space.

In this work we have provided two additional approaches to an analysis of the instability.
These are both based not on any particular definition of particles but on the study of perturbations
of the maximally symmetric $|\ups\rag$ states and the conserved currents they produce. In the
electric field background this state is constructed in Sec.\,\ref{Sec:ConstantE} and is just as well
a self-consistent solution of the semiclassical Maxwell eqs. (\ref{Max}) as the $O(4,1)$ CTBD
state is a self-consistent solution of the semiclassical Einstein eqs. (\ref{scE}), with a shifted
cosmological `constant.' In each case there are large classes of perturbations of the symmetric
state that produce an electric current $\lag j_z\rag$ or stress tensor $\lag T_{ab}\rag$ that are initially
zero or negligibly small, but which grow larger than any prescribed finite value. In each case
this is due to the blueshifting of field modes to ultrarelativistic energies.
In de Sitter space this occurs clearly in the contracting phase $u < 0$, and requires that backreaction of the
energy-momentum through the semiclassical Einstein eqs. (\ref{scE}) be taken into account.
Hence the assumption of a fixed de Sitter background is violated, `eternal' de Sitter space is
unstable to perturbations satisfying (\ref{bound}), which produce large backreaction
effects, and the classical $O(4,1)$ symmetry is broken by quantum fluctuations.

The second approach to instability of the maximally symmetric state in both the electric
field and de Sitter backgrounds is through the relation to a quantum anomaly. The
chiral anomaly and bosonization method in the 2D Schwinger model shows that
the invariance of the electric field background is broken by quantum effects.
In the approximation of a fixed background field the solutions of the anomaly eq.
(\ref{anom2d}) for the effective boson field which are spatially homogeneous
predict the linear growth with time (\ref{Schj}), found also by direct study of the perturbations
of the symmetric state. Even at the level of the effective action (\ref{2danomact}),
the appearance of the Green's function $\sq^{-1}$ of the 2D wave operator makes
it clear that there will be infrared sensitivity to boundary and/or initial conditions
associated with the anomaly.

In the gravitational case it is the conformal trace anomaly which produces long lived
infrared effects sensitive to either boundary or initial conditions. It is important that
the kinematics of the persistent de Sitter background will always produce
ultrarelatvistic energies for large enough $k$ so that the stress tensor eventually
behaves like that of a massless conformal field which is described by the stress
tensor of the anomaly. From the non-local form (\ref{Sanom}) and the infrared properties
of the conformal operator $\Delta_4$ and its inverse, it is already clear without detailed
calculation that de Sitter invariance is broken. As in the 2D chiral anomaly one can
introduce a composite effective bosonic field $\varphi$ whose eq. of motion (\ref{phieq})
has a constant source and therefore possesses no de Sitter invariant solutions. Since
the eq. (\ref{phieq}) is de Sitter invariant but none of its solutions are, the anomaly
provides a mechanism for spontaneous breaking of de Sitter symmetry \cite{MazMotdS}.
The behavior of the $O(4)$ symmetric solutions which break de Sitter invariance is easily
found and the same conclusion of the instability of global de Sitter space to these state
perturbations follows. The anomaly approach is quite general and shows that the same
large backreaction effect and instability to initial state perturbations occurs in de Sitter space
for fields of any spin.

In both cases it is essential that moderate or small physical momenta are blueshifted
to very large physical momenta, arbitrarily large if backreaction is turned off  and the background
field persists indefinitely. This unboundedness and relation to anomalies is a direct consequence
of the infinite reservoir of arbitrarily high momentum or short distance modes in any vacuum state
of QFT with no UV cutoff. The physical momentum which first produces large backreaction effects 
is of order $\sqrt{HM_{Pl}}$. There is thus an interesting interplay of UV and IR physics in these 
effects, as has been noticed by other authors \cite{PolydS,KroPoly}.

The instability of the de Sitter invariant state to large backreaction effects shows that
the $O(4,1)$ symmetry of global de Sitter space is  broken by quantum perturbations of
the state. One can construct fully $O(4,1)$ invariant theories in eternal fixed de Sitter
spacetime mathematically by continuation from the Euclidean ${\mathbb S}^4$, order by
order in perturbation theory \cite{MarMor}. By this construction the very state perturbations
responsible for the instability of de Sitter space in real time are disallowed by the Euclidean
regularity conditions. If one requires these regularity conditions explicitly or implicitly by analytic
continuation from ${\mathbb S}^4$, and fixes the geometry to be de Sitter exactly,
also disallowing the possibility of dynamical backreaction through the semiclassical
Einstein eqs., it is not surprising then to find no sign of the instability we have discussed
in this paper, which makes neither of those assumptions. This also shows that it is not
matter self-interactions {\it per se} which are critical for the instability, but rather the boundary
conditions on states and Green's functions used. If Euclidean boundary conditions are
required, interactions lead to no apparent instability \cite{MarMor}. Conversely, instability
is seen immediately those restricted boundary/initial conditions are relaxed (even in
free QFT) and the backreaction effects of the energy-momentum tensor on the background
de Sitter geometry are considered.

The anomaly stress tensor shows further that there are also spatially inhomogeneous
perturbations of the initial state which break $O(4)$ symmetry and which vanish more
rapidly in the infinite past and grow to larger values at later times than the
$O(4)$ symmetric ones. This shows that there is no $O(4)$ invariant stable vacuum state
in de global de Sitter space either. In the particle creation language this is quite
natural since localized particles and fluctuations about the mean are always
spatially inhomogeneous. It will be necessary to study these fluctuations and
spatial inhomogeneities in order to describe accurately the dissipation of
either electric field or cosmological vacuum energy into particle modes.

The spatially inhomogeneous perturbations are particularly relevant for inflation
and cosmological models more generally, since they will occur even in the
monotonically expanding background of de Sitter space in the flat Poincar\'e
sections (\ref{flatFRW}). Indeed because of the homogeneity of classical
de Sitter space all points on the de Sitter manifold are locally equivalent,
so that particle creation and the spatially inhomogeneous fluctuations
in the geometry they cause are not limited to the global hyperboloid
coordinates (\ref{hypermet}). In fact, the same anomaly stress tensor
evaluated in the static coordinates (\ref{static}) of de Sitter space shows
that it also describes sensitivity to boundary conditions on the cosmological
horizon $r=H^{-1}$ which is entirely contained within the flat Poincar\'e
patch \cite{DSAnom}. These new cosmological horizon modes associated
with the anomaly scalar effective field and stress tensor have potentially large
effects on the de Sitter horizon. Their fluctuations are capable of generating
and describing the anisotropies in the CMB \cite{AMMCMB}. This suggests
that spatial inhomogeneous models of cosmological dark energy within a
Hubble horizon volume which have only rotational $O(3)$ symmetry are
relevant and may be required for a resolution of the problem of cosmological 
dark energy in the present universe.

\centerline{\large{\bf Acknowledgements}}
P. R. A. would like to thank Dillon Sanders for help with the early stages of this project.
This work was supported in part by the National Science Foundation under Grant Nos. PHY-0856050
and PHY-1308325.  The numerical computations herein were performed on the WFU DEAC cluster;
we thank WFU's Provost Office and Information Systems Department for their generous support.

\end{document}